\documentclass[acmtog,screen,nonacm]{acmart}

\AtBeginDocument{%
  }

\setcopyright{acmlicensed}
\copyrightyear{2018}
\acmYear{2018}
\acmDOI{XXXXXXX.XXXXXXX}

\acmConference[Conference acronym 'XX]{Make sure to enter the correct
  conference title from your rights confirmation emai}{June 03--05,
  2018}{Woodstock, NY}
\acmISBN{978-1-4503-XXXX-X/18/06}

\DeclareMathOperator{\Tr}{Tr}

\usepackage{derivative}
\usepackage{algorithm}
\usepackage{algorithmic}
\DeclareMathOperator*{\argmin}{arg\,min}
\usepackage[per-mode = symbol]{siunitx}

\begin{document}

\title{Energy-Controllable Time Integration for Elastodynamic Contact}




\author{Kevin You}
\affiliation{%
  \institution{Carnegie Mellon University}
  \country{USA}
}
\affiliation{%
  \institution{New York University}
  \country{USA}
}
\email{ky2765@nyu.edu}

\author{Juntian Zheng}
\affiliation{%
  \institution{Carnegie Mellon University}
  \country{USA}
}
\email{juntianz@andrew.cmu.edu}

\author{Minchen Li}
\affiliation{%
  \institution{Carnegie Mellon University}
  \country{USA}
}
\affiliation{%
  \institution{Genesis AI}
  \country{USA}
}
\email{minchernl@gmail.com}







\newcommand{\R}{\mathbb{R}}
\newcommand{\Q}{\mathbb{Q}}
\newcommand{\Z}{\mathbb{Z}}
\renewcommand{\C}{\mathbb{C}}
\newcommand{\Ec}{\mathcal{E}}
\newcommand{\vphi}{\varphi}
\newcommand{\vt}{\vert}
\newcommand{\Vt}{\Vert}
\newcommand{\vp}{\mathbf{p}}
\newcommand{\vq}{\mathbf{q}}
\newcommand{\vw}{\mathbf{w}}
\newcommand{\vx}{\mathbf{x}}
\newcommand{\vz}{\mathbf{z}}
\renewcommand{\vv}{\mathbf{v}}
\newcommand{\vu}{\mathbf{u}}
\newcommand{\vm}{\mathbf{m}}
\newcommand{\vk}{\mathbf{k}}
\newcommand{\vell}{\mathbf{\ell}}
\newcommand{\vb}[1]{\mathbf{#1}}
\newcommand{\RE}{\mathrm{Re}}
\newcommand{\IM}{\mathrm{Im}}

\begin{abstract}
Dynamic simulation of elastic bodies is a longstanding task in engineering and computer graphics. 
In graphics, numerical integrators like implicit Euler and BDF2 are preferred due to their stability at large time steps, but they tend to dissipate energy uncontrollably.
In contrast, symplectic methods like implicit midpoint can conserve energy but are not unconditionally stable and fail on moderately stiff problems.
To address these limitations, we propose a general class of numerical integrators for Hamiltonian problems which are symplectic on linear problems,
yet have superior stability on nonlinear problems.
With this, we derive a novel energy-controllable time integrator, 
A-search, a simple modification of implicit Euler that can follow user-specified energy targets, 
enabling flexible control over energy dissipation or conservation while maintaining stability and physical fidelity.
Our method integrates seamlessly with barrier-type energies and allows for inversion-free and penetration-free guarantees, 
making it well-suited for handling large deformations and complex collisions. 
Extensive evaluations over a wide range of material parameters and scenes demonstrate that A-search has biases to keep energy in low frequency motion rather than dissipation, and A-search outperforms traditional methods such as BDF2 at similar total running times by maintaining energy and leading to more visually desirable simulations.
\end{abstract}


\begin{teaserfigure}
  \includegraphics[width=\textwidth]{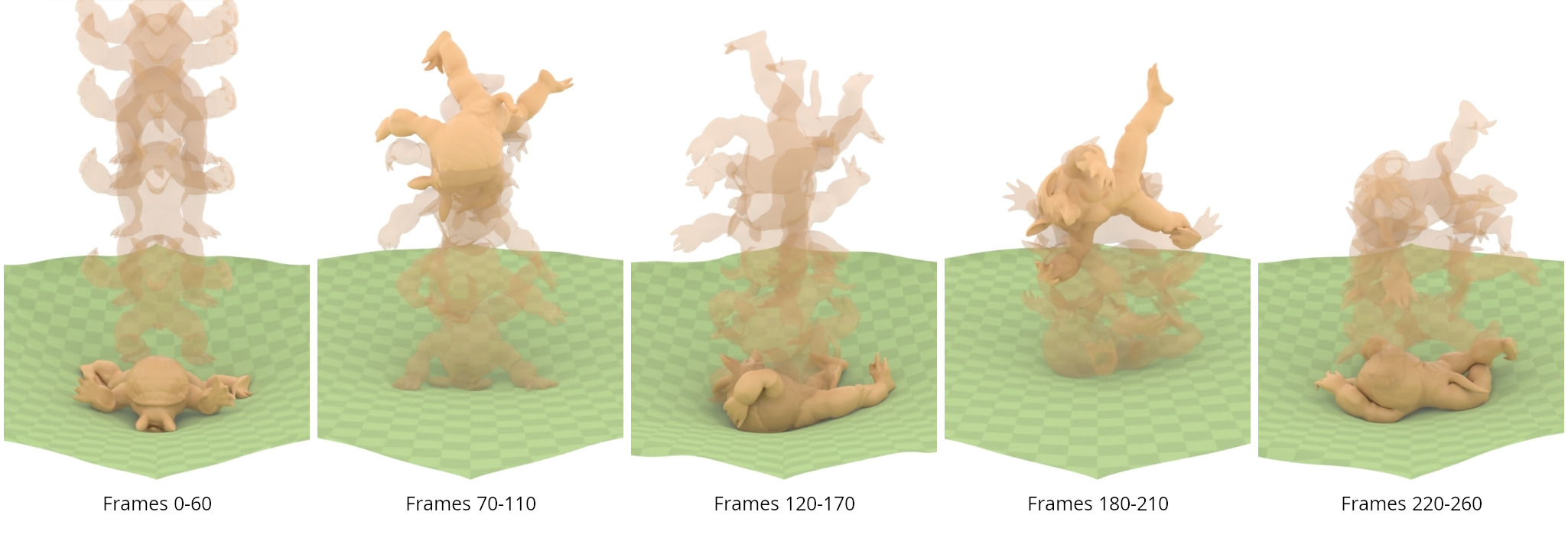}
  \caption{\textbf{Armadillo trampoline.} Soft elastic armadillo dropped on elastic thin shell trampoline. Every 10th frame is shown partially transparent,
  the last frame in each group is solid. 
  Our time integrator enables energetic and controllable simulation of elastodynamics with contact.}
  \label{fig:teaser}
\end{teaserfigure}


\maketitle

\section{Introduction}

Physics-based animation of deformable objects is numerically challenging due to a fundamental mismatch between continuum elasticity and the temporal and spatial resolutions used in computer graphics. Elasticity is governed by hyperbolic equations, implying wave propagation at characteristic speeds that, for most real materials, far exceed what can be resolved at frame-scale time steps. In the nonlinear regime, elastic systems further exhibit a transfer of energy toward increasingly high frequencies, leading to steep gradients and possible shock formation. As a result, stable simulation at practical resolutions necessarily relies on regularization, and commonly used discretizations combined with implicit integration define a dissipative surrogate of the underlying continuum dynamics.

In graphics applications involving contact and large deformation, one is typically interested in coarse-scale motion and settling behavior rather than accurate elastic wave propagation. High-frequency motion is strongly attenuated in real materials due to viscous and frictional effects, and its detailed dynamics are both difficult to model and largely irrelevant to the desired macroscopic behavior. Consequently, suppressing unresolved high-frequency dynamics is consistent with both physical intuition and animation goals.

At the same time, elastic objects are discretized using meshes that are fine relative to the deformations of interest, since deformation and contact locations are not known a priori and sharp geometric features require sufficient resolution. This introduces many degrees of freedom associated with high-frequency modes that are numerically stiff but perceptually unimportant. Implicit time integration methods, most notably implicit Euler, are therefore widely adopted in graphics as an effective form of regularization: they provide robust stability at large time steps while preferentially damping high-frequency motion and driving systems toward equilibrium,
mimicking viscous effects,
albeit in an uncontrollable manner.

However, physics-based animation often seeks to retain more dynamic behavior than that produced by strongly dissipative integrators. Rather than eliminating all transient motion, the goal is to limit energy transfer toward unresolved modes while preserving controllable dynamics in the low-frequency degrees of freedom. Reduced subspace methods partially address this objective. But ultimately, regardless of the spatial discretization, the time integrator plays a central role: a suitable method must offer stability comparable to implicit Euler while enabling explicit control over energy dissipation and modal bias.

To elaborate on time integration, upon spatial discretization, 
the dynamics of elastic bodies are governed by a system of nonlinear first-order differential equations.
These equations are stiff, 
which makes many numerical methods unstable. 
Explicit methods, such as the explicit Euler or explicit Runge-Kutta methods,
are especially prone to numerical explosions.
Symplectic methods, such as symplectic Euler, implicit midpoint, or Newmark-beta methods,
while better, also suffer from numerical instabilities.
Instead,
implicit methods are commonly used in graphics due to their ability to handle large time steps stably. 
While implicit methods like implicit Euler and 2-step backward differentiation formula (BDF2) are almost always stable, 
they tend to dissipate energy, causing the scene to lose motion over time, and eventually come to rest. 
This property is desirable in many cases in physics-based simulation and animation.
However, the amount of dissipation induced by the implicit method depends not just on the material or model parameters, 
but rather on the time step.
Selecting an appropriate time step becomes crucial,
small enough to preserve motion, yet not so small as to waste computational resources. 
This selection relies on heuristics,
and it is difficult to exercise manual control over the amount of dissipation.

In this paper, we propose a novel numerical integrator that allows control of system energy, 
enabling the simulation of dynamic scenes over extended durations. 
To achieve this, we first introduce a new general class of time integrators,
the \textit{decoupled symplectic methods},
which are \textit{linearly symplectic} yet more stable than alternative symplectic methods.
We also introduce a mechanism of interpolation,
leading to the \textit{decoupled $\alpha$-methods},
which can lead to energy conservation while maintaining momentum conservation and feasibility of positional constraints.
From this general class, we pick out the \textit{A-1} and \textit{A-search} methods,
which are simple and can be implemented as minor modifications of the implicit Euler method.

We prove that \textit{A-1} is \textit{linearly symplectic} and has unconditional linear stability,
and also demonstrate that \textit{A-1} is significantly more stable than other symplectic methods like implicit midpoint or trapezoidal in the presence of nonlinearity.
From \textit{A-1}, we derive \textit{A-search}, 
an integrator that adaptively interpolates velocities between \textit{A-1} and implicit Euler by dynamically selecting quadrature points for force evaluation. 
\textit{A-search} allows flexible user control over energy dissipation, 
between perfect energy conservation and the dissipation behavior of implicit Euler,
without modifying the time step.
Unlike existing energy-conserving methods based on solution interpolation (e.g., \citet{dinev2018stabilizing}), \textit{A-search} maintains inversion-free and interpenetration-free guarantees.
We also show how our framework supports friction and other dissipative forces.

In our experiments, we demonstrate mentioned features of our method using a comprehensive benchmark formed by examples involving soft and stiff materials with large deformations or collisions.
In these scenes, energy conservation is critical for realistic behavior. 
Our results demonstrate that \textit{A-search} significantly outperforms implicit Euler and BDF2 under equivalent time steps and surpasses BDF2 at comparable total runtime in terms of energy conservation. Moreover, our method eliminates the need for manual time step tuning to achieve desired energy dissipation profiles. 

We quantitatively confirm that at large time steps, our method naturally suppresses high-frequency oscillations common in soft materials, and instead channeling energy into low-frequency motion and large deformations, which is an advantageous feature for animation purposes.
We also discuss some potential issues regarding interpolation and noise neglected by previous methods, and show that our method is able to avoid these issues.

In summary, we present a novel energy-controllable time integrator that excels in complex dynamic scenarios, providing stable and physically realistic animations without the need for meticulous time step adjustments.

%



\section{Related Works}
Since the pioneering work of \citet{terzopoulos1987elastically}, elastic object simulation has become a major research topic in computer animation. Efforts in this area have largely focused on fast numerical solvers \cite{macklin2016xpbd,bouaziz2023projective,overby2017admm,liu2017quasi,lan2023second,chen2024vertex}, material modeling \cite{romero2021physical,smith2018stable,li2015stable,grinspun2003discrete,martin2011example}, collision handling \cite{Li2020IPC,verschoor2019efficient,harmon2009asynchronous,muller2015air,kaufman2008staggered}, and coupled simulations with rigid bodies and fluids \cite{chen2022unified,xie2023contact,takahashi2022elastomonolith,robinson2008two} (for comprehensive reviews, see \citet{kim2022dynamic,sifakis2012fem}). However, relatively fewer studies have explored advancements in time integration methods, which will be the focus of our review in this section.

\paragraph{Time Integration}
Common numerical integrators for first-order initial value problems are often categorized into two classical families: Runge–Kutta methods and linear multistep methods, and they both can be viewed as special cases of the general linear methods \cite{hairer1}.
The former involves taking intermediate sub-steps within a time step,
while the later involves using information from multiple previous steps.
Among the two families there exists both explicit and implicit methods.
For explicit methods, the new estimate of the solution can be computed directly in terms of function evaluation of previous estimates.
However, for implicit methods the new estimate requires solving system of equations.
For this reason, explicit methods are much faster to compute.
However, their stability is conditional, with admissible time step sizes tightly constrained by the dynamics of the system.
Many partial differential equations in physics-based simulation,
including elasticity, are hyperbolic,
and Courant–Friedrichs–Lewy (CFL) conditions \cite{courant1928partiellen} limit the time step size for explicit methods 
in terms of the mesh size and wave speed of the material.

Within the field of computer graphics, and specifically for elastic objects, 
Baraff and Witkin \cite{10.1145/280814.280821} pioneered the use of the implicit Euler method.
Its robustness, nearly guaranteed stability, and arguably its dissipative behavior,
has attracted its use for animation.
The two-step BDF2 generalizes implicit Euler to second-order and preserves dynamics much better than implicit Euler,
and since introduced in the graphics community \cite{10.1145/566654.566624},
has become a popular replacement of implicit Euler \cite{chen2022unified,loschner2020higher}.
Higher-order implicit Runge-Kutta methods have received less attention in the field of computer animation,
again due to the priority of computational costs and stability over accuracy for practical simulations.
\citet{loschner2020higher} and \citet{ascher} compared various diagonally implicit Runge Kutta (DIRK) methods,
including singly diagonally implicit Runge Kutta (SDIRK) methods 
and the explicit first stage TR-BDF2 method.
Among these, second-order methods such as SDIRK2 \cite{alexander1977diagonally} and TR-BDF2 have been shown to offer improved accuracy and reduced dissipation compared to implicit Euler. 
However, it is also mentioned that these DIRK methods are not as good as BDF2 for stiff problems.
Additionally, \citet{10.1145/3730872} suggested that,
in the context of rigid body dynamics,
the single-step TR-BDF2 is worse than the two-step BDF2 method.

Taking a different direction, \citet{michels2014exponential} introduced exponential integrators to the graphics community for simulating corotated linear elastic materials, leveraging the analytic solution of first-order ODE systems via matrix exponentials. This approach was later extended to support nonlinear materials by \citet{michels2017stiffly} and \citet{chen2017exponential}, who applied numerical quadrature to approximate the integral of nonlinear terms. Building on these ideas, \citet{chen2020siere} proposed a hybrid scheme that combines semi-implicit Euler with an exponential integrator, applying them to low- and high-frequency modes respectively to improve both scalability and stability.

\begin{figure*}[h]
  \centering
  \includegraphics[width=0.49\linewidth]{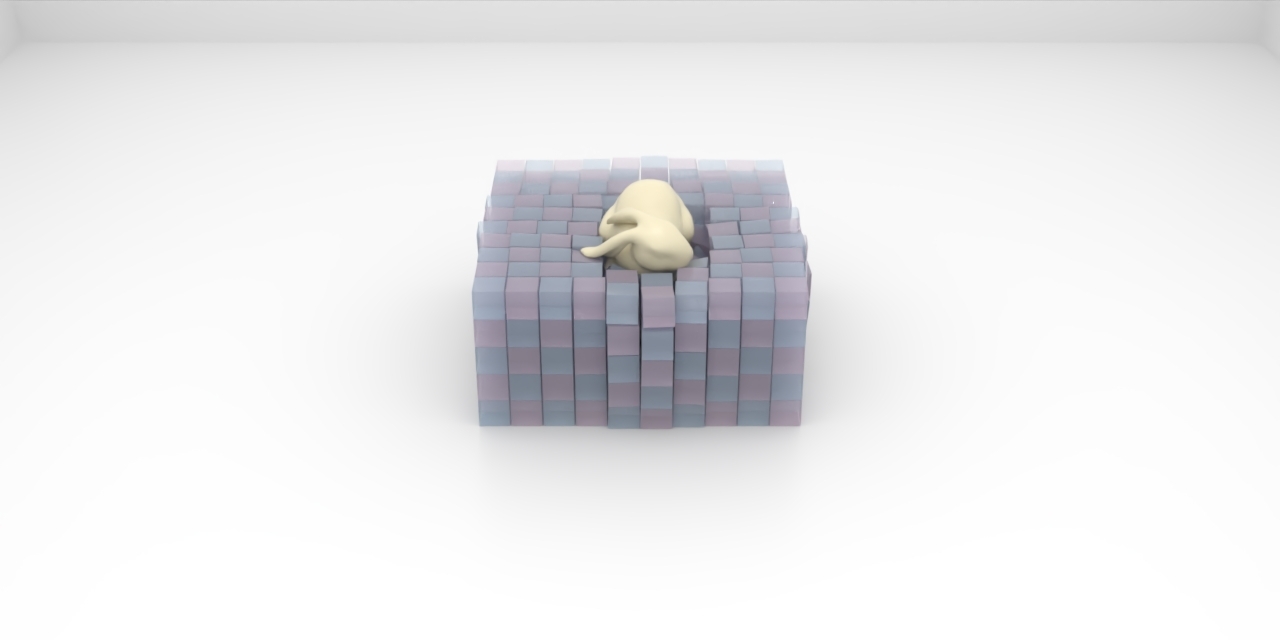}
  \includegraphics[width=0.49\linewidth]{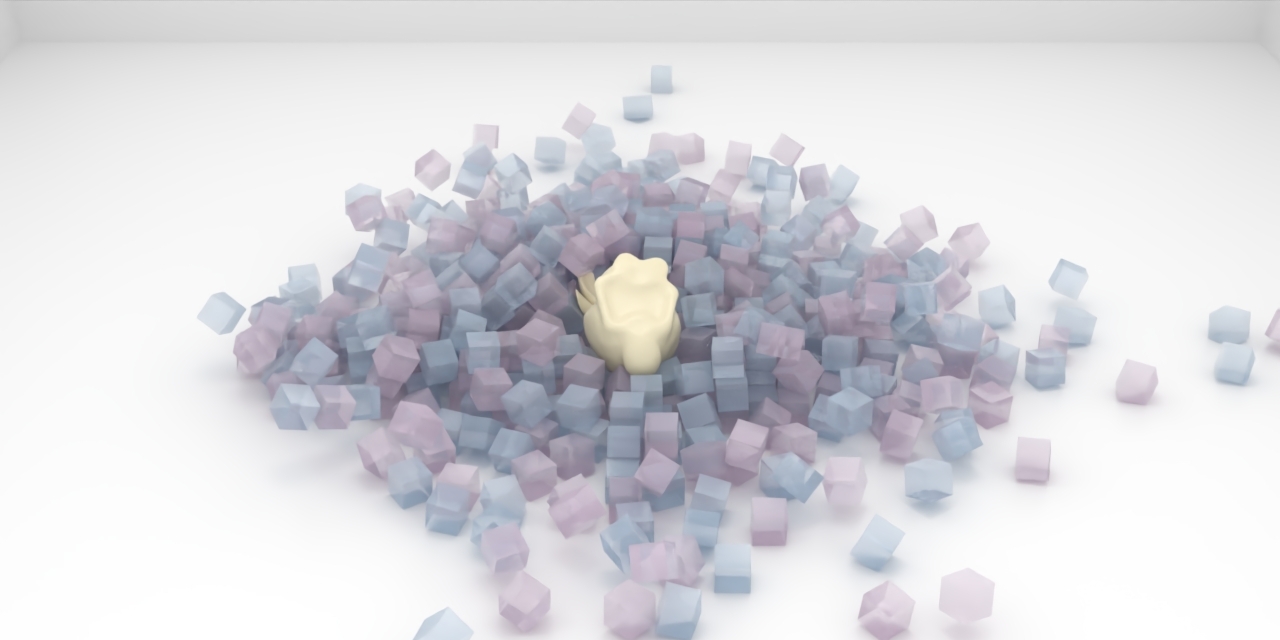}
  \includegraphics[width=0.49\linewidth]{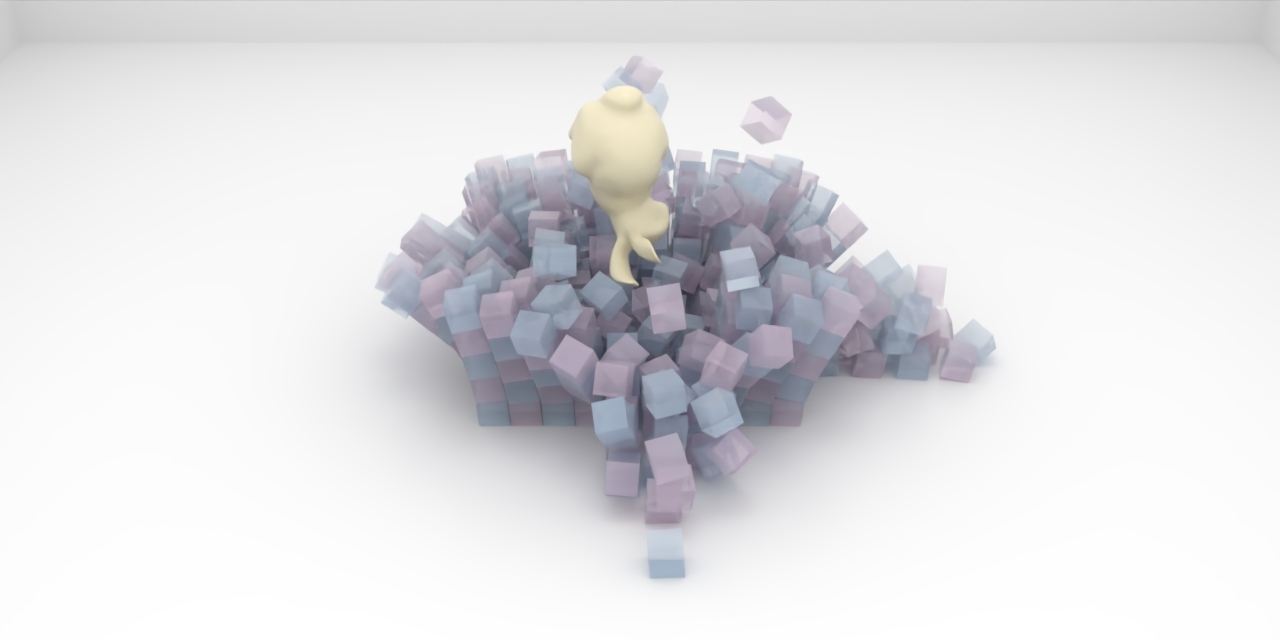}
  \includegraphics[width=0.49\linewidth]{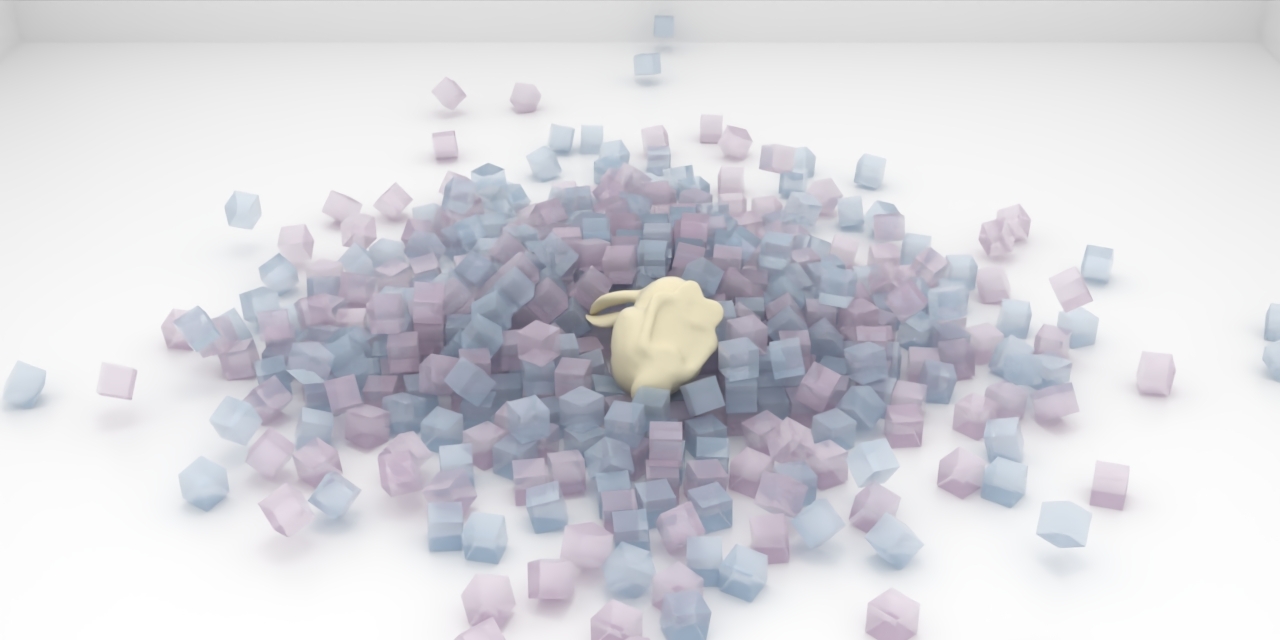}
  \includegraphics[width=0.49\linewidth]{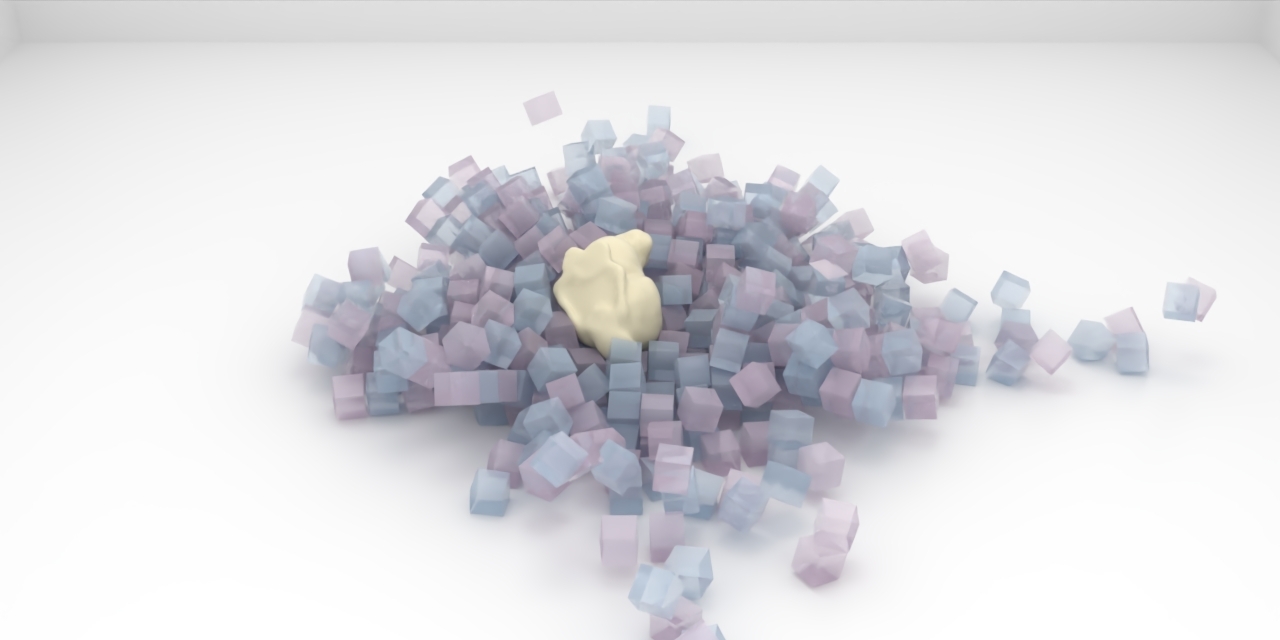}
  \includegraphics[width=0.49\linewidth]{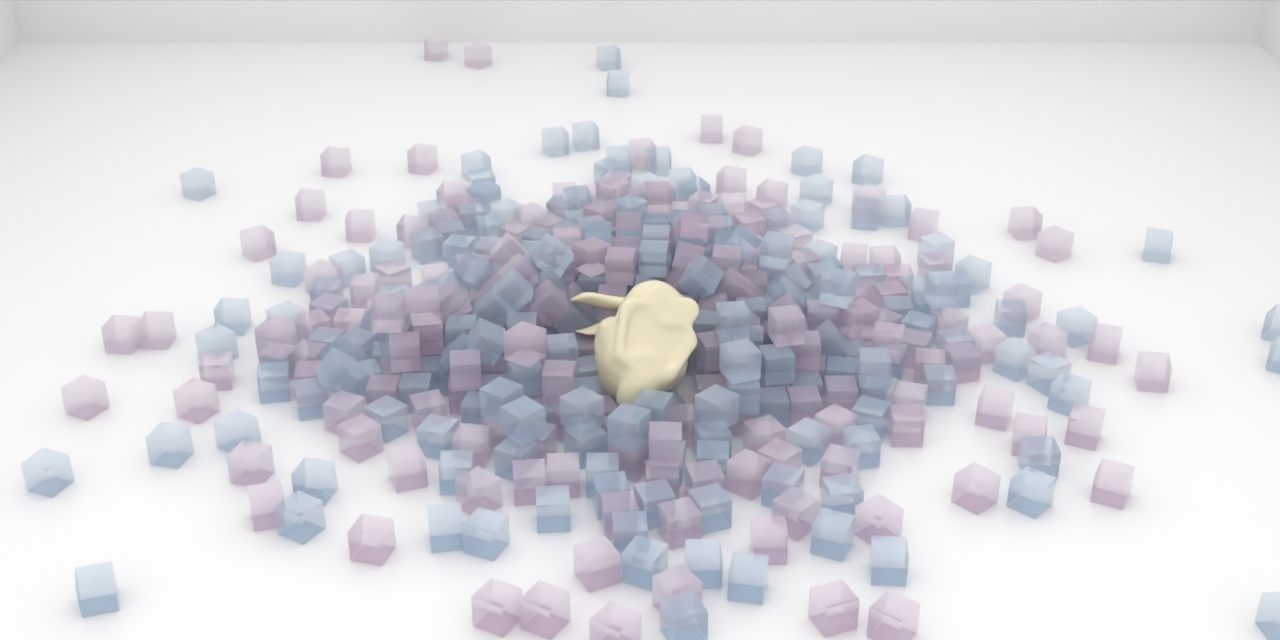}
  \caption{\textbf{Bunny splash.} Soft elastic bunny dropped on soft elastic cubes with friction. Static friction holds columns of cubes together when falling. A-search with decay gives dynamical results while respecting static and dynamic frictional contact. }
  \label{fig:splash}
\end{figure*}

\paragraph{Energy Conservation}
To achieve improved energy behavior, symplectic time integration methods have been widely studied in the numerical analysis literature. Symplectic schemes, such as symplectic Euler, implicit midpoint, and Newmark-type methods, are designed to preserve geometric structure and exhibit favorable long-term energy properties for Hamiltonian systems. However, their applicability in physics-based animation is limited. Explicit symplectic methods remain subject to conditional stability and severe time step restrictions, while implicit symplectic schemes are only guaranteed to be stable for linear or mildly nonlinear systems. In the context of nonlinear elasticity, these methods can still suffer from numerical instabilities and, more importantly, faithfully propagate energy toward high-frequency modes. While such behavior is consistent with the underlying hyperbolic dynamics, it is generally undesirable in graphics applications, where unresolved high-frequency motion leads to visual noise and numerical stiffness.

One of the most immediate alternatives is to explicitly enforce energy conservation by formulating time integration as a constrained optimization problem with Lagrange multipliers \cite{10.1115/1.3424303}. However, such constraints effectively introduce additional forces and do not necessarily yield physically meaningful dynamics. Other approaches project the system state onto energy- and momentum-conserving configurations \cite{dinev2018FEPR}, but the correctness and convergence properties of these projections are difficult to establish in the presence of strong nonlinearity and contact.
 
Another idea is to limit the class of physical materials and utilize clever quadrature point selection. 
\citet{rojas2018average} expresses the update rule as an integral equation involving the forces that conserves energy perfectly.
The integrals can be evaluated exactly via Simpson's rule, if the force is a degree-3 polynomial, such as that of St. Venant-Kirchhoff materials.
Of course, this method is limited to specific materials and cannot handle collisions.

Finally, \citet{dinev2018stabilizing} proposed such a mechanism of energy-conserving time integration, by interpolating between explicit and implicit numerical schemes and choosing the blending that preserves energy, and this has shown decent results. 
By interpolating between two convergent numerical schemes,
the resulting method is convergent, and conserves momentum exactly.
However, this method keeps the high frequency artifacts induced by implicit midpoint, 
and by interpolation it can violate constraints such as nonpenetration and noninversion. 

Our method is similar to the interpolation approach proposed by \citet{dinev2018stabilizing}, 
but it differs by modifying only the velocities while preserving the implicit solution for the positions, thereby naturally supporting positional constraints.
We demonstrate that our method can be interpreted as utilizing adaptive quadrature to enable energy controls.
Our method builds on first-order methods for their simplicity and efficiency,
though we mention possible extensions to higher-order Runge-Kutta methods.

\paragraph{Adaptive Time Integration}
Adaptively redistributing computational efforts to different (sub)steps of time integration is a widely used strategy to enable stable and efficient simulation. However, strict time step size restrictions still apply, making significant speedups challenging to achieve.
The simplest form of temporal adaptivity is to integrate PDEs with adaptive time step sizes. \citet{kavetski2001adaptive,soderlind2002automatic} analyze the local truncation error of the time integration and adaptively control the time step sizes to maximize efficiency while keeping the estimated error within a specified tolerance. Similar adaptive strategies are also applied to ensure the stability of explicit time integration \cite{trias2011self}. \citet{sun2020effective,bai2022stability} propose novel and efficient ways to compute near-optimal time step sizes for stable material point method (MPM) simulation.
\citet{fang2018temporally,logg2004multi} combine temporal and spatial adaptivity and apply adaptive time step sizes for different subdomains to more efficiently simulate systems exhibiting spatially varying stiffness and dynamic behaviors. 
These strategies can be more efficient and stable than simulations with fixed time steps, but their capability is still limited by the underlying time integration scheme. Instead, we develop a new time integrator that centers adaptivity on quadrature point selection rather than adjusting time steps.

\section{Background}

Consider an initial value problem based on a vector-valued autonomous ordinary differential equation of the form $\dot{\vz} = f(\vz)$.
In the case of Lagrangian mechanics with a separable Lagrangian of form
\[ L(\vx,\vv) = \frac{1}{2} \Vt \vv \Vt_{\vm}^2 - P(\vx) \]
the equations of motion are given by the system
\[ (\dot{\vx}, \dot{\vv}) = (\vv, - \vm^{-1} \nabla P(\vx))\] 
where $P$ is the potential energy function and $\vm$ is the mass matrix. 
Hamiltonian mechanics are an alternative formulation of Lagrangian mechanics,
which instead tracks the momentum $\vp = \vm \vv$ rather than the velocity via the Legendre transform
\[ H(\vx,\vp) = L^*(\vx,\vv)= \frac{1}{2} \Vt \vp \Vt_{\vm^{-1}}^2 + P(\vx), \]
with equations of motion
\[ (\dot{\vx}, \dot{\vp})  = \left( \pdv{H}{\vp},-\pdv{H}{\vx} \right) = (\vm^{-1} \vp, -\nabla P(\vx)). \] 
Numerical methods applied to the Lagrangian and Hamiltonian formulation are equivalent,
and typically using the velocity variable is more natural,
though for some theoretical analysis we need the Hamiltonian formulation.

\subsection{Common Numerical Methods}

For an overview of numerical methods, see \citet{hairer1}.
The $\theta$-method is a single stage Runge-Kutta method of form
\begin{align} \label{eq:inner-theta}
    \vz_{n+1} = \vz_n + h f((1-\theta) \vz_n + \theta \vz_{n+1}).
\end{align}
We interpret the point $(1-\theta) \vz_n + \theta \vz_{n+1}$ as a collocation or quadrature point for numerical integration.
When $\theta = 0, 1$ or $1/2$, 
we recover the familiar explicit Euler, implicit Euler, and implicit midpoint methods, respectively,
which are commonly known to be unstable, stable but dissipative, and barely stable.
We can also move the interpolation outside $f$ to obtain,
sometimes also referred to as $\theta$-method,
\begin{align} \label{eq:outer-theta}
    \vz_{n+1} = \vz_n + h  \left( (1-\theta)f( \vz_n) + \theta f(\vz_{n+1}) \right).
\end{align}
Here, $\theta = 0, 1$ is still explicit Euler and implicit Euler,
but $\theta = 1/2$ is the Crank–Nicolson or trapezoidal method.
The outer $\theta$-method (\ref{eq:outer-theta}) is both a linear $1$-step method,
and a two stage stiffly accurate Runge-Kutta method due to the $f(\vz_{n+1})$ term.
See Appendix \ref{appendix:RK} for more details,
including a definition for Runge-Kutta methods.

We also consider multi-step methods, and relevant to us is the BDF2 method:
\begin{align} \label{eq:bdf2}
     \vz_{n+1} = \vz_n + \frac{2}{3} h f(\vz_{n+1}) + \frac{1}{3} (\vz_n - \vz_{n-1}).
\end{align}
BDF2 is a two step extension of implicit Euler.
All of the numerical integrators above are applied to $\vx$ and $\vv$ in the same manner.
However, it is possible to treat $\vx$ and $\vv$ differently. 
The most well known examples include the symplectic Euler method and Stormer-Verlet method,
which are partitioned Runge-Kutta methods.
Again, see Appendix \ref{appendix:RK} for a general definition.
The symplectic Euler method takes form
\begin{align} \label{eq:symplectic}
     \vx_{n+1} &= \vx_n + h \vv_{n+1}, \\
     \vv_{n+1} &= \vv_n - h \vm^{-1} \nabla P(\vx_{n}).
\end{align}
It applies implicit Euler to $\vx$ and explicit Euler to $\vv$ in a coupled manner,
and the resulting scheme can be computed explicitly.

\subsection{Symmetric and Symplectic Methods}

For an overview of numerical methods respecting geometric properties of the underlying system, see \citet{hairer3}.
The autonomous ordinary differential equation $\dot{\vz} = f(\vz)$ 
has a corresponding flow map $\vphi_h$ 
such that $\vphi_h(\vz(t)) = \vz(t+h)$.
The true flow map is symmetric in the sense that $\vphi_h \circ \vphi_{-h} = id$.
First order numerical methods also lead to a flow map $\Phi_h$ defined by $\Phi_h(\vz_n) = \vz_{n+1}$.
Given a method $\Phi_h$, 
its adjoint method $\Phi_h^*$ satisfies 
\[ \Phi_{h} \circ \Phi_{-h}^* = id. \]
For the inner or outer $\theta$-method, its adjoint is the $1-\theta$-method, e.g. implicit Euler is adjoint to explicit Euler.
For any Runge-Kutta method with Butcher Tabelau $(A,b)$,
its adjoint can be explicitly written as $(eb^T - A, b)$ where $e = (1,\ldots,1)$.
Symmetric methods are self-adjoint, and considered desirable since they respect the natural symmetry of the continuous system.

Hamiltonian dynamics form a special class of autonomous systems with qualitative behavior distinct from that of generic dissipative dynamics. While general autonomous systems may exhibit asymptotically stable equilibria or limit cycles, Hamiltonian systems do not admit attracting states. In one spatial dimension, the flow map $\vphi_h$ preserves phase-space area, satisfying $\det D\vphi_h = 1$. In higher dimensions, the flow preserves symplectic area more generally, and $\vphi_h$ is a symplectic transformation. Conversely, up to topological obstructions, a symplectic flow map characterizes an underlying Hamiltonian system. Additionally, the Hamiltonian $H(\vx(t), \vp(t))$ is conserved along trajectories, and motion persists in the absence of external dissipation.

Therefore, for Hamiltonian problems, other than being symmetric,
there are two more desirable properties that a numerical method can have.
One is being symplectic, preserving the phase-space area. The other is conserving the Hamiltonian.
For a Runge-Kutta method to be symplectic,
it is sufficient to have $b^TA + A^Tb = bb^T$.
However, there are not many such methods.
If a diagonally implicit Runge-Kutta method is symplectic, 
it is in fact a composition of implicit midpoint methods \cite{hairer3}.
Even the trapezoidal method is not symplectic.
This suggests that symplectic is an overly strict criteria.

More interesting is to consider partitioned Runge-Kutta methods,
which apply a different Runge-Kutta scheme $(A,b)$ to $\vx$ 
and $(C,d)$ to $\vp$ in a coupled manner.
Such a method is symplectic if $b = d$ and $b^TC + A^Td = bd^T$.
The Runge-Kutta method $(C,d)$ has been referred to as symplectic adjoint of $(A,b)$ \cite{sun2022}, analogous to the symmetric adjoint defined earlier.
The symplectic Euler method uses the implicit and explicit Euler pair,
and the Stormer-Verlet method uses the midpoint and trapezoidal pair.
The later is a special case of the Lobatto IIIA-IIIB pairs.

In the simplified scenario of linear problems, symmetric, symplectic and Hamiltonian conserving methods coincide.  
A Runge-Kutta method is symmetric when applied to any linear problem $\dot{\vz} = L \vz$
if and only if it is symplectic when applied to the quadratic Hamiltonian $P(\vx) = \frac{1}{2} \Vt x \Vt_{K}^2$ for any symmetric $K$.
Moreover, the Hamiltonian $H$ will be exactly conserved. 
But in general, the three properties are not equivalent.

\subsection{Stability}

For graphics, stability of the numerical method independent of the material parameters and time step $h$ is crucial. 
One way to measure stability is that of linear stability \cite{hairer2}.
For a linear problem $\dot{\vz} = L \vz$,
the update rule of a one-step method is linear: $\vz_{n+1} = \Phi_h \vz_n$.
Usually $\Phi_h$ is assumed to be diagonalizable,
and it suffices to examine the properties of the stability function 
$R(h \lambda) = \Phi_h$ when applied to $\dot{z} = \lambda z$ for $\lambda \in \C$.
A method is A-stable if $\vt R(z) \vt \leq 1$ for all $\RE(z) \leq 0$,
which means that the the numerical solution is bounded if the true solution bounded.
Moreover, a method is L-stable if $R(z) \rightarrow 0$ as $z \rightarrow \infty$,
so the numerical solution approaches zero in a single step as $h \rightarrow \infty$.
It is well-known that for $\theta$-methods, $\theta \geq 1/2$ is A-stable,
and only implicit Euler is L-stable.

To analyze stability of integrators for Hamiltonian dynamics,
we may consider applying a numerical integrator against the test system where $P = \frac{1}{2} kx^2$ for various $k > 0$.
For non-partitioned Runge-Kutta methods, up to rescaling this leads to the study of the stability function $R(z)$ for strictly imaginary $z$.
Here, $\Phi_h$ is interpreted as a 2x2 matrix.
We say that the method is \textit{linearly symplectic} if $\det(\Phi_h) = 1$,
corresponding to area preserving for the linear one-dimensional problem.
Moreover, we say that it is \textit{stably symplectic} if additionally eigenvalues of $\Phi_h$ have absolute value $1$,
or equivalently $\vt \Tr \Phi_h \vt \leq 2$,
thus leading to a numerical solution that is bounded and non-dissipative in the long term.
Implicit midpoint and trapezoidal methods are both symplectic and A-stable, hence stably symplectic.
On the other hand, symplectic Euler is linearly symplectic,
but stably symplectic only for small $h$, preserving energy in the long run if $h$ is sufficiently small and otherwise leading to an increase in energy.
See Appendix \ref{appendix:stability} for more details.

\begin{figure}[h]
  \centering
  \includegraphics[width=1.0\linewidth]{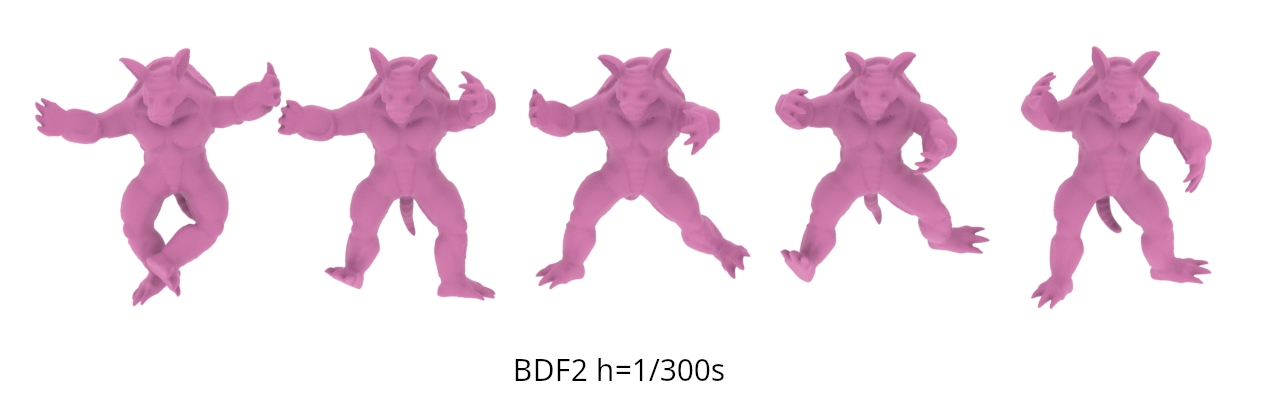}
  \includegraphics[width=1.0\linewidth]{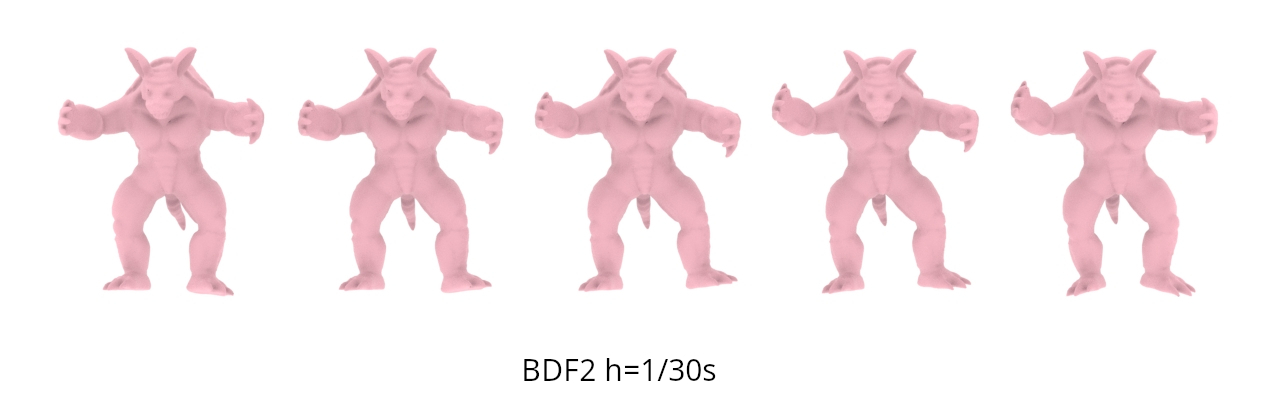}
  \includegraphics[width=1.0\linewidth]{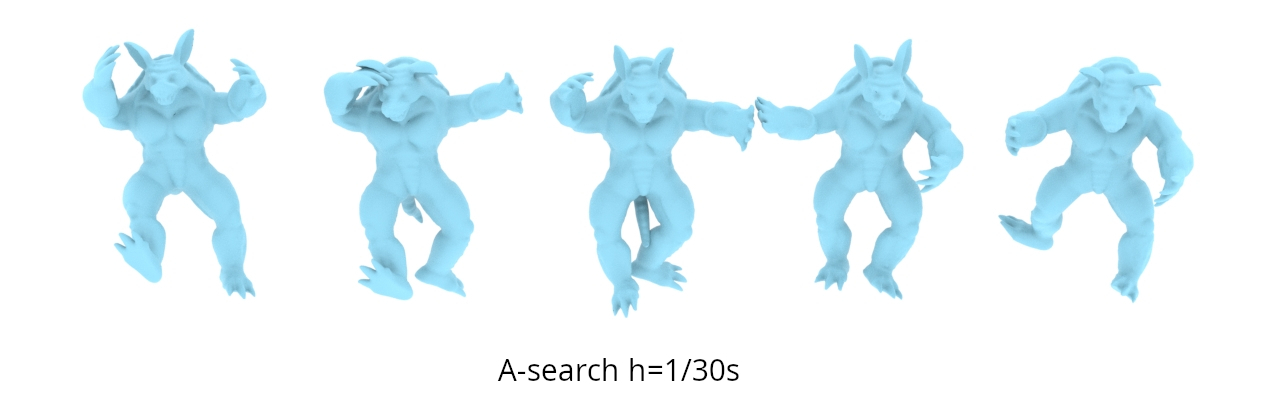}
  \includegraphics[width=1.0\linewidth]{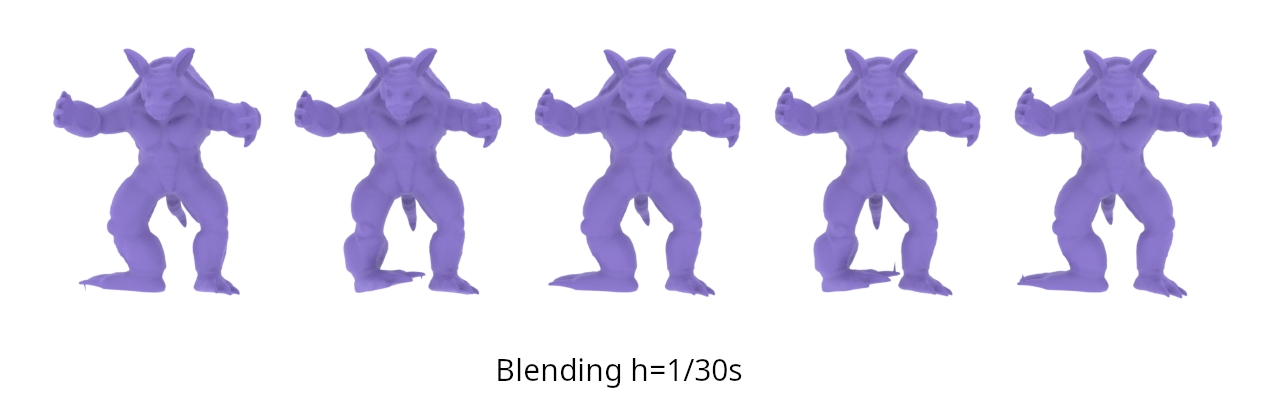}
  \caption{\textbf{Suspended armadillo.} Five frames between 3s and 4s.
  A-search at $h=1/30s$ sees slightly more motion than BDF2 at $h=1/300s$,
  and much more than BDF2 at $h=1/30s$. Blending has motion only in the foot and artifacts (see video).}
  \label{fig:suspended}
\end{figure}

\subsection{Blending} 
The starting observation is that for an arbitrary convex potential $P(\vx)$,
explicit Euler always gains energy and implicit Euler always loses energy \cite{dinev2018stabilizing}.
In practice, even when $P$ is not convex, the above observation almost always holds. 
Thus, there almost always exist some good quadrature point $(1-\theta) \vz_n + \theta \vz_{n+1}$ for the theta method (\ref{eq:inner-theta}) that leads to energy preservation.
However, this value of $\theta$ is difficult to find.
Searching for $\theta$ by binary search requires solving the implicit system many times and is thus prohibitively expensive.

In a different approach, \citet{dinev2018stabilizing} attempted to interpolate between the results of two time integrators as a form of post processing.
Specifically, assuming that implicit midpoint leads to an energy increase and implicit Euler leads to energy decrease, 
the results from implicit Euler
\begin{align*}
    \vz_{n+1}^{imp} = \vz_n + h f(\vz_{n+1})
\end{align*}
and implicit midpoint
\begin{align*}
    \vz_{n+1}^{mid} = \vz_n + h f \left( \frac{1}{2} \vz_{n+1} + \frac{1}{2} \vz_n \right)
\end{align*} 
are computed separately.
Then, they would find some $\alpha$ such that if 
\begin{align*}
    \vz_{n+1} = \alpha \vz_{n+1}^{mid} + (1 - \alpha) \vz_{n+1}^{imp}, 
\end{align*}
then the interpolated $\vz_{n+1}$ attains energy conservation with 
\[ H(\vx_{n+1},\vp_{n+1}) = H(\vx_n,\vp_n). \] 
This method, which we refer to as blending, can indeed attain energy conservation, 
and requires two implicit solves but none while performing binary search for $\alpha$. 
However, this approach has a tendency to suffer from high frequency artifacts due to the use of implicit midpoint.
Additionally, interpolation is problematic when non-convex constraints exist, such as with contact. 

\section{Method}

\subsection{Decoupled Method}

Our new method combines two ideas, 
the decoupling introduced by \citet{dinev2018stabilizing},
and the partitioned Runge-Kutta methods.
We first take some underlying Runge-Kutta method $\Phi$, 
and perform a step
\begin{align} \label{eq:original-update}
 (\tilde{\vx}_{n+1},\tilde{\vp}_{n+1}) = \Phi_h(\vx_{n},\vp_{n}). 
\end{align}
Independently, using the adjoint method $\Phi^*$, 
perform a step
\begin{align} \label{eq:adjoint-update}
(\vx_{n+1}^*,\vp_{n+1}^*) = \Phi_h^*(\vx_{n},\vp_{n}). 
\end{align}
Finally, our result is interpolated only on the momentum,
\begin{align}\label{eq:decoupled}
    \vx_{n+1} = \tilde{\vx}_{n+1}\text{ and } \vp_{n+1} = \alpha \vp_{n+1}^* + (1-\alpha) \tilde{\vp}_{n+1}. 
\end{align}
We call this method (\ref{eq:decoupled}) the decoupled $\alpha$-method based on $\Phi$,
denoted as $\vx_{n+1}, \vp_{n+1} = \Xi_\alpha (\vx_n, \vp_n)$.
In the case that we fix $\alpha = 1$, the momentum only depends on $\Phi^*$, thus
\[ \vx_{n+1},\_ = \Phi_h(\vx_{n},\vp_{n}) \text{ and } \_, \vp_{n+1} = \Phi_h^*(\vx_{n},\vp_{n}), \]
we call this the decoupled symplectic method based on $\Phi$, denoted as $\Xi_1$.
Both the $\alpha$-method and symplectic method have the same order as $\Phi$,
and if $\alpha$ is taken in a bounded range, the method is convergent.
In fact, $\Xi_1$ is technically is a partitioned Runge-Kutta method with twice the number of stages and a block structure for the Butcher Tabelau,
and $\Xi_{\alpha}$ has one more extra stage with weights $\alpha, 1-\alpha$ to do interpolation.

If the underlying method $\Phi$ was implicit Euler, writing the decoupled $\alpha$-method out fully gives
\begin{align}
    \vx_{n+1} &= \vx_n + h \vw_{n+1}  \\
    \vw_{n+1} &= \vv_n - h \vm^{-1} \nabla P(\vx^{n+1}) \label{eq:imp} \\
    \vv_{n+1} &= (1-\alpha) \vw_{n+1} + \alpha \vv_{n+1}^{exp} \label{eq:friction1} \\
    &= \vv_n - h \vm^{-1} \left( (1 - \alpha ) \nabla P(\vx_{n+1}) + \alpha \nabla P(\vx_{n}) \right) \\
    &= \vw_{n+1} - \alpha h \vm^{-1} ( \nabla P(\vx_{n})  -  \nabla P(\vx_{n+1})).  \label{eq:friction2}
\end{align}
If $\alpha = 1$ is fixed, we call the resulting decoupled symplectic method integrator \textit{A-1}.
To achieve energy conservation or control the energy more finely, 
$\alpha$ can be varied every time step.
In this case, we call the integrator \textit{A-search}. 
We see that the first two steps of A-search is identical to implicit Euler,
and from (\ref{eq:friction2}) \textit{A-search} corrects the implicit velocity $\vw_{n+1}$ by a force difference scaled by $\alpha$.

\subsection{Decoupled Symplectic Method}

\begin{theorem}
    The decoupled symplectic method $\Xi_1$ is linearly symplectic for any underlying Runge-Kutta method $\Phi$.
    Moreover, when $\Phi$ is implicit Euler,
    \textit{A-1} is unconditionally stably symplectic.
\end{theorem}

See Appendix \ref{appendix:stability} for the proof.
Our decoupled symplectic method $\Xi_1$ clearly resembles the partitioned Runge-Kutta methods based on coupled adjoint pairs.
Though being symplectic in general is too much to ask, 
we do recover linear symplecticity.
Moreover, the decoupled symplectic method appears to be often more stable than the coupled counterpart. 
Therefore, the decoupled symplectic method is a useful recipe to cook up various linearly symplectic methods based on arbitrary Runge-Kutta methods.

For example, \textit{A-1} is unconditionally \textit{stably symplectic}, 
as $\Tr \Xi_1 = 1 + 1/(1+\hbar)$ is monotonically decreasing from $2$ to $1$ as a function of $\hbar = h^2k/m$.
But if $\Psi$ denoted symplectic Euler,
then $\Tr \Psi = 2 - \hbar$ is not stable for $\hbar > 4$.
Similarly, if the underlying method was SDIRK2,
then its decoupled symplectic method has $\Tr \Xi_1$ bounded, 
with $\vt \Tr \Xi_1 \vt \leq 2$ for all $\hbar$ outside a bounded interval.
But if $\Psi$ was the partitioned Runge-Kutta method based on SDIRK2,
then $\Tr \Psi$ is unbounded as $\hbar \rightarrow \infty$.

We focus on \text{A-1} and consequently \textit{A-search} in experimental results due to its simplicity and good stability properties. 
For a discussion of decoupled methods using other underlying Runge-Kutta methods, see Appendix \ref{appendix:higher-order}.

\begin{figure}[h]
  \centering
  \includegraphics[width=1\linewidth]{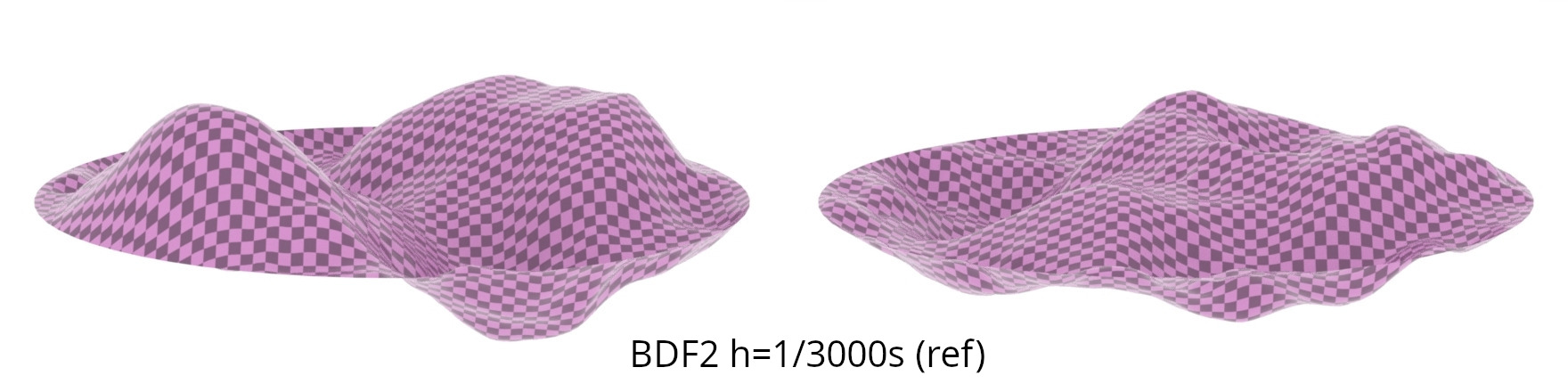}
  \includegraphics[width=1\linewidth]{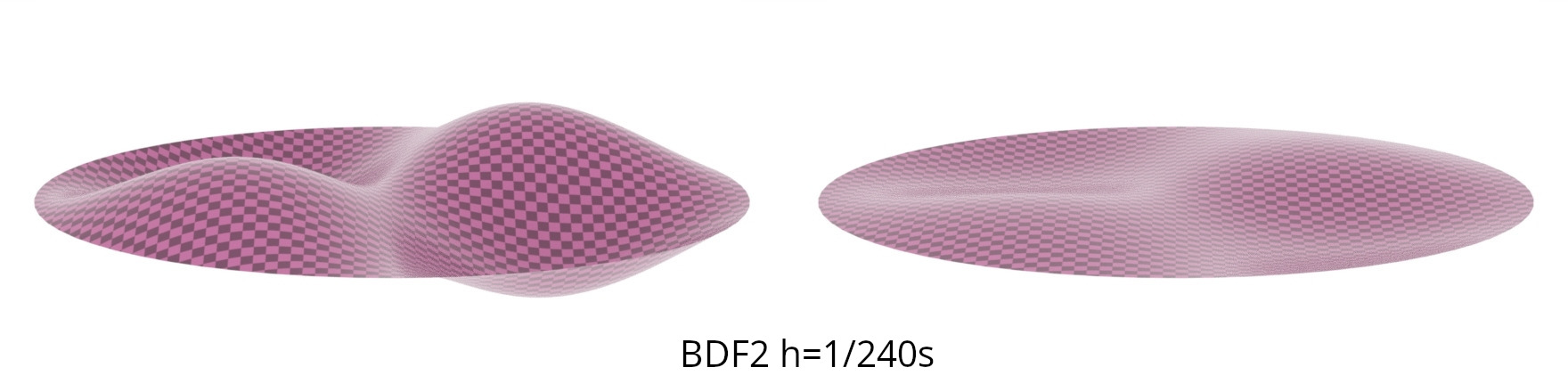}
  \includegraphics[width=1\linewidth]{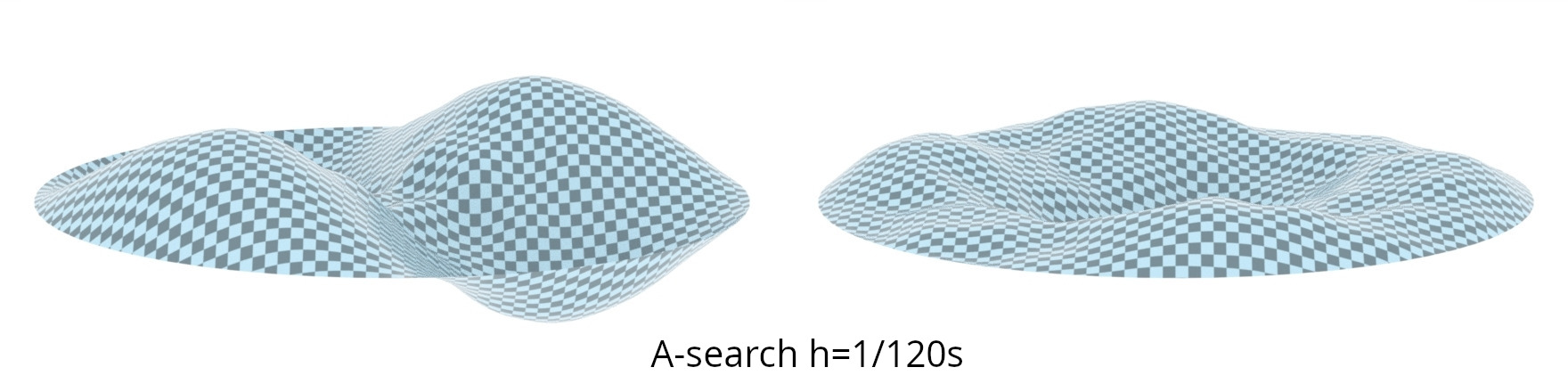}
  \caption{\textbf{Vibrating membrane}, initialized to $u_{21}$ mode.
  Left shows frame with largest $u_{21}$ amplitude in the final $0.5s$,
  right shows frame with most wrinkles.
  Amplitudes are magnified by 10x for visual clarity.
  A-search both kept more energy in the $u_{21}$ mode than BDF2,
  and captured some of wrinkling modes exhibited by the reference.}
  \label{fig:membrane}
\end{figure}

\subsection{A-1}

Unlike other symplectic and A-stable methods,
$\Xi_1$is resilient against numerical explosions if $\Psi$ is chosen as so.
For \text{A-1}, where $\Psi$ is implicit Euler,
energy gains can slowly accumulate over time, but never has a blow-up been encountered the way commonly exhibited by implicit midpoint or trapezoidal.
Heuristically, the potential energy term $P(\vx)$ in the Hamiltonian is stiff,
and the stable implicit Euler update guarantees that the potential energy is well controlled.
The kinetic energy term $\frac{1}{2} \Vt v \Vt_{\vm}^2$ is not stiff, so using the unstable explicit Euler method is safe,
and the kinetic energy is mediated in the long term through the joint implicit Euler solve.
This also suggests that \textit{A-1} can dampen high frequency modes on the position side with implicit Euler,
while redistributing the lost energy through the velocity in the next step.

\textit{A-1} handles collisions well.
Specifically, for a single particle against a one-sided quadratic function of the form 
\begin{align}\label{eq:quadratic-barrier}
    P(x) = \frac{1}{2} k x^2 1_{x > 0},
\end{align}
in the limit when $\hbar = h^2k/m \rightarrow \infty$,
that is, a very stiff wall or a very large time step size ,
\textit{A-1} always resolves the collision while preserving the speed of the particle,
independent of the initial position $x$ and the phase of the collision.
This is not true for other symplectic methods like trapezoidal,
which usually lead to gaining speed. 
See Appendix \ref{appendix:collision} for a proof.
The same qualitative behavior also appears to extend to other forces, like the IPC barrier,
and for an elastic body rather than a point mass.

A distinctive property of \textit{A-1} is that, as $\hbar \rightarrow \infty$, the eigenvalues of $\Xi_1$ approach $e^{\pm i\pi/3}$, in contrast to other symplectic methods, including implicit midpoint, whose eigenvalues typically approach $-1$. This  behavior is particularly beneficial in the presence of collisions. For the quadratic barrier model in (\ref{eq:quadratic-barrier}), the trapezoidal rule typically interacts with the barrier for only a single time step; the resulting abrupt transition leads to noticeable energy error and poor conservation. In contrast, \textit{A-1} naturally distributes the interaction over three time steps, allowing the collision response to be resolved more smoothly and enabling effective energy preservation. In practice, this behavior is well matched to simulation pipelines that advance the dynamics at a modest multiple of the rendering frame rate, ensuring that collisions remain visually smooth.

\begin{figure}[h]
  \vspace{-0.2cm}
  \centering
  \includegraphics[width=0.49\linewidth]{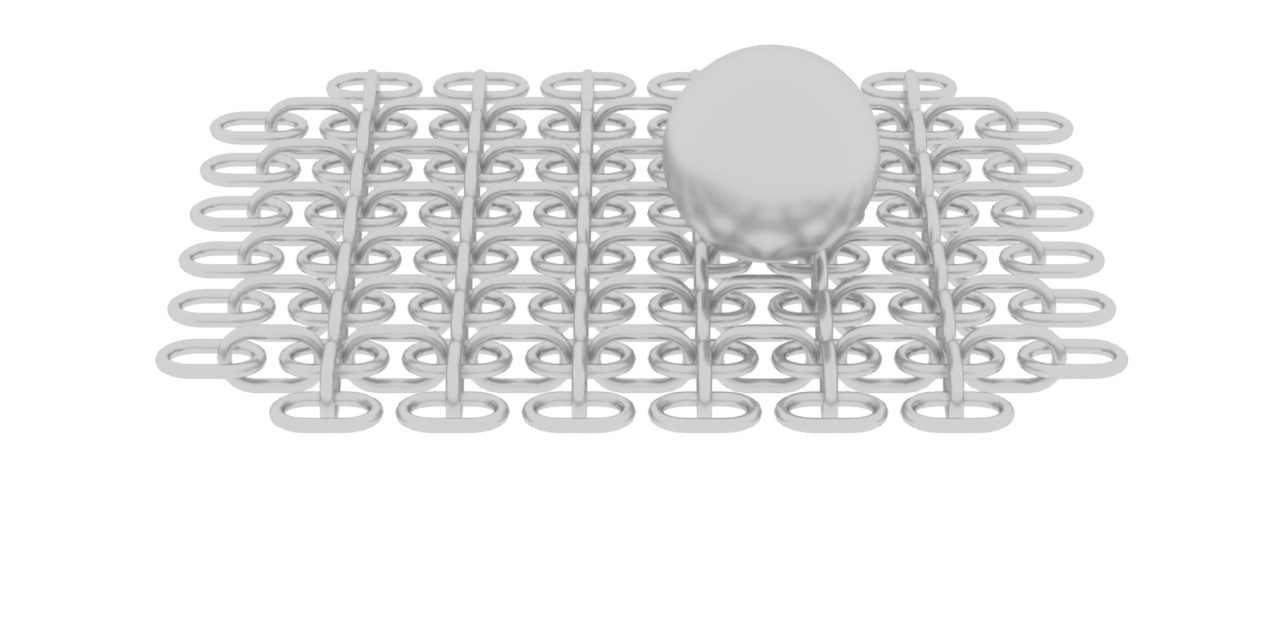}
  \includegraphics[width=0.49\linewidth]{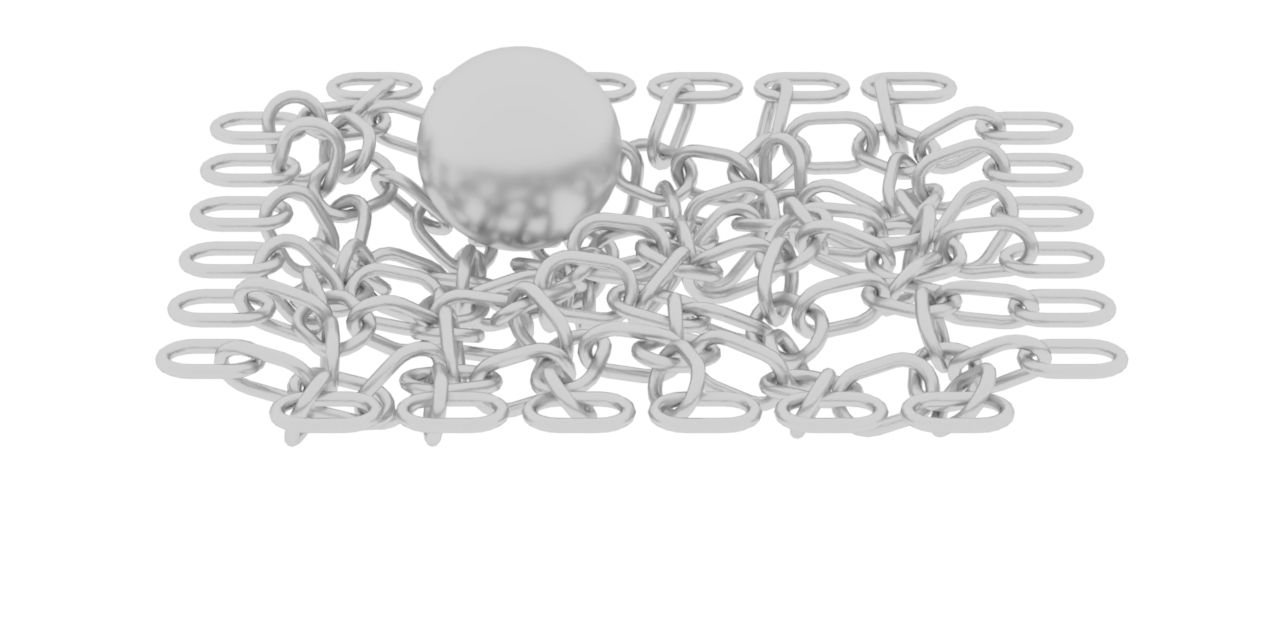}
  \includegraphics[width=0.49\linewidth]{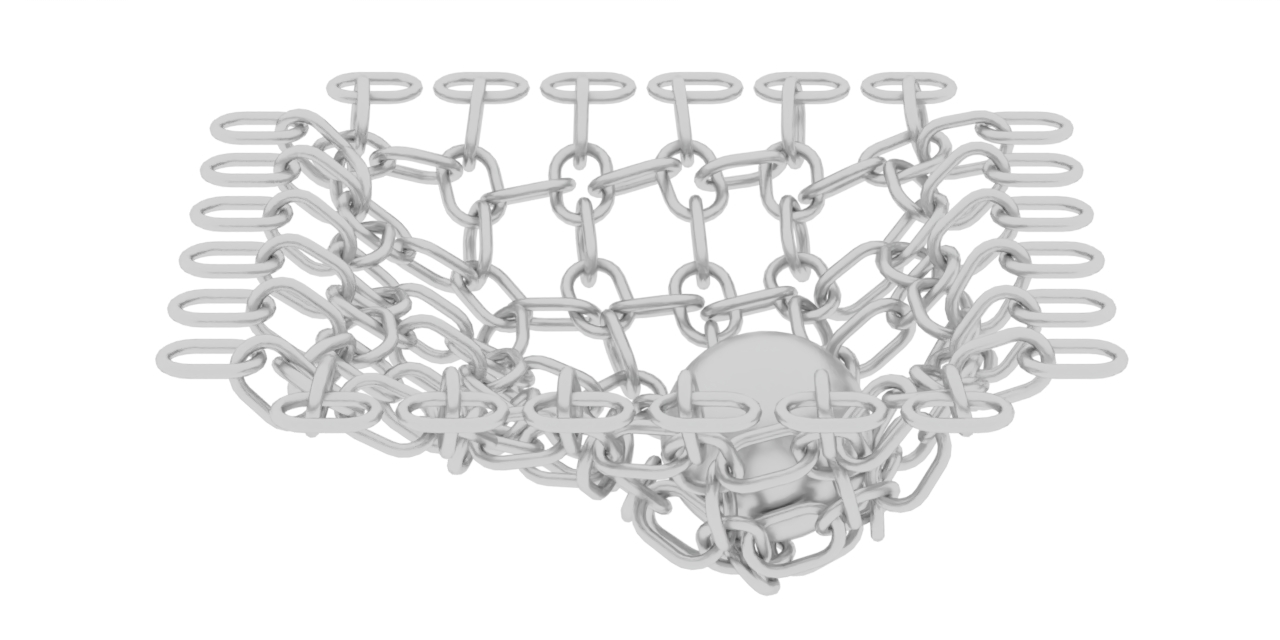}
  \includegraphics[width=0.49\linewidth]{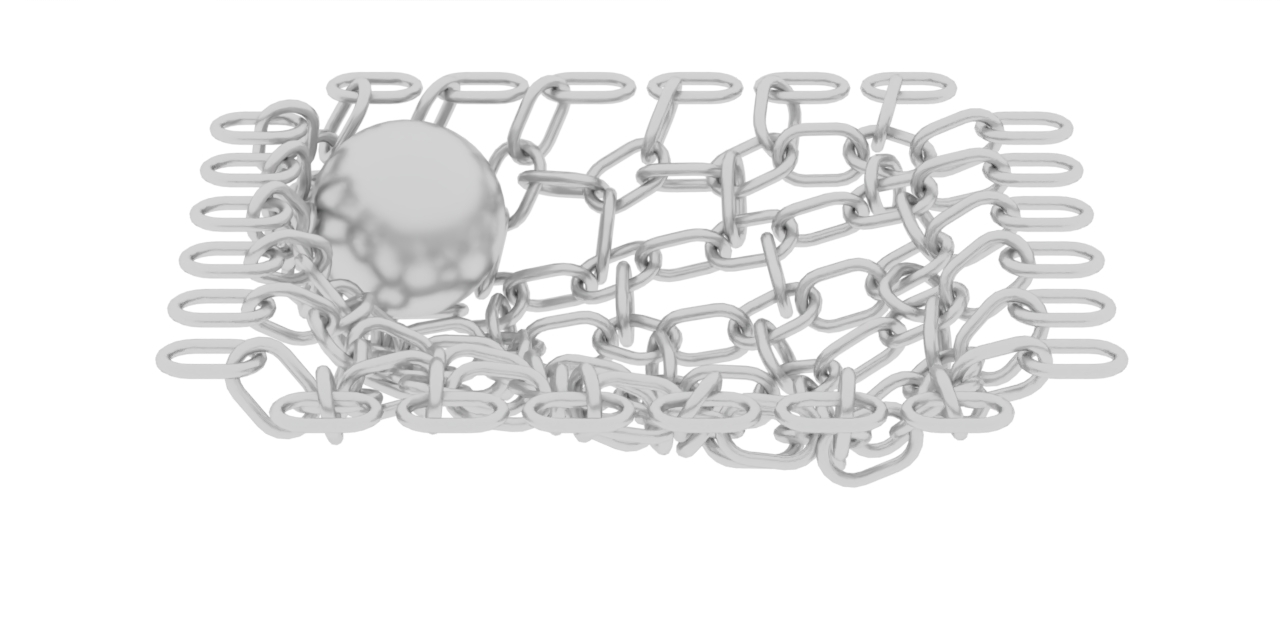}
  \includegraphics[width=0.49\linewidth]{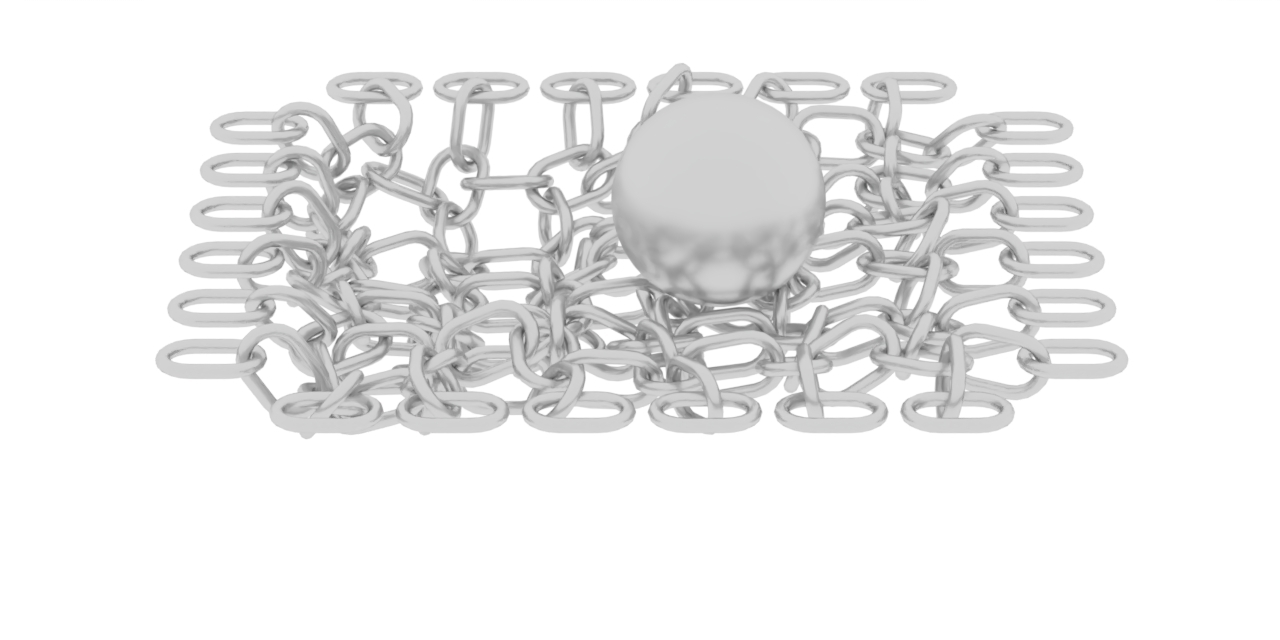}
  \includegraphics[width=0.49\linewidth]{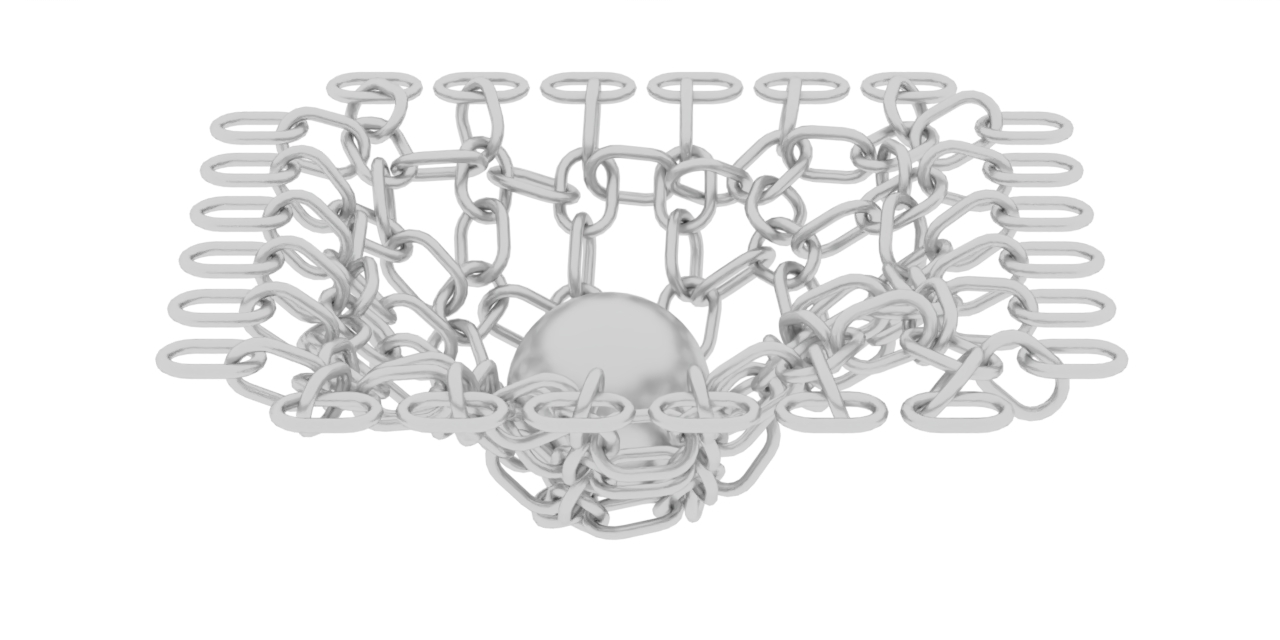}
  \includegraphics[width=0.49\linewidth]{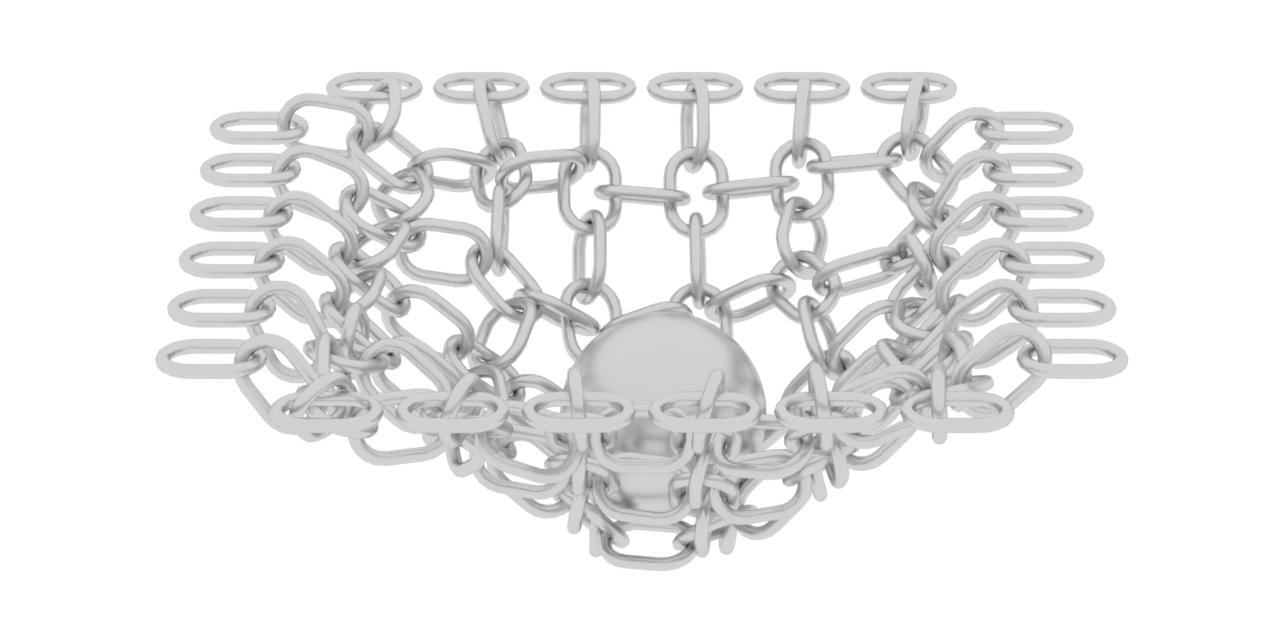}
  \includegraphics[width=0.49\linewidth]{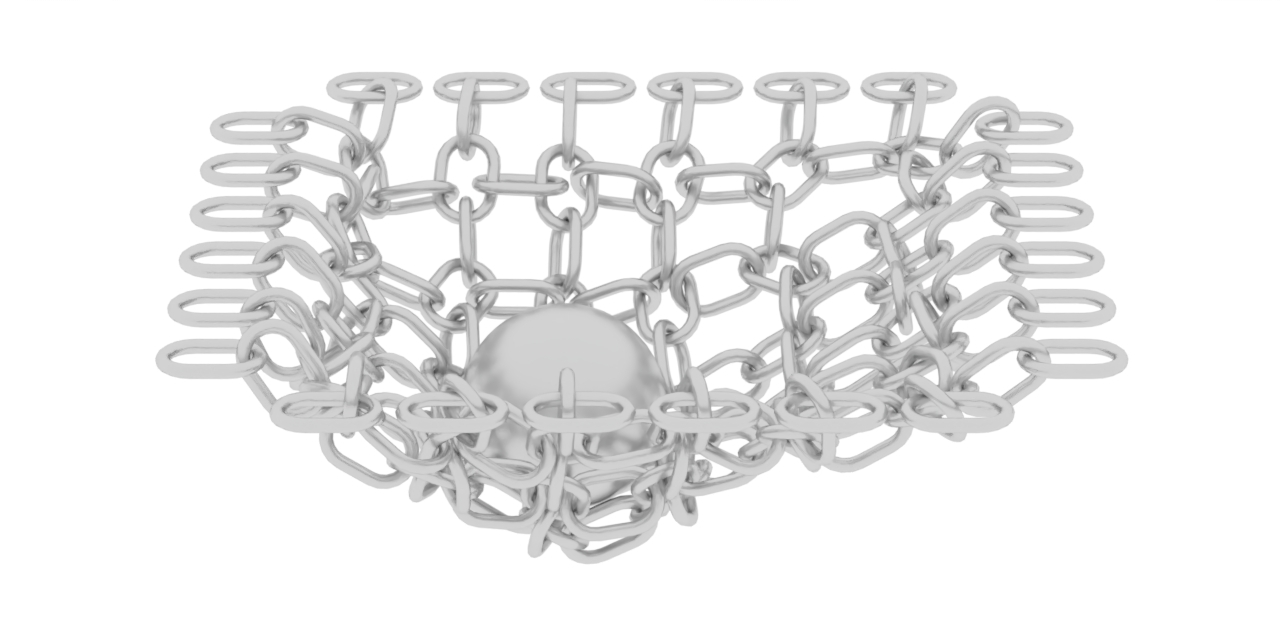}
  \vspace{-0.3cm}
  \caption{\textbf{Chain net} with Affine Body Dynamics (ABD). 
  A-search handles the complex collisions between chains,
  while enabling the ball to bounce multiple times.
  Various frames, top to bottom, then left to right.}
  \label{fig:chain}
\end{figure}

\subsection{A-search}

For the decoupled $\alpha$-method (\ref{eq:decoupled}),
it is generally possible to compute the energy 
\[ H_{n+1} := H(\vx_{n+1},\vp_{n+1}) = H(\tilde{\vx}_{n+1},\tilde{\vp}_{n+1},\vp_{n+1}^*,\alpha) \]
in terms of results from $\Phi, \Phi^*$ and the choice of $\alpha$ in closed form with
\begin{align}
    H_{n+1} = P(\vx_{n+1}) &+ \frac{1}{2} \Vt \tilde{\vp}_{n+1} \Vt_{\vm^{-1}}^2 + \alpha \langle \tilde{\vp}_{n+1} , \vp_{n+1}^* - \tilde{\vp}_{n+1} \rangle_{\vm^{-1}} \\
    &+ \alpha^2 \frac{1}{2} \Vt \vp_{n+1}^* -  \tilde{\vp}_{n+1} \Vt_{\vm^{-1}}^2 
\end{align} 
A natural candidate for $\alpha$ is the one that leads to $H_{n+1} = H_0$,
which means energy conservation.
Such a method can be seen as affine interpolation with respect to the constant $H$ submanifold on each step. 
Often the phase space is the cotangent bundle of an underlying manifold of admissible positions
whose admissibility is enforced through the potential energy,
while the velocity and momentum are unconstrained.
Since the projection only occurs along the momentum,
the position due to \textit{A-1} and \textit{A-search} is still admissible.

Since $\alpha = 1$ is symplectic,
we expect that $\alpha = 1$ can lead to either energy increase or decrease.
Thus, to achieve energy conservation,
it is necessary for $\alpha > 1$ on some time steps.
Thus, we consider affine combinations in the momentum by letting $\alpha \in [\alpha_{min},\alpha_{max}]$ to obtain energy conservation.
Some finite interval is necessary for time integrator consistency.
We find that a simple choice of $\alpha \in [0.0,1.1]$ works fairly well in practice over a range of material parameters.
It is essential to note that due to the cropping at $\alpha_{max}$ and the natural variation of energy due to symplectic methods,
$H_{n+1}$ may not be $H_0$ at all steps.
In particular, temporary energy lost is inevitable for stiff collisions (Appendix \ref{appendix:collision}).
Nevertheless,
since $\alpha = \alpha_{max}$ almost always gains energy in the long run,
we almost always have $H_{n+1} \approx H_0$ in the long term.
See subsection \ref{subsec:asearch} for more details,
including an ablation study on $\alpha_{min}, \alpha_{max}$.

In addition, we may not want $H_{n+1} \approx H_0$ for all time steps,
and we may prefer some amount of energy decay, imitating natural dissipation.
This is trivial on top of what we already have earlier.
For any specified desired energy function $E_n$ that is decreasing in time,
the same procedure can be performed while targeting $H_{n+1} \approx E_{n+1}$.
Then, $\alpha$ will tend to be biased to be smaller than $1$, 
and it can go as low as $\alpha = 0$ to reduce the energy.
Therefore, \textit{A-search} can dissipate energy as much as implicit Euler does, 
if strong energy dissipation is desirable,
independent of the scene parameters. To achieve this, we implemented exponential decay with a half-life time scale $\tau$ relative to a ground energy level $E_g$, with 
\[ E_{n+1} - E_g = e^{-(n+1)h/\tau} (E_0 - E_g). \]

\begin{figure}[h]
  \centering
  \includegraphics[width=1.0\linewidth]{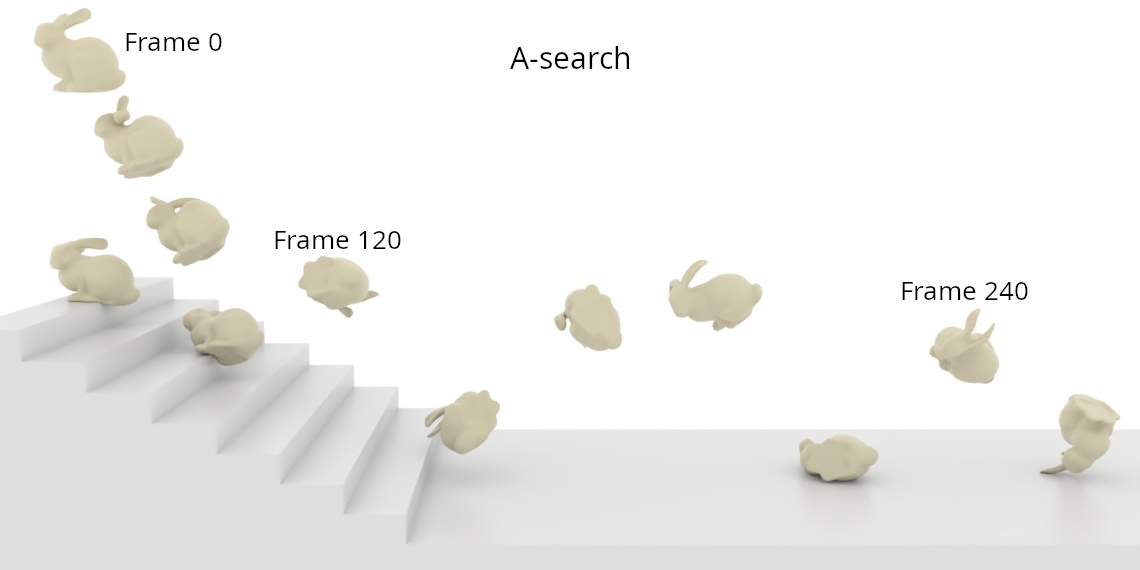}
  \includegraphics[width=1.0\linewidth]{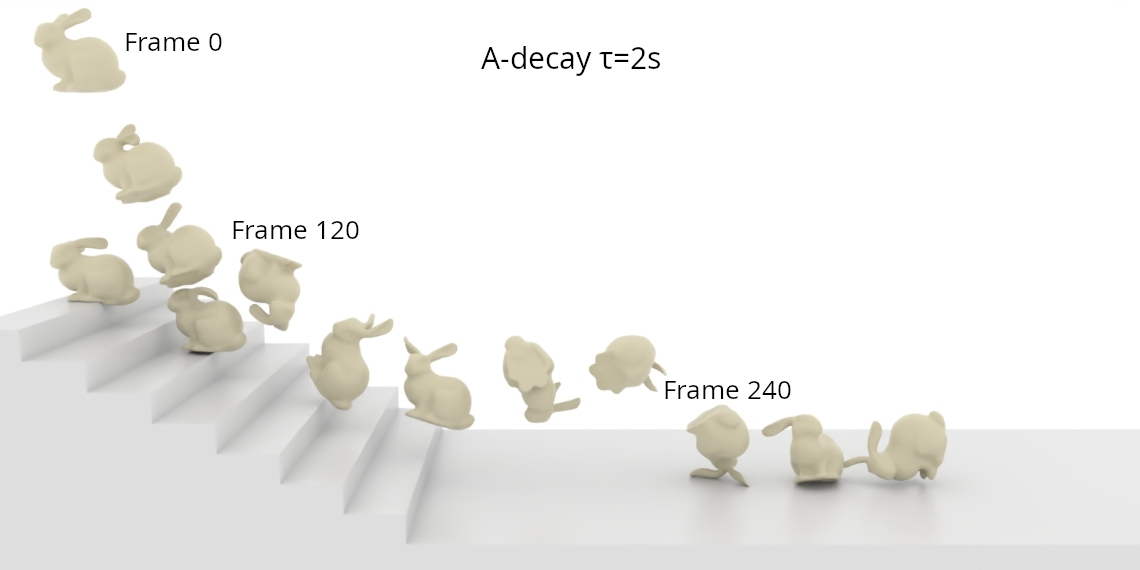}
  \includegraphics[width=1.0\linewidth]{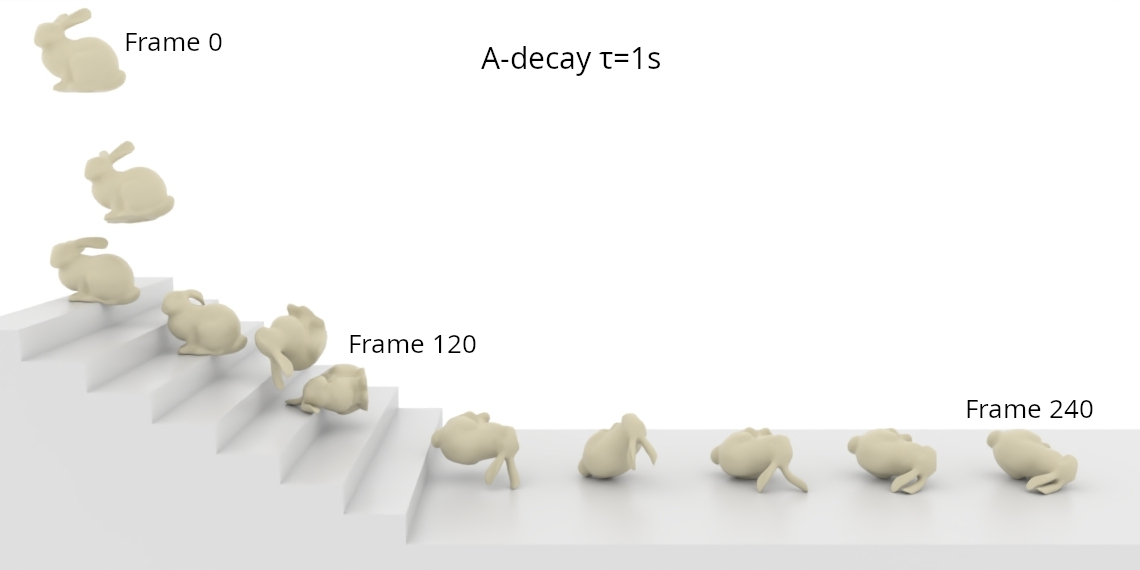}
  \caption{\textbf{Bunny stair drop}, first five frames showing bounce, then frames in regular $0.2s$ intervals starting from $1s$.
  A-search at three different energy profiles lead to a natural interpolation between energy conservation and close to implicit Euler dissipation.}
  \label{fig:stair}
\end{figure}

\subsection{Friction}

When friction is present, the system is no longer Hamiltonian, and our integrator is less interpretable.
Nonetheless, we still have an ordinary differential equation with an additional friction term,
\[ f(\vx,\vv) = (\vv, -\vm^{-1} (\nabla P(x) - g(\vx,\vv)), \]
and the application of the Runge-Kutta methods (\ref{eq:original-update}) and (\ref{eq:adjoint-update}) still make sense.
However, to actually solve the equations associated with the Runge-Kutta method may not be possible in general.

Even if the system is solvable, we may lose stability.
Consider the simple damped harmonic oscillator $m\dot{v} = -kx -cv$.
If $c = 0$, we know that \textit{A-1} and \textit{A-search} is stable independent of $h$ despite having an explicit velocity update,
since there is another implicit position update before the position can effect the velocity.
But instead if there is damping, 
$c$ can be large relative to $m$,
then $m \dot{v} = -cv$ with explicit velocity update is certainly unstable.
The velocity dependent dissipative force feeds into itself directly.
Therefore, a more reasonable solution is to use (\ref{eq:friction2}).
The friction is included in the implicit Euler solve to obtain $
\vx_{n+1}$ and $\vw_{n+1}$,
but excluded from the final $\alpha$ correction.
The interpretation is that energy correction is only performed on the conservative forces,
and energy dissipation due to implicit Euler is allowed on the non-conservative forces.

We implement two forms of semi-implicit friction/damping, thus recovering a separable Hamiltonian system throughout the duration of a time step,
which can be solved within the optimization time integration framework.
For 1D convergence studies,
we use a semi-implicit Rayleigh damping described in \citet{gast2015}:
\begin{align}
     g_{n+1} = -\mu \pdv[2]{\Psi}{\vx} (\vx_n) \vv_{n+1}.
\end{align}
On the other hand, for 3D scenes,
we use the IPC \cite{Li2020IPC} Coulomb friction 
\begin{align}
     g_{n+1} = - \mu \lambda_n \vv_{n+1},
\end{align} 
where $\lambda_n$ is sparse and only depends on the contact pairs and normal forces at $\vx_n$.
In both cases, we modify the implicit solve (\ref{eq:imp}) into
\[ \vw_{n+1} = \vv_n - h \vm^{-1} \nabla P(\vx^{n+1}) + g_{n+1}, \]
while ignoring friction/damping for the explicit velocity correction (\ref{eq:friction2}).
In both cases, we can write 
\[ g_{n+1} = -\nabla_x P_n(\vx_{n+1})
\text{ where } P_n(\vx_{n+1}) = \frac{\mu h}{2} \Vt \frac{\vx_{n+1} - \vx_n}{h} \Vt_{\lambda_n}^2. \]

\begin{algorithm}
\caption{A-Serach with friction}\label{alg:cap}
\begin{algorithmic}[1]
\REQUIRE $\vx_n, \vv_n, E_{n-1}, \Ec_{n-1}$
\STATE Update energy target $E_{n} \gets E(E_{n-1}, n, \Ec_{n-1})$ 
\STATE Solve 
\[ \vx_{n+1} \gets \argmin_{\vx} \frac{1}{2} \Vert \vx - \vx_n - h \vv_n \Vert^2_m + h^2 P(\vx) + h^2 P_n(\vx) \] 
via Newton's method with backtracking line search.
\STATE Update frictional losses $\Ec_{n+1} \gets 2P_n(\vx)$
\STATE Compute implicit velocity $\vw_{n+1} \gets (\vx_{n+1} - \vx_n) / h$
\STATE $\Delta \vv \gets h \vm^{-1} ( \nabla P(\vx_{n}) - \nabla P(\vx_{n+1})) $
\STATE Solve $\alpha$ by taking the root closer to $1$ of the quadratic \[ E_{n} - \Ec_{n} = H_{n+1} (\alpha) = P(\vx_{n+1}) + \frac{1}{2} \Vert \vw_{n+1} - \alpha \Delta v \Vert_\vm^2\] 
\STATE $\alpha \gets \text{clip}(\alpha_{min}, \alpha_{max}, \alpha)$
\STATE Perform A-search correction $\vv_{n+1} \gets \vw_{n+1} - \alpha \Delta v $
\end{algorithmic}
\end{algorithm}

When friction is present, energy is no longer conserved.
Therefore, it is necessary to estimate the energy dissipated due to friction $\Ec_{n+1}$,
which we do as $\Ec_{n+1} \approx g_{n+1} \cdot (\vx_{n+1} - \vx_n) = 2 P_n(\vx_{n+1})$.
For energy conservation modulo friction, 
our energy target is updated as $E_{n+1} = E_n - \Ec_n$.
When applying energy decay with friction, we define
\[ E_{n+1} - E_g = e^{-h/\tau} (E_n - E_g - \Ec_{n}). \]
Usually, the target energy $E_0$ is understood to be the true energy $H_0$ in the system at time zero.
However, since usually in animation scenes are initialized kinematically,
there may be spurious forces at the beginning which we prefer to dampen out.
Therefore, in practice it is useful to initialize $E_0$ to be slightly smaller than $H_0$.
See Algorithm \ref{alg:cap} for the complete algorithm for integrating one time step using our method.

\section{Evaluation}

Unless otherwise noted, forces can include elasticity, gravity, and frictional contact \cite{Li2020IPC} forces.
We use the standard linear tetrahedral mesh and lumped mass matrix discretization.
The implicit systems arising from time integration are formulated as an optimization problem and solved using Newton's method with backtracking line search, 
where each linear system is solved using Jacobi-preconditioned conjugate gradient and the tolerance on the Newton step is $0.01h$. The line search is filtered by the additive continuous collision detection method \cite{li2021codimensional}.

We use BDF2 as the baseline integrator and compare it against A-search under two settings, 
at matching time step sizes $h$, and at comparable total runtime. 
In principle, A-search is a lightweight correction to implicit Euler and incurs only slightly higher per-step cost. 
However, because it better preserves energy, A-search often resolves more dynamic motion, which can increase the computational effort per time step. 
For this reason, comparisons are also made at roughly equal total runtimes while fixing the time step for A-search.
We will also use BDF2 at a much smaller time step and much more total run time as a reference solution.

\begin{figure}[h] 
  \centering
  \includegraphics[width=1.0\linewidth]{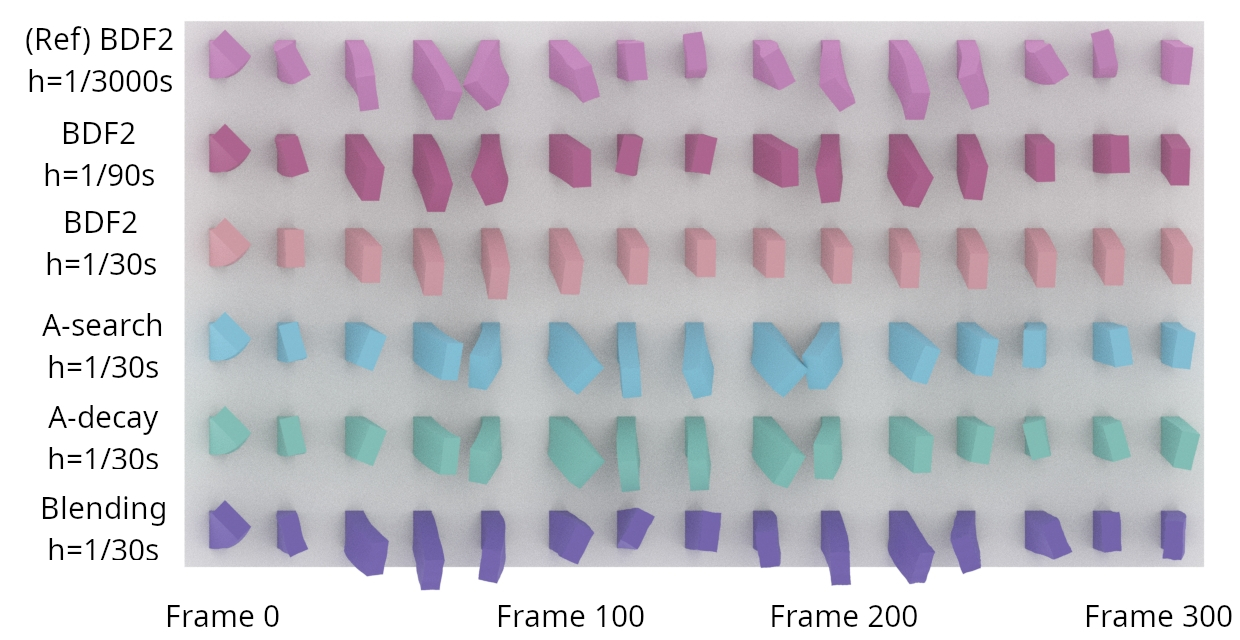}
  \caption{\textbf{Twisted bar.} Selected frames over 10 s. At $h=1/30,\mathrm{s}$, \textit{A-search} achieves motion magnitude comparable to the BDF2 reference while retaining more motion than BDF2 at similar runtime.}
  \label{fig:bar}
\end{figure}

\subsection{Comparisons}

In the following deformation-only comparison tests, 
we use corotated elasticity,
and by default a uniform simulation time step $h = 1/30$s matching the frame rate.
We additionally compare with the blending integrator proposed by \citet{dinev2018stabilizing}.
See Appendix \ref{appendix:material} for material and scene parameters.

\begin{figure}[h] 
  \centering
  \includegraphics[width=0.95\linewidth]{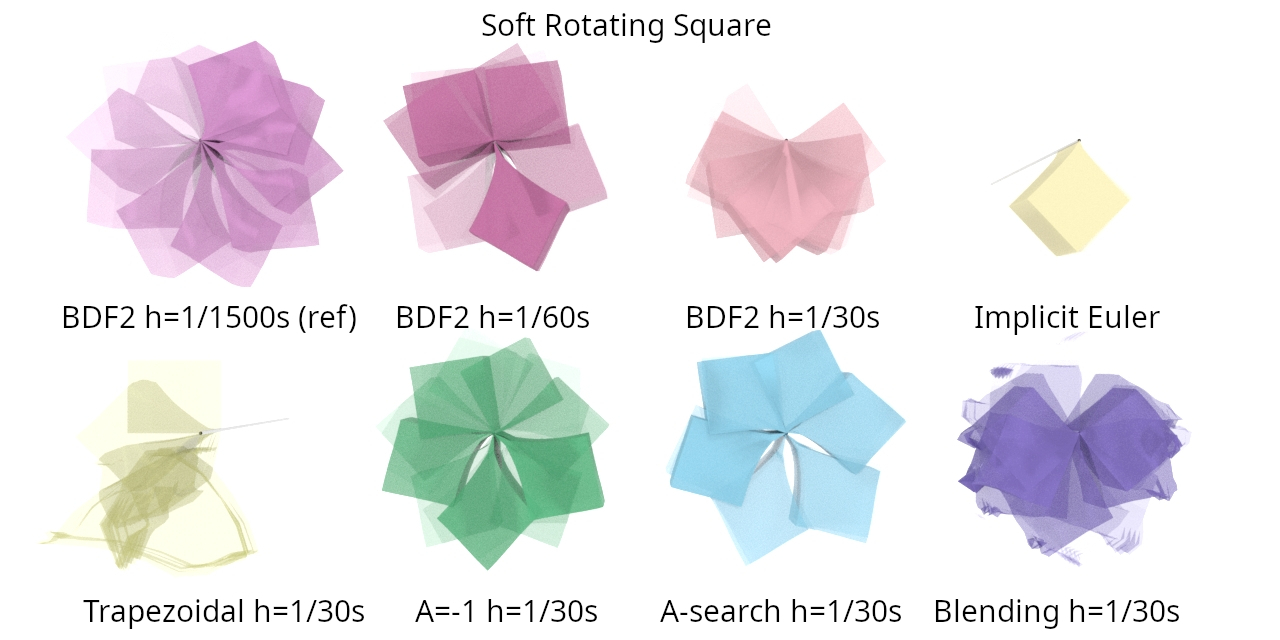}
  \includegraphics[width=0.95\linewidth]{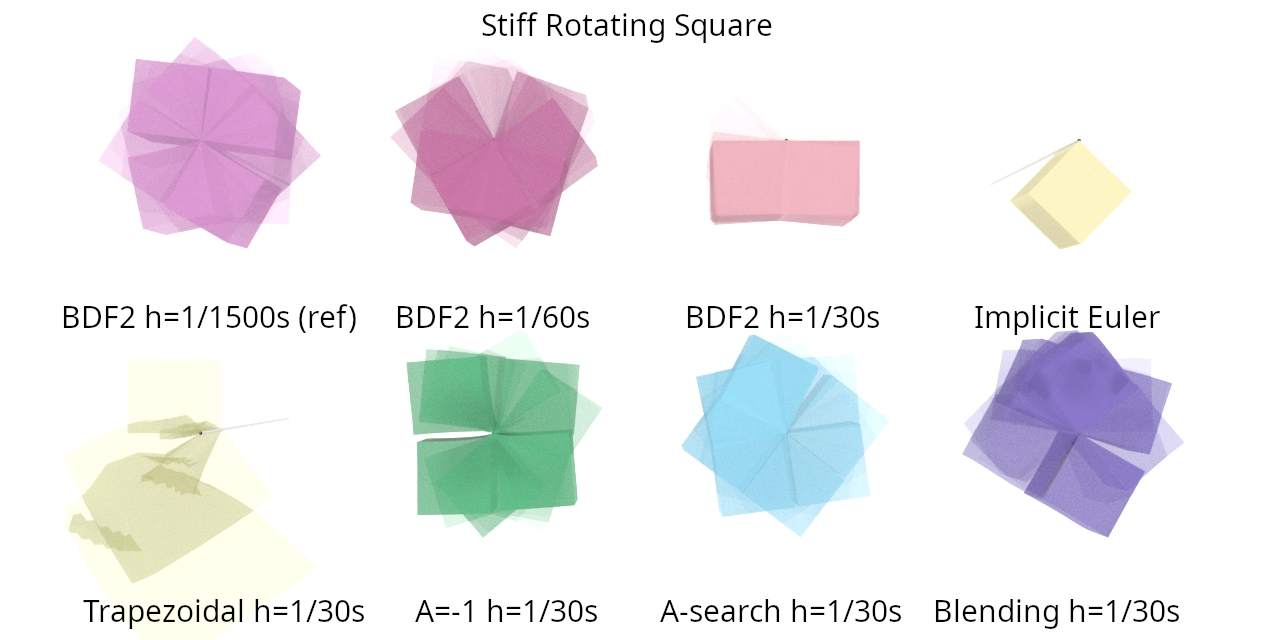}
  \vspace{-0.3cm}
  \caption{\textbf{Soft and stiff rotating square}, 20 frames over 10s, superimposed.
  For implicit Euler and BDF2, the cube eventually settles in the bottom,
  while for A-1 and A-search, the cube continues to make full rotations about the pivot, matching the reference behavior.}
  \label{fig:rotating}
\end{figure}

\paragraph{Rotating cube}
We rotate a soft and stiff cube around a fixed edge for 10s by a prescribed initial angular velocity, similar to the test from \citet{dinev2018stabilizing},
while also allowing the cube to interact with gravity.
In both the soft and stiff case, implicit Euler and BDF2 at the baseline $h=1/30$s lead to significant energy and angular momentum loss
(Fig. \ref{fig:rotating}).
Similarly, BDF2 at $h=1/60$s, with similar total runtime as A-search at $h=1/30$s, 
saw noticeable energy loss, especially in the stiffer case.
Implicit midpoint is unstable and explodes.

\begin{figure*}[!bt]
  \centering
  \includegraphics[width=0.49\linewidth]{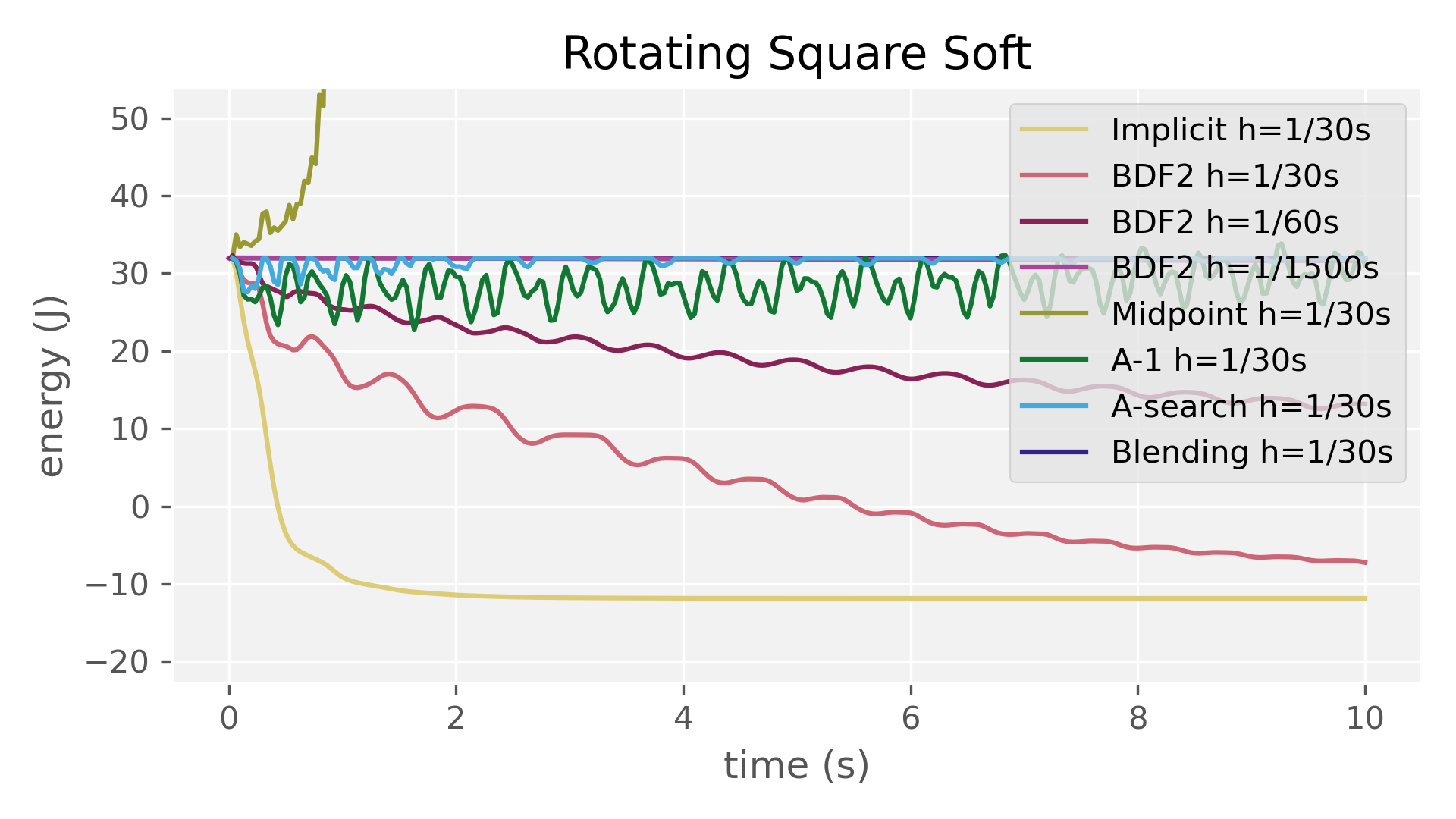}
  \includegraphics[width=0.49\linewidth]{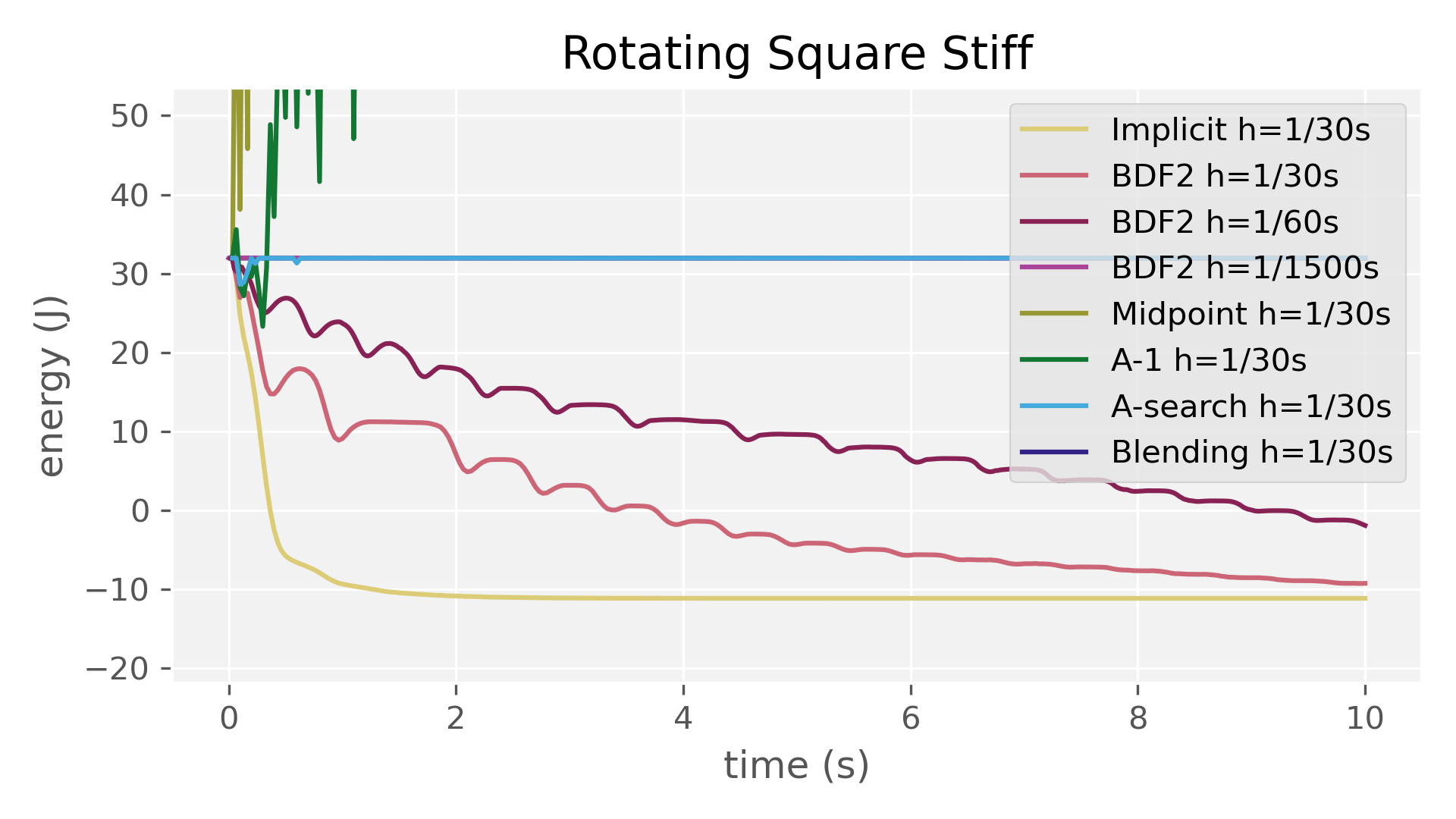}
  \includegraphics[width=0.24\linewidth]{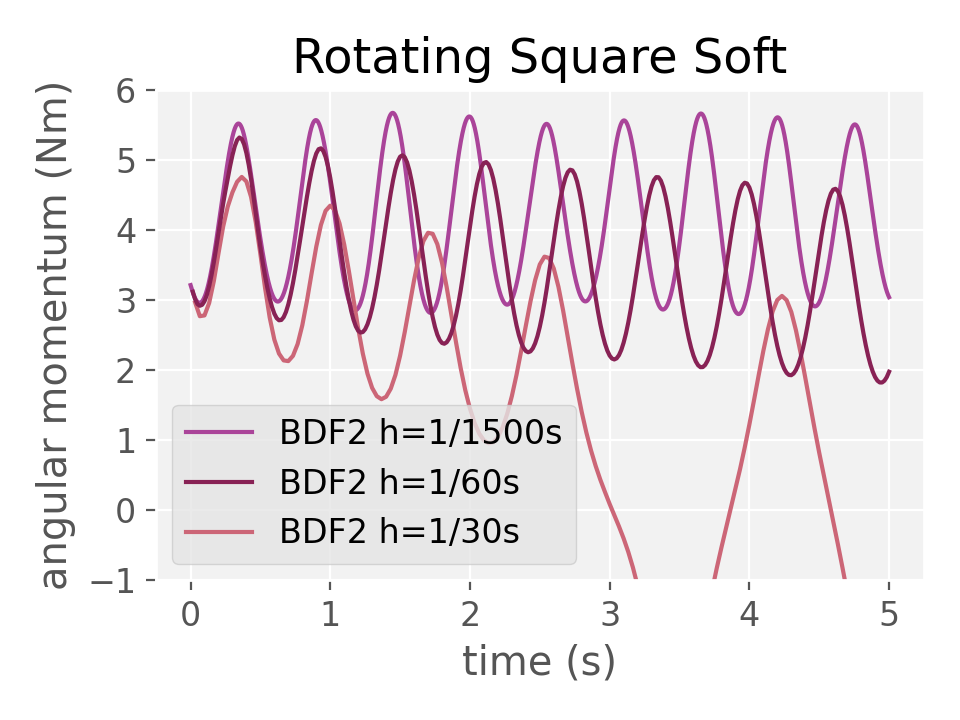}
\includegraphics[width=0.24\linewidth]{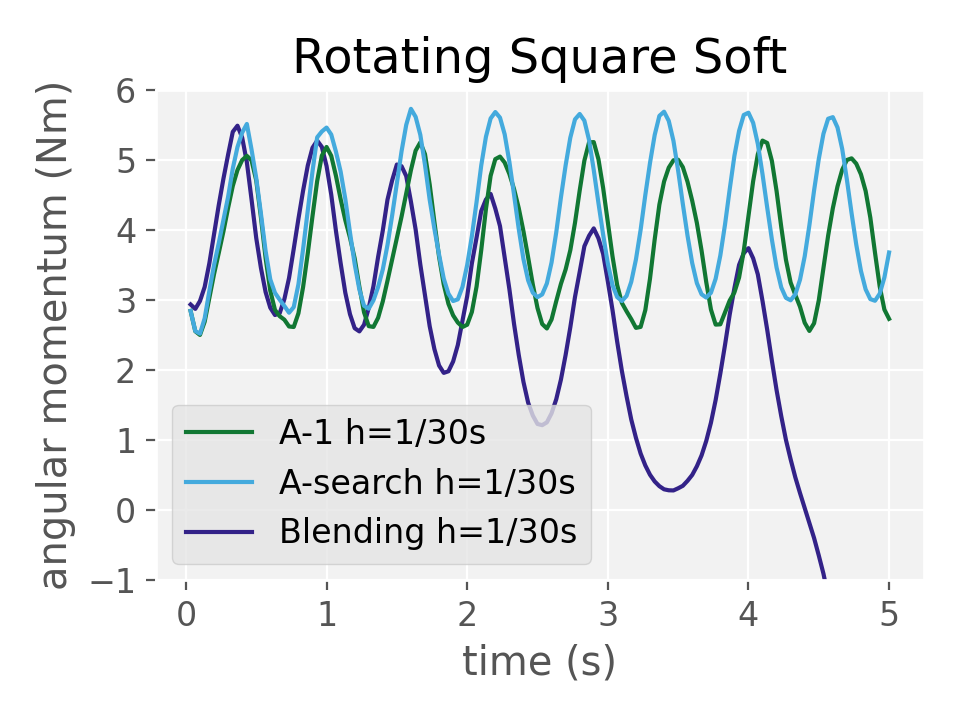}
\includegraphics[width=0.24\linewidth]{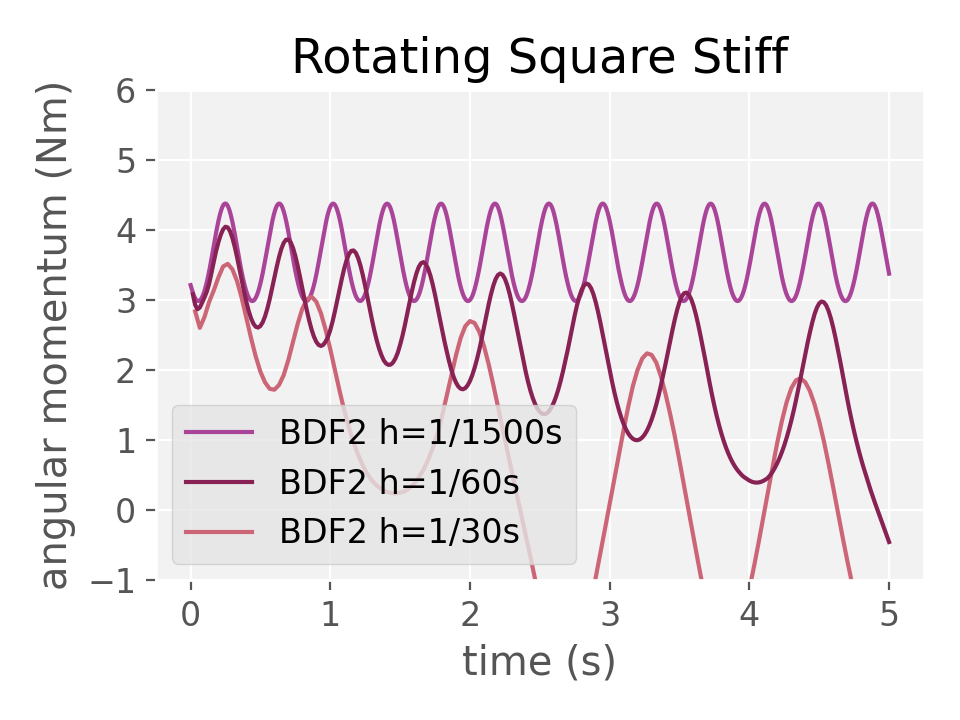}
\includegraphics[width=0.24\linewidth]{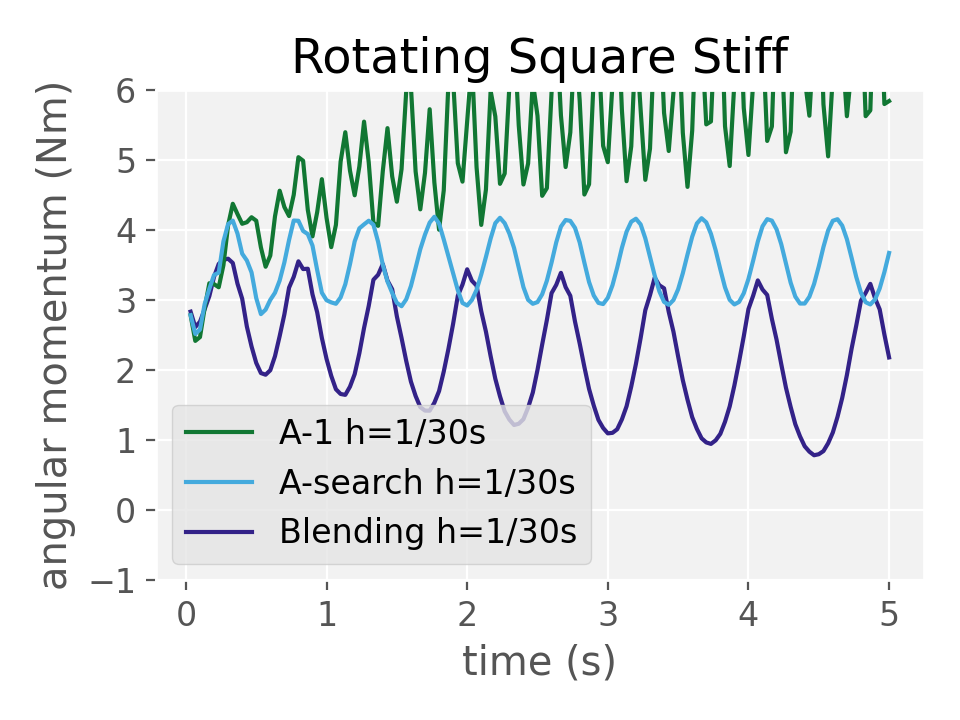}
  \caption{\textbf{Rotating square, total energy
  and angular momentum plots.}
  All three of A-search, blending, and the reference BDF2 $h=1/1500$s kept approximately constant total energy.
  A-search has the most accurate angular momentum compared to the reference.}
  \label{fig:90003}
\end{figure*}

While blending does not explode and has perfect energy conservation, 
it inherited some high frequency noises from using midpoint, 
noticable in the softer case.
A-1 and A-search at $h=1/30$s are stable and provide consistent, smooth-looking rotation.
A-1 does lose or gain energy over time,
though the change in energy is gradual and leads to smooth motion,
and far outperforms implicit midpoint as a symplectic integrator.
A-search best preserves the angular momentum and is the closest to the reference solution (Fig. \ref{fig:90003}).

\paragraph{Twisted bar} 
We fix one shortest side of a rectangular bar, twist and bend the bar, and let it fall under gravity for 8s.
For an object bending in only one direction, 
BDF2 performs quite well, far better than implicit Euler, 
but with more complicated twisting motion and gravity, BDF2 leads to energy lost.
Blending once again has visible high-frequency noises.
Our A-search method has more dynamical horizontal swinging and twisting compared to BDF2 at similar total runtime (Fig. \ref{fig:bar}).
Additionally, our solution with a $\tau = 10$s exponential decay damps the energy in a reasonable manner,
with decreasing amplitudes of oscillation towards the end.

\paragraph{Suspended armadillo} 
We hang an armadillo by the back under the influence of gravity,
and prescribe a downwards initial velocity to the armadillo.
(Fig. \ref{fig:suspended}).
In the baseline BDF2 solution at $h=1/120$s, the motion is dissipated near the end of the solution.
Blending got stuck in a motion with all the energy and dynamics concentrated on one foot, with clear artifacts.
For A-search at $h=1/120$s and BDF2 at $h=1/600$s,
the limbs, tail, ears, and head of the armadillo continue to wiggle in a smooth manner.
The BDF2 solution has more shaking,
but A-search had a wider range of motion,
including arms in front of the body, crossing legs, and larger tail motion,
that is more visually pleasing for animation.
Moreover, the appropriate value of $h=1/600$s for BDF2 relies on trial and error to determine,
while A-search outputs a qualitatively similar output at just $h=1/120$s.

\subsection{Validation}

In all following scenes, we use Neo-Hookean elasticity and IPC contact to ensure inversion- and penetration-free behavior. Here, we take the default of $h = 1/120$s due to the stiffness of contact and the potential aliasing effect due to \textit{A-search} for collisions. Nevertheless, we demonstrate that \textit{A-search} works across a range of material parameters at such a fixed time step, whereas traditional methods lead to significant energy dissipation.

In this subsection, we evaluate three key claims: that \textit{A-1} and \textit{A-search} robustly resolve contact interactions; that they preferentially retain energy in low-frequency deformation modes rather than dissipating it; and that they remain stable and effective in the presence of frictional contact.

\begin{figure}[]
  \centering
  \includegraphics[width=\linewidth]{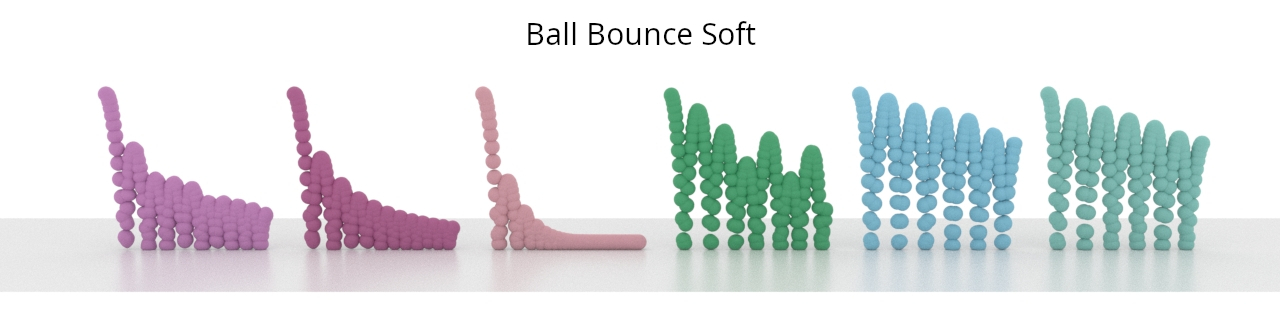}
  \includegraphics[width=\linewidth]{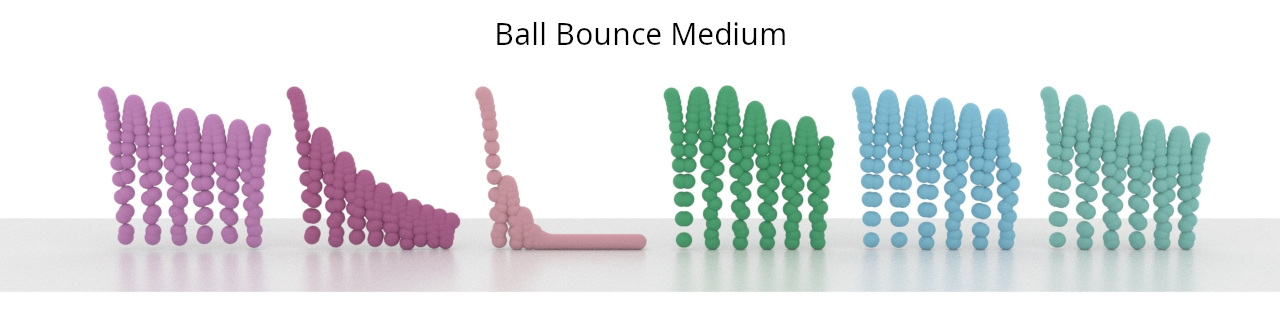}
  \includegraphics[width=\linewidth]{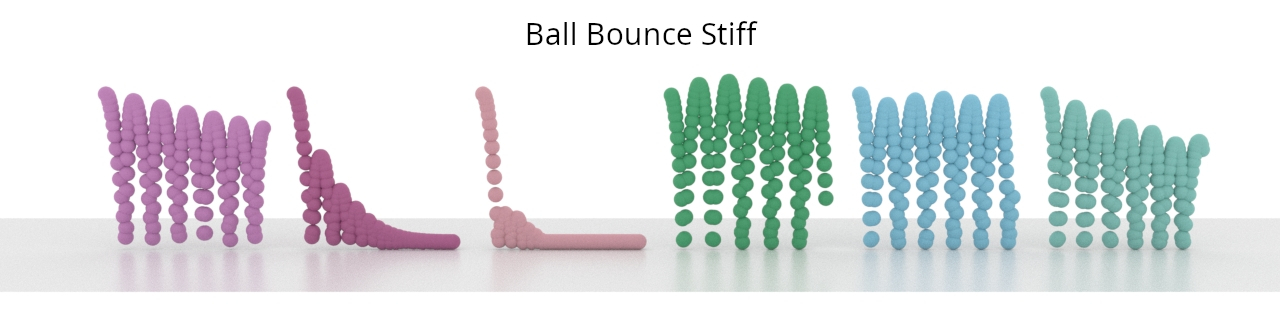}
  \includegraphics[width=\linewidth]{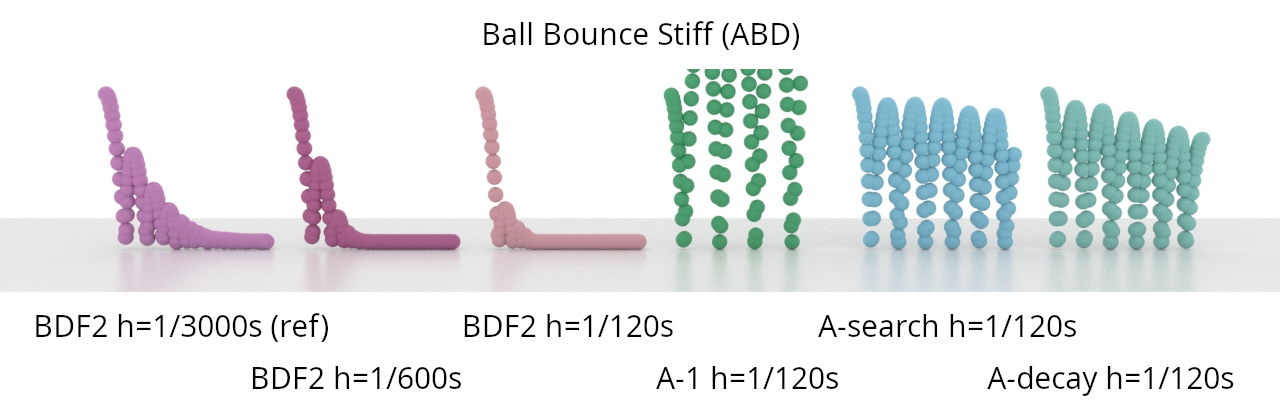}
  \caption{\textbf{Bouncing Ball.} Selected frames over the first 5 s. Across a range of material stiffnesses, \textit{A-search} and \textit{A-1} retain visually plausible rebound behavior at large time steps, while BDF2 exhibits increased damping unless significantly smaller time steps are used.}
  \label{fig:bounce}
\end{figure}

\paragraph{Bouncing ball}
We drop balls of three varying stiffness,
from a height of 1m and allow it to bounce for 10s (Fig. \ref{fig:bounce}).
Moreover, we repeat the stiffest case with ABD.
As expected, BDF2 is overly dissipative, 
while trapezoidal leads to immediate instabilities.
On the contrary, \textit{A-1} is stable,
though it may lead to energy loss or energy gain.
A-search at $h=1/120$s significantly outperforms BDF2 at both the same time step $h=1/120$s and similar runtime $h=1/600$s and leads to good restitution.
Compared to the reference solution, \textit{A-search} has more motion and less higher frequency vibration.
This also causes the full resolution stiff and reduced-order stiff ABD simulation to be similar for \textit{A-search} , 
despite not being similar for the reference solution.
This affirms that \textit{A-search} naturally prefers the low-frequency motion.

\begin{figure*}[h]
  \centering
  \includegraphics[width=0.49\linewidth]{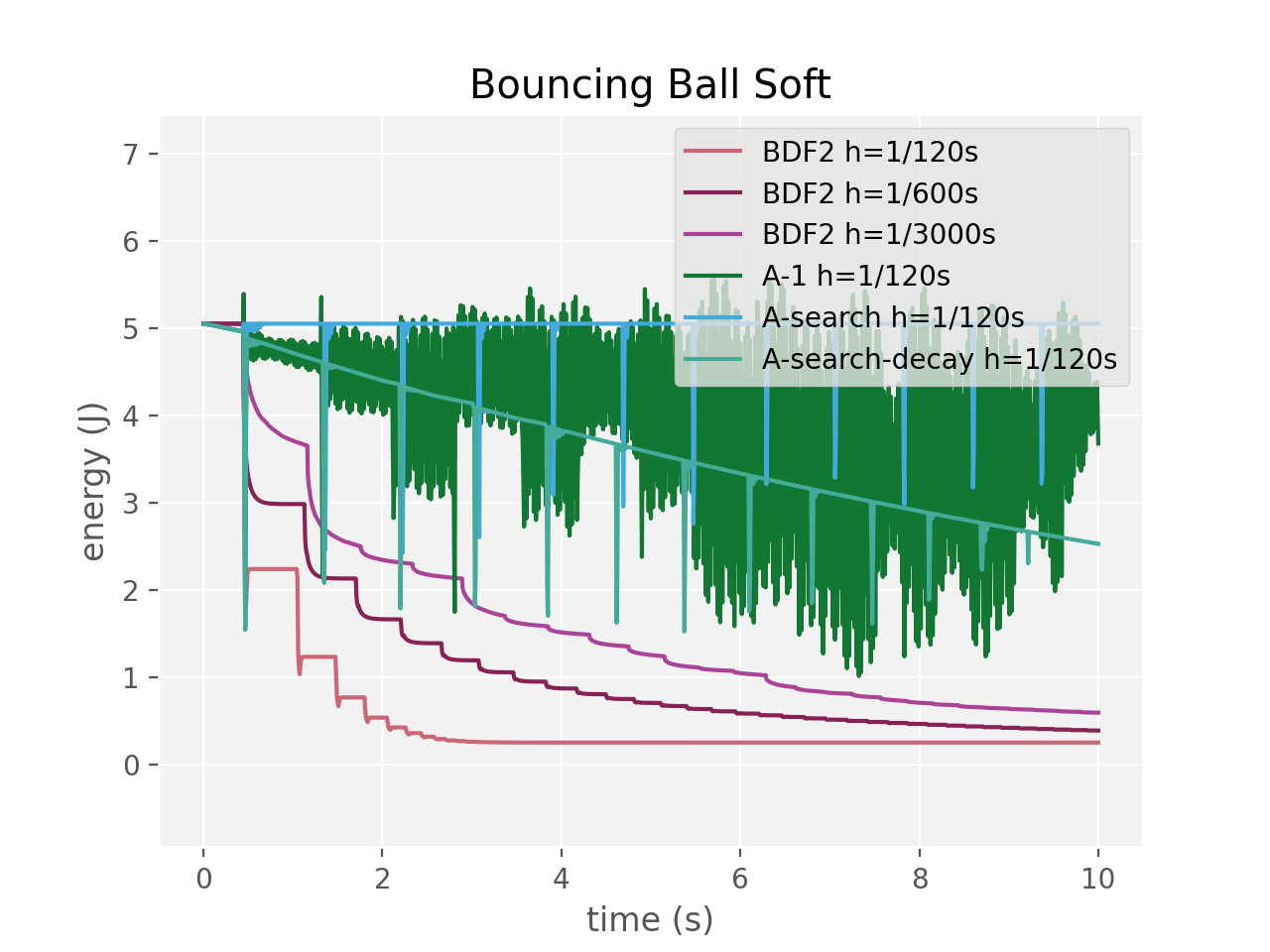}
   \includegraphics[width=0.49\linewidth]{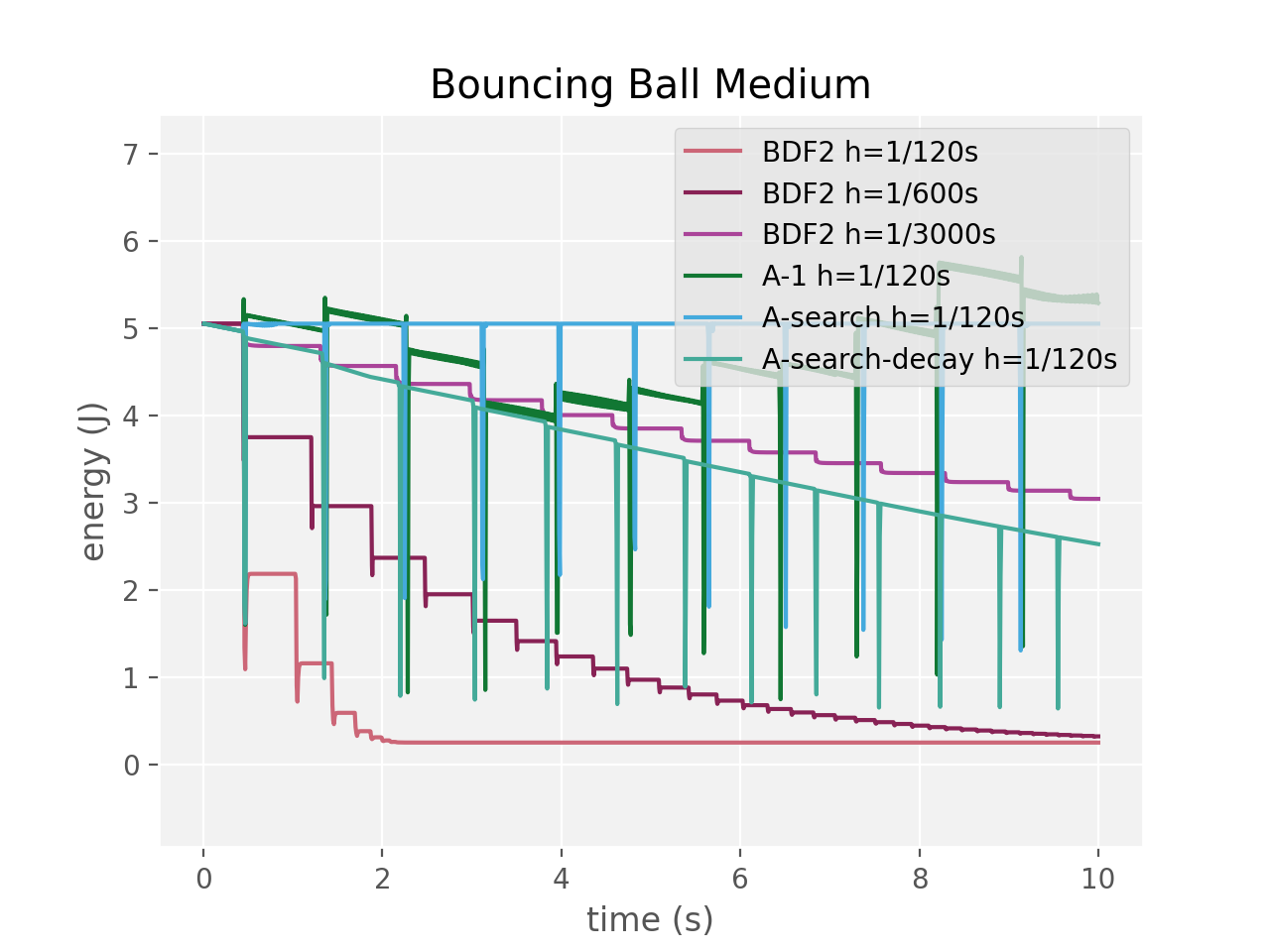}
  \caption{\textbf{Bouncing ball, total energy over time.}
  BDF2 dissipates energy at all tested time steps. \textit{A-1} shows significant energy fluctuations, while \textit{A-search} maintains nearly constant energy except for transient collision-induced spikes. Explicit decay produces controlled energy loss.}
  \label{fig:90011}
\end{figure*}

The above observations are also shown in Fig \ref{fig:90011}. 
Do note the spikes in the energy shown by both A-search and BDF2,
which are inevitable energy losses due to the stiff collision.
A-search is able to recover the energy after the collision,
maintaining an otherwise constant energy level.

In the softest case,
with most oscillations in the reference solution,
energy decay prefers dissipating the oscillations while keeping the bounce,
as can be seen in the video.
In the medium and stiff case,
there is less oscillations, and energy decay instead dissipated the overall bounce.

\paragraph{Vibrating membrane}
We take a vibrating circular elastic membrane with fixed boundary.
Ordinarily, as in a drum membrane,
the tension in the membrane is sufficiently large and the orthogonal displacement is sufficiently small,
so that the motion is very close to linear,
and the linear modes are preserved under time evolution.
However, we take small tension and large displacement to simulate a highly nonlinear scene.
We initialize the velocity to the $u_{21}$ vibration mode with two peaks and two troughs in opposite quadrants. 

While BDF2 at $h=1/120$s or $h=1/240$s resulted in quick dissipation,
in the reference solution BDF2 at $h=1/3000$s,
we find that over time, the initial $u_{21}$ mode disintegrates into other modes of motions in the low and medium frequencies,
as can be seen visually in the many wrinkles.
A-search at $h=1/120$s preserves the $u_{21}$ motion well.
It also exhibits some but not as many wrinkles (Fig. \ref{fig:membrane}).

\paragraph{Dice roll}
We roll eight elastic die with friction coefficients quadratically increasing from $0$ to $1$ and allow rolling for 6s.
We find that BDF2 at $h=1/120$s lead to significant numerical dissipation and not enough bouncing.
On the other hand, \textit{A-search} at $h=1/120$s enables more bouncing and motion,
while having similar total runtime as BDF2 at $h=1/120s$.
It must be acknowledged that \textit{A-search} has less rotational motion compared to the reference,
due to the semi-implicit friction solve at large time steps,
but ultimately its behavior better resembles the reference solution,
especially at low friction coefficients (see video).
If one is willing to decrease the time step for \textit{A-search},
then \textit{A-search} at $h=1/240$s has better rotation, 
and also compares favorably against BDF2 $h=1/240$s while running in less total time.

Like other symplectic integrators,
\textit{A-search} tend to keep the jiggly motion as the die makes small bounces off the ground with negligible friction loss.
Such motion can also be found to a smaller extent in the reference solution,
but nevertheless such behavior may be undesirable.
To avoid this, a small amount of energy decay can help put the dice to rest. 
We use an exponential decay with $\tau = 6$s.

\begin{figure*}[!ht]
  \centering
  \includegraphics[width=0.49\linewidth]{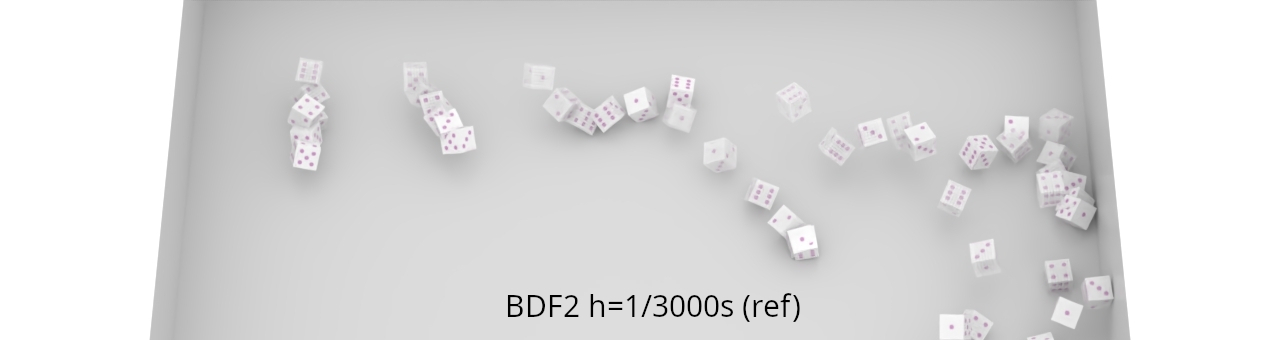}
  \includegraphics[width=0.49\linewidth]{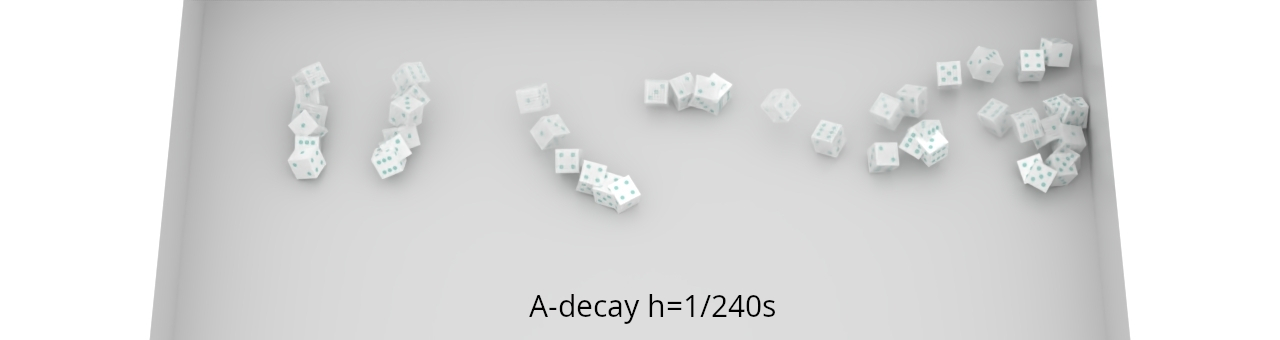}
  \includegraphics[width=0.49\linewidth]{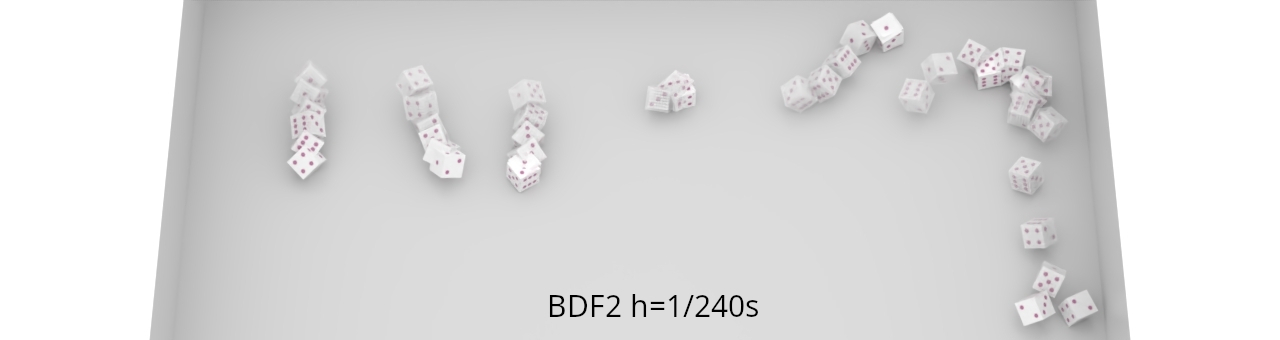}
  \includegraphics[width=0.49\linewidth]{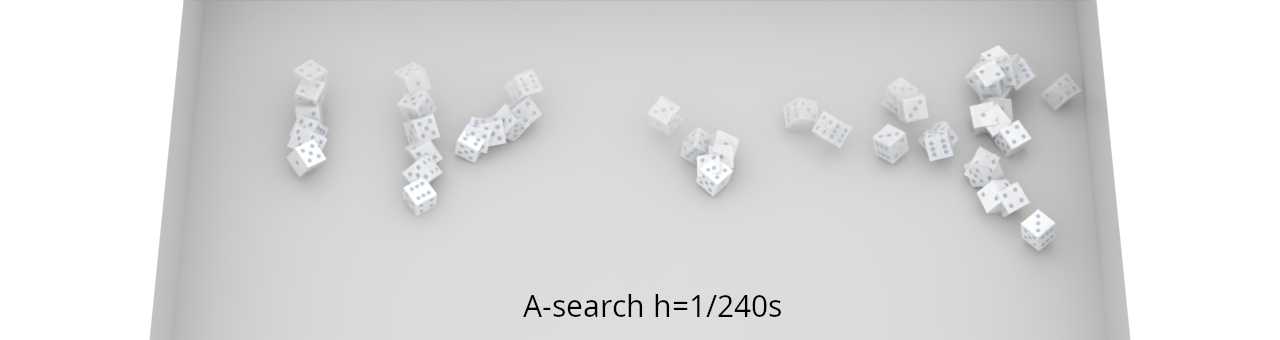}
  \includegraphics[width=0.49\linewidth]{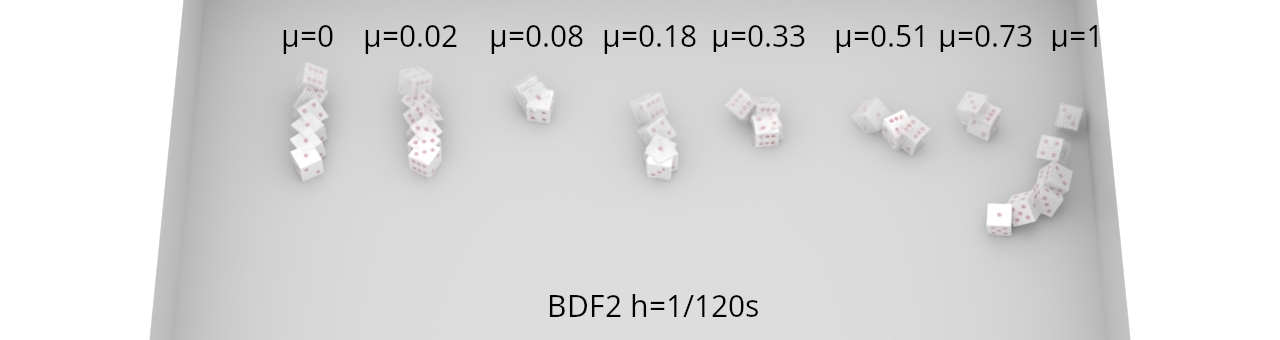}
  \includegraphics[width=0.49\linewidth]{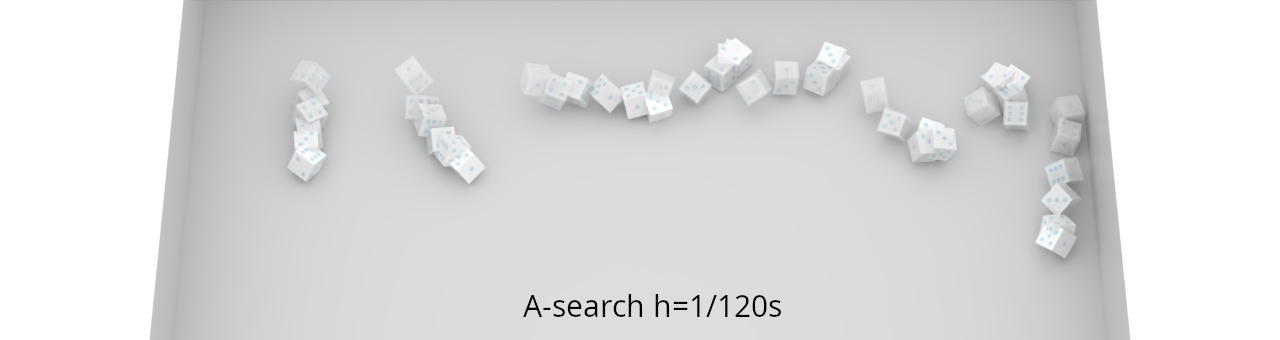}
  \caption{\textbf{Dice roll}, six frames between $0.75s$ and $2s$, superimposed. 
  The dice have increasing coefficient of friction from left to right. 
  A-search has noticeably more rolling and lateral motion than BDF2.}
  \label{fig:dice}
\end{figure*}

\subsection{Demonstrations}

\begin{figure}[h]
  \centering
  \includegraphics[width=0.49\linewidth]{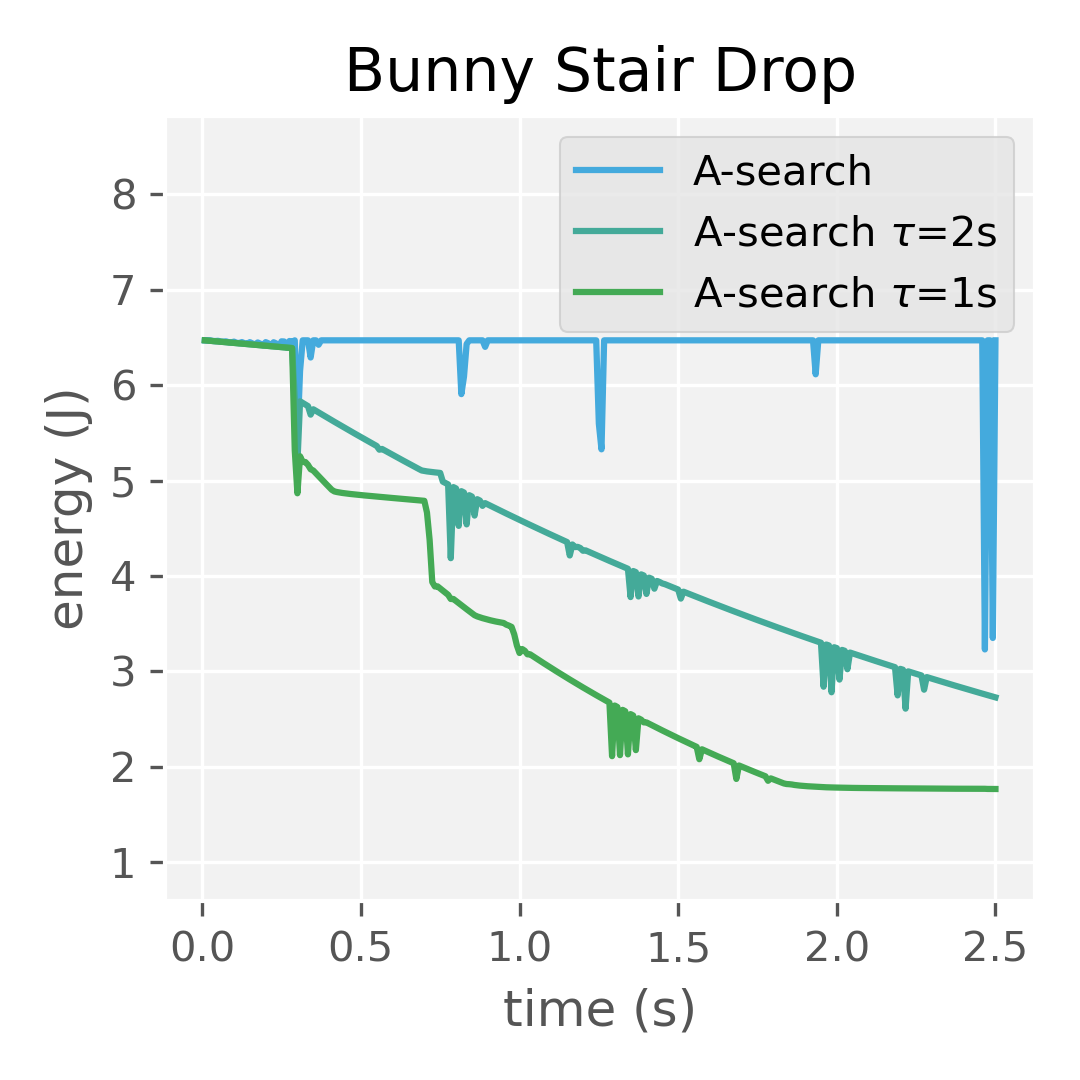}
  \includegraphics[width=0.49\linewidth]{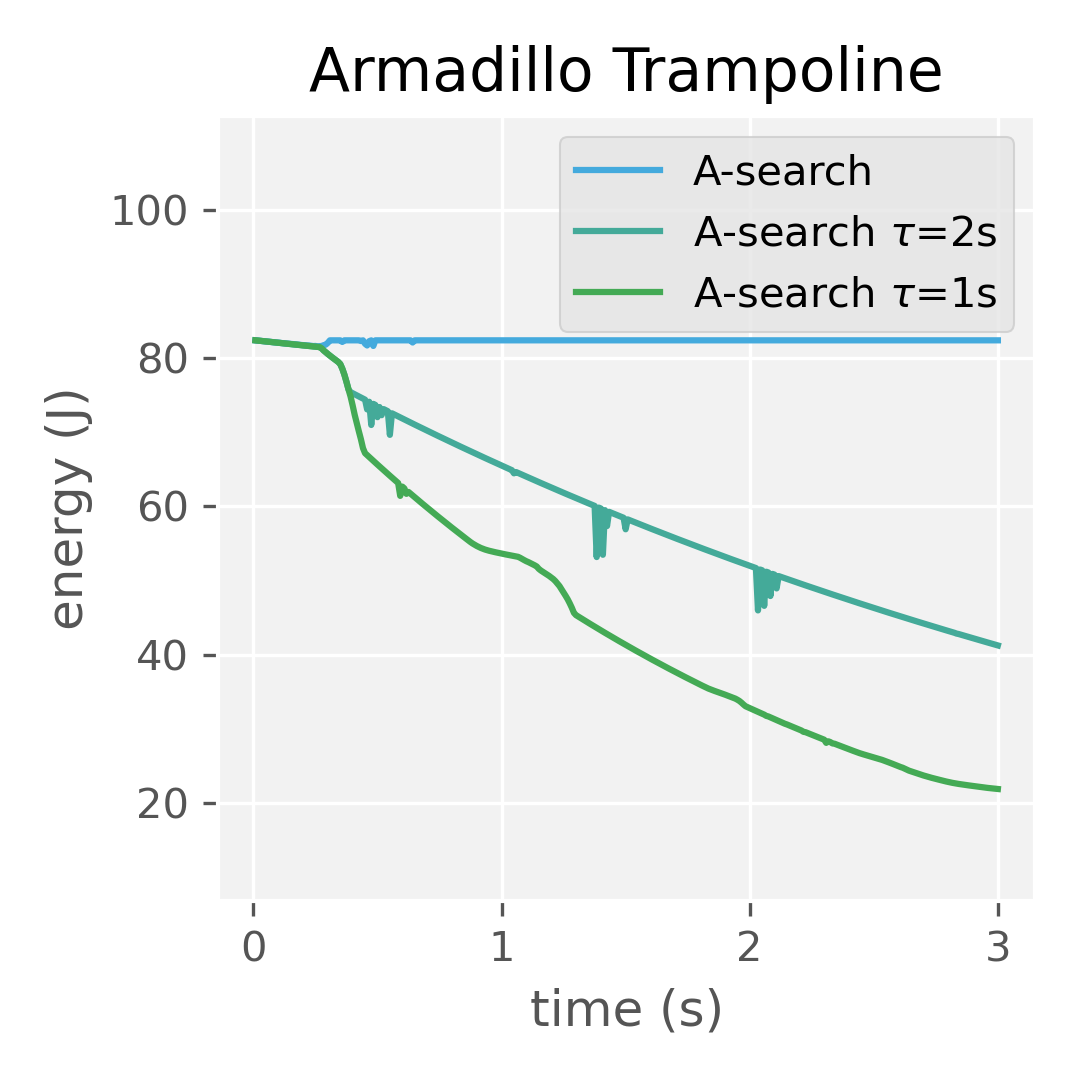}
  \caption{\textbf{Energy plot for bunny stair drop and armadillo trampoline.}
  The exponential decay model leads to a natural loss of energy over time,
  and A-search closely adheres to the desired decay.}
  \label{fig:decay}
\end{figure}

\paragraph{Bunny Stair Drop}
We drop a soft bunny onto a set of stairs and simulate for $2.5s$,
demonstrating A-search's ability to handle non-convex collisions.
By colliding against the stair edges,
the bunny's momentum is converted from the vertical to the horizontal direction.
We repeated the A-search simulation with no energy decay,
$\tau = 2s$ decay, and $\tau = 1s$ decay.
The no decay solution shows a wide range of motion (Fig. \ref{fig:stair}),
difficult to capture with implicit Euler or BDF2. 
Moreover, the two decayed solutions also shows reasonable decay.
The $\tau = 2s$ solution continued to bounce,
while the $\tau = 1s$ decay fell back onto implicit Euler ($\alpha = 0$, since the bunny sliding has more energy than ground energy) at around $1.8s$ and stopped bouncing in a natural manner (Fig. \ref{fig:decay}).

\begin{figure}[h]
  \centering
  \includegraphics[width=0.49\linewidth]{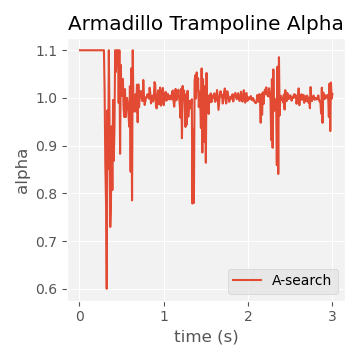}
  \includegraphics[width=0.49\linewidth]{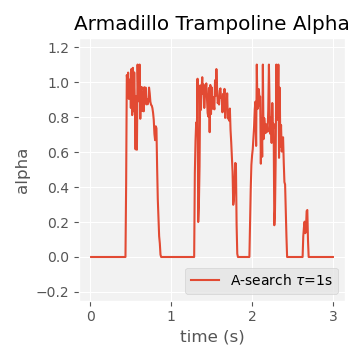}
  \caption{\textbf{Alpha plot for armadillo trampoline.}
  The values of $\alpha$ generally stay near $1$ for energy conservation, though it can go as low as $0$ for energy dissipation. Spikes may occur in $\alpha$ during collision,
  similar to energy.}
  \label{fig:alpha}
\end{figure}

\begin{figure*}[!ht]
  \centering
  \includegraphics[width=0.43\linewidth]{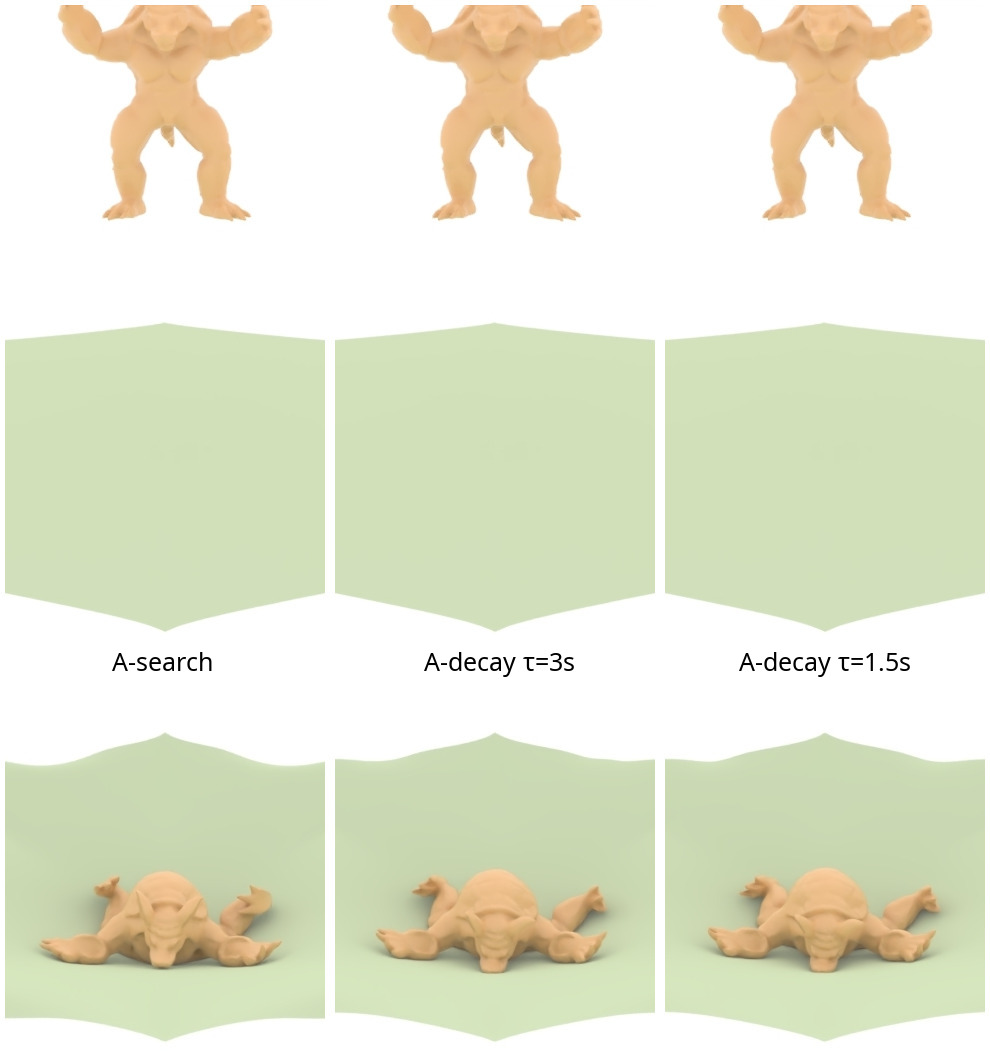}
  \hspace{0.03\linewidth}
  \includegraphics[width=0.43\linewidth]{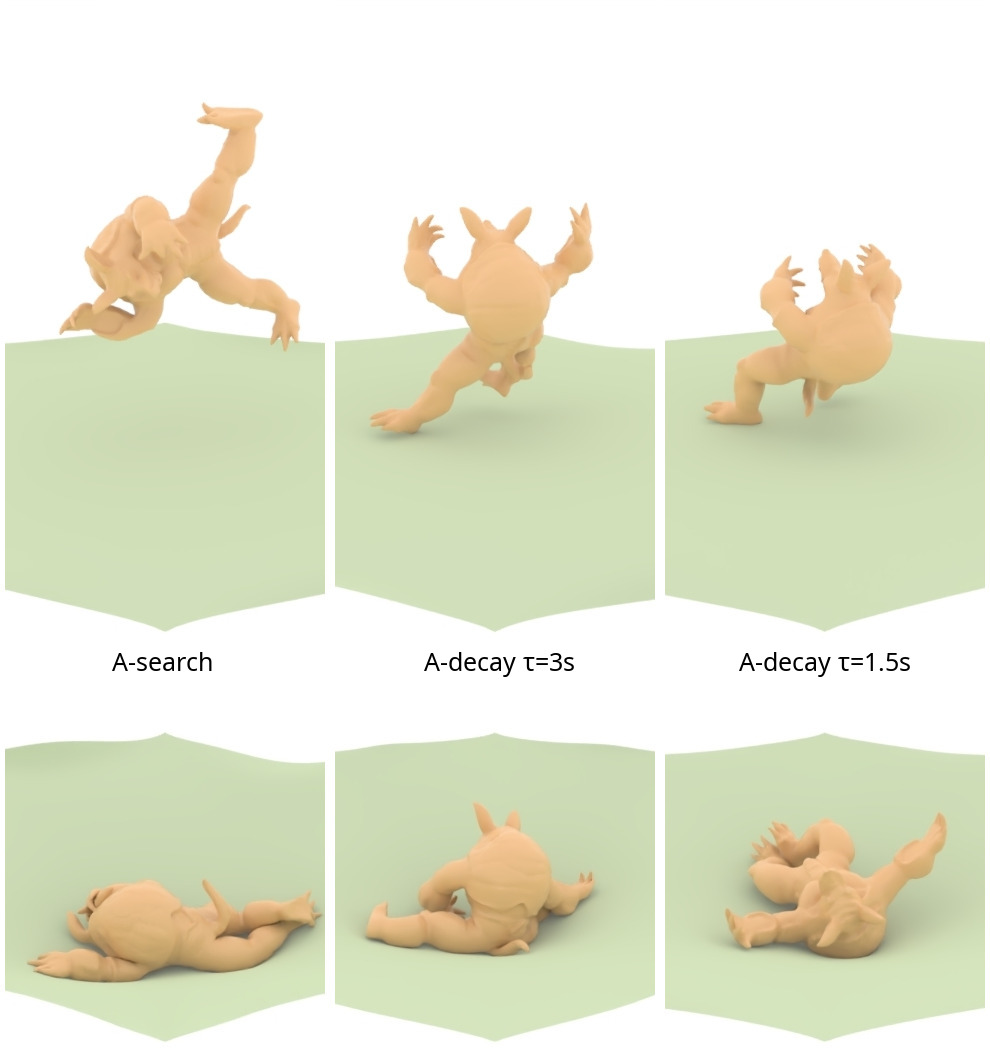}
\includegraphics[width=0.43\linewidth]{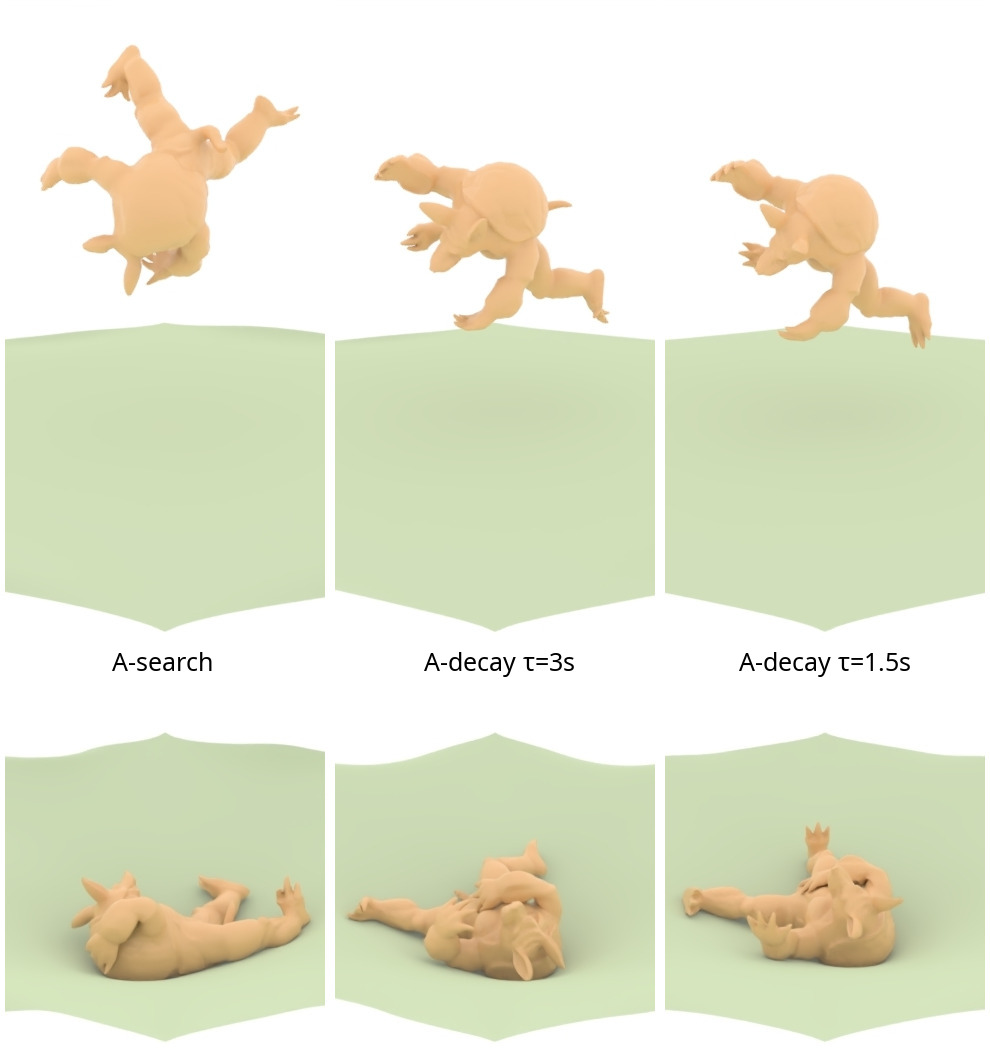}
  \hspace{0.03\linewidth}
  \includegraphics[width=0.43\linewidth]{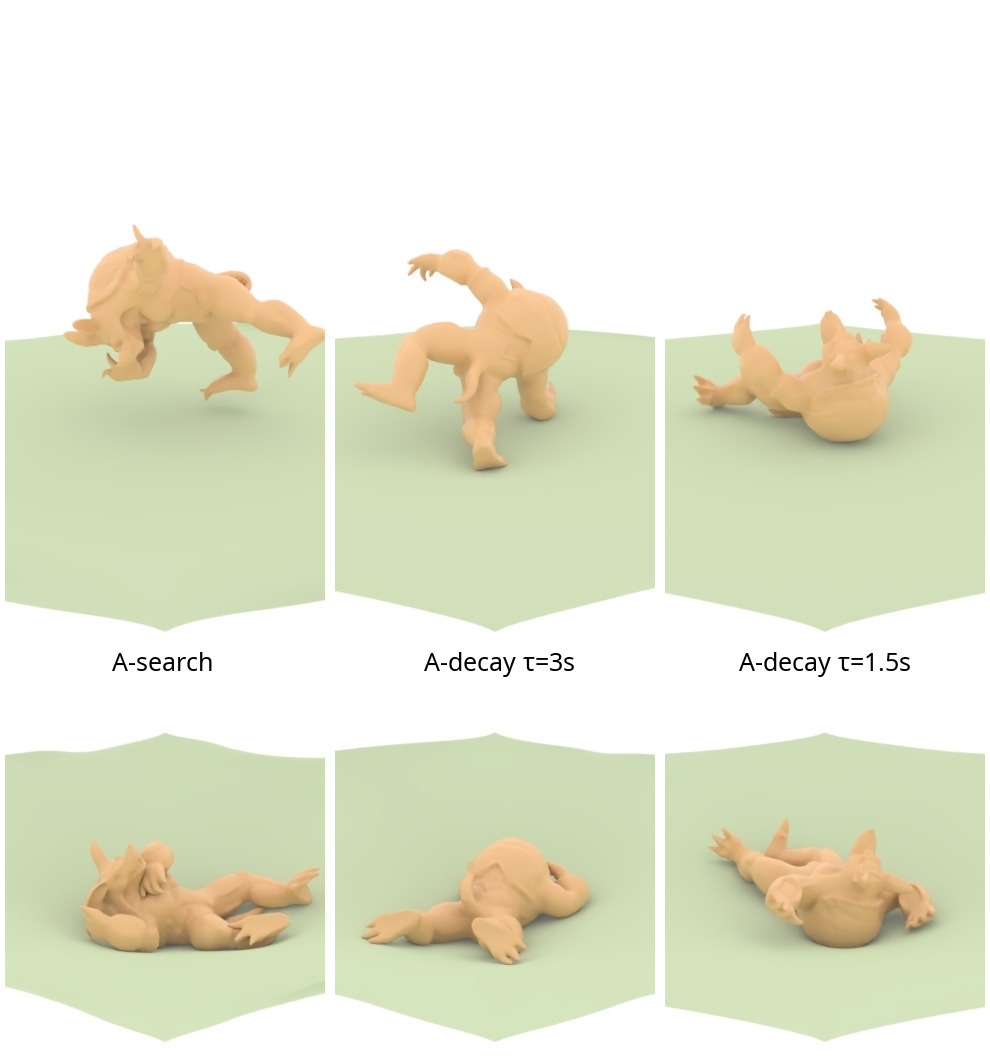}
  \caption{\textbf{Armadillo trampoline, different decays.} Three successive bounces shown for \textit{A-search} and \textit{A-decay} with different decay times $\tau$. \textit{A-search} maintains consistent rebound behavior, while \textit{A-decay} increasingly suppresses motion as the decay rate increases.}
  \label{fig:tramp}
\end{figure*}

\paragraph{Armadillo Trampoline}
We drop a soft armadillo onto an elastic thin-shell trampoline fixed at 8 corners and simulate for $3s$,
examining A-search's behavior for scenes with coupled co-dimensional objects. 
We repeated the A-search simulation with no energy decay,
$\tau = 3s$ decay, and $\tau = 1.5$s decay.
The no decay solution shows a wide range of humorous motion (Fig. \ref{fig:tramp}),
difficult to capture with implicit Euler or BDF2. 
Moreover, the two decayed solutions also lead to reasonable behavior with less motion.

It is worthwhile to examine the typical values of $\alpha$ (Fig. \ref{fig:alpha}).
During the initial free fall, $\alpha = 1.1$ for energy conservation, since free fall loses energy,
but no force differences are present for energy correction.
Correction only occurs after contact.
Generally $\alpha$ oscillates near $1$, 
and $\alpha$ tends to spike the most during collisions,
which is expected.
With decay, $\alpha$ averages to be smaller than $1$.
It gracefully falls to $0$ in the later portions of each bounce,
where all motion has been dissipated to implicit Euler, yet our energy curve demands even more nonphysical dissipation.

\paragraph{Chain Net}
We drop a stiff ball on a stiff chain net with ABD and run for $3s$, similar to the test described in \citet{lan2022affine}.
The energy target is initialized to $E_0 = 0.95 H_0$,
with an exponential decay $\tau = 1s$ starting after $t=1s$.
A-search appropriately resolves the complicated contact behavior in this scene,
giving multiple good ball bounces back above the net in roughly the first half of the simulation (Fig. \ref{fig:chain}),
while slowly resting the ball to the bottom in the second half.

\paragraph{Bunny Splash}
We drop a soft bunny into an array of soft cubes for $3s$.
The energy target is initialized to $E_0 = 0.95 H_0$,
with an exponential decay $\tau = 1s$ starting after $t=1s$ 
and increased to $\tau = 0.5s$ after $t=1.5s$.
Due to the large deformations attained by both the bunny and cubes, 
this scene is difficult to simulate while conserving energy, though A-search is able to transfer the energy from the bunny's motion into the cubes' motion and keep the cubes bouncing for a short while.
Additionally, this scene includes static and kinetic friction, with friction coefficient $\mu = 0.25$, 
which A-search is able to capture (Fig. \ref{fig:splash}).

\subsection{Performance}
All experiments except bunny splash were conducted on a machine with an Intel Xeon w5-2445 CPU, 32 GB of memory, and an NVIDIA RTX A1000 GPU.
Bunny splash was on a machine with an Intel Core i9-14900KF CPU, 32 GP of memory,
and an NVIDIA RTX 4080 GPU.
Our performance statistics is shown in Table \ref{fig:runtime}.
The material parameters can be found in Appendix \ref{appendix:material}.

In general, the relation between the simulation time and energy behavior is complex. Traditionally, e.g. BDF2, the timestep controls the runtime, 
but the energy profile is unpredictable. 
For A-search, the user doesn’t need to tune the timestep,
and instead controls the energy profile.
The runtime is unpredictable, but we can still believe from presented experiments that the runtime would compete well against BDF2 if we require good energy conservation. 
There is a tradeoff between being able to control the runtime and energy profile, and BDF2 and A-search are on opposite ends. We believe that A-search is especially useful because it is controllable and thus reduces the time the user must spend tuning the timestep parameter to obtain desired results. 

It is also worth mentioning that, at equal runtimes,
the total Newton iterations for A-search tend to be less than that of BDF2,
or alternatively, the time spent per Newton iteration via conjugate gradient is larger.
This means that the linear systems for A-search are worse conditioned,
which is to be expected, as A-search must resolve new high frequency data every frame.
By using alternative means of solving the linear system, 
A-search has the potential to receive a better speedup than BDF2.

\begin{table*}[h]
\begin{tabular}{l|l|l|rr|rr|rr|rr} \toprule
                    \multicolumn{1}{c|}{Performance}  & &   & \multicolumn{2}{c}{} & \multicolumn{2}{|c}{BDF2 (equal} &  \multicolumn{2}{|c}{BDF2 (similar} &  \multicolumn{2}{|c}{BDF2}  \\ 
                   \multicolumn{1}{c|}{Runtime / NIters}  & Nodes, Tets, Faces  & \multicolumn{1}{c|}{Time step sizes}           & \multicolumn{2}{c}{A-search} & \multicolumn{2}{|c}{time step size)} &  \multicolumn{2}{|c}{runtime)} &  \multicolumn{2}{|c}{(reference)}  \\ \midrule
Soft square         & 1.3k, 6.0k, 1.2k  & 1/30s,1/60s,1/1500s   & 27.6                         & 902                  & 20.3                                & 661                  & 28.3                                & 1203                 & 160.0                        & 16143                \\
Stiff square        & 1.3k, 6.0k, 1.2k  & 1/30s,1/60s,1/1500s   & 43.8                         & 1500                 & 27.3                                & 822                  & 43.0                                & 1349                 & 282.8                        & 30000                \\
Twisted bar         & 4.8k, 24k, 3.2k   & 1/30s,1/90s,1/1500s   & 113.5                        & 947                  & 60.8                                & 409                  & 98.6                                & 1546                 & 184.5                        & 1632                 \\
Suspended arm. & 28k, 110k, 30k    & 1/30s,1/300s,N/A        & 1458.7                       & 1870                 & 396.6                               & 453                  & 1020.9                              & 5750                 & \multicolumn{1}{l}{}         & \multicolumn{1}{l}{} \\
Soft ball           & 1.6k, 7.1k, 0.7k  & 1/120s,1/600s,1/3000s & 75.0                         & 2681                 & 36.3                                & 868                  & 137.5                               & 8342                 & 581.8                        & 60097                \\
Medium ball         & 1.6k, 7.1k, 0.7k  & 1/120s,1/600s,1/3000s & 68.3                         & 1683                 & 34.6                                & 423                  & 163.8                               & 6955                 & 516.2                        & 32637                \\
Stiff ball          & 1.6k, 7.1k, 0.7k  & 1/120s,1/600s,1/3000s & 94.2                         & 2391                 & 37.3                                & 326                  & 213.5                               & 5739                 & 657.6                        & 30674                \\
Stiff ball (ABD)      & 0.6k, 0, 1.3k     & 1/120s,1/600s,1/3000s & 224.9                        & 8902                 & 48.4                                & 1221                 & 227.7                               & 6064                 & 1252.8                       & 32514                \\
Membrane            & 16k, 0, 31k       & 1/120s,1/240s,1/3000s & 286.5                        & 954                  & 185.9                               & 442                  & 284.0                               & 804                  & 1213.2                       & 12762                \\
Dice roll\textsuperscript{*}               & 5.8k, 24.6k, 6.1k & 1/120s,N/A,1/3000s      & 326.6                        & 4659                 & 307.3                               & 5744                 &          & & 2413.3                       & 146443  \\ \midrule
 & &   & \multicolumn{2}{c}{} & \multicolumn{2}{|c}{A-search} &  \multicolumn{2}{|c}{A-search} &  \multicolumn{2}{|c}{}  \\ 
                    &   &     & \multicolumn{2}{c}{A-search} & \multicolumn{2}{|c}{(weak decay)} &  \multicolumn{2}{|c|}{(strong decay)} &  \\ \midrule
Bunny stairs         & 7.5k, 19k, 12k    & 1/120s                & 530.9                        & 4534                 & 342.1                               & 2248                 & 369.1                               & 2092                 &                       &                \\
Arm. trampoline & 33k, 110k, 43k    & 1/120s                & 7570.9                       & 13145                & 6502.7                              & 10975                & 4498.0                              & 7411                 &                       &                \\
Chain net            & 47k, 0, 93k       & 1/120s                & 6612.0                       & 21683                &                              &                &                              &                &                       &                \\
Bunny splash         & 19k, 43k, 31k     & 1/120s                & \textsuperscript{\textdagger}1303.1                          & 5799                    &                              &                &                              &                &                       &               \\ \bottomrule
\end{tabular}
\caption{\textbf{Runtime statistics.} Reported is total runtime in seconds and
total Newton iterations. The three step sizes refer to that of
1. A-search and BDF2 with equal time step size; 2. BDF2 with runtime comparable to A-search at the time step size in 1; 3. BDF2 reference. 
*At $h=1/240s$, BDF2 had runtime 307.2s, 5744 iterations, A-search had runtime 335.3s, 7224 iterations.
\textdagger On a different machine.}
\label{fig:runtime}
\end{table*}

\subsection{Convergence}

\begin{figure*}[p]
  \centering
  \includegraphics[width=0.31\linewidth]{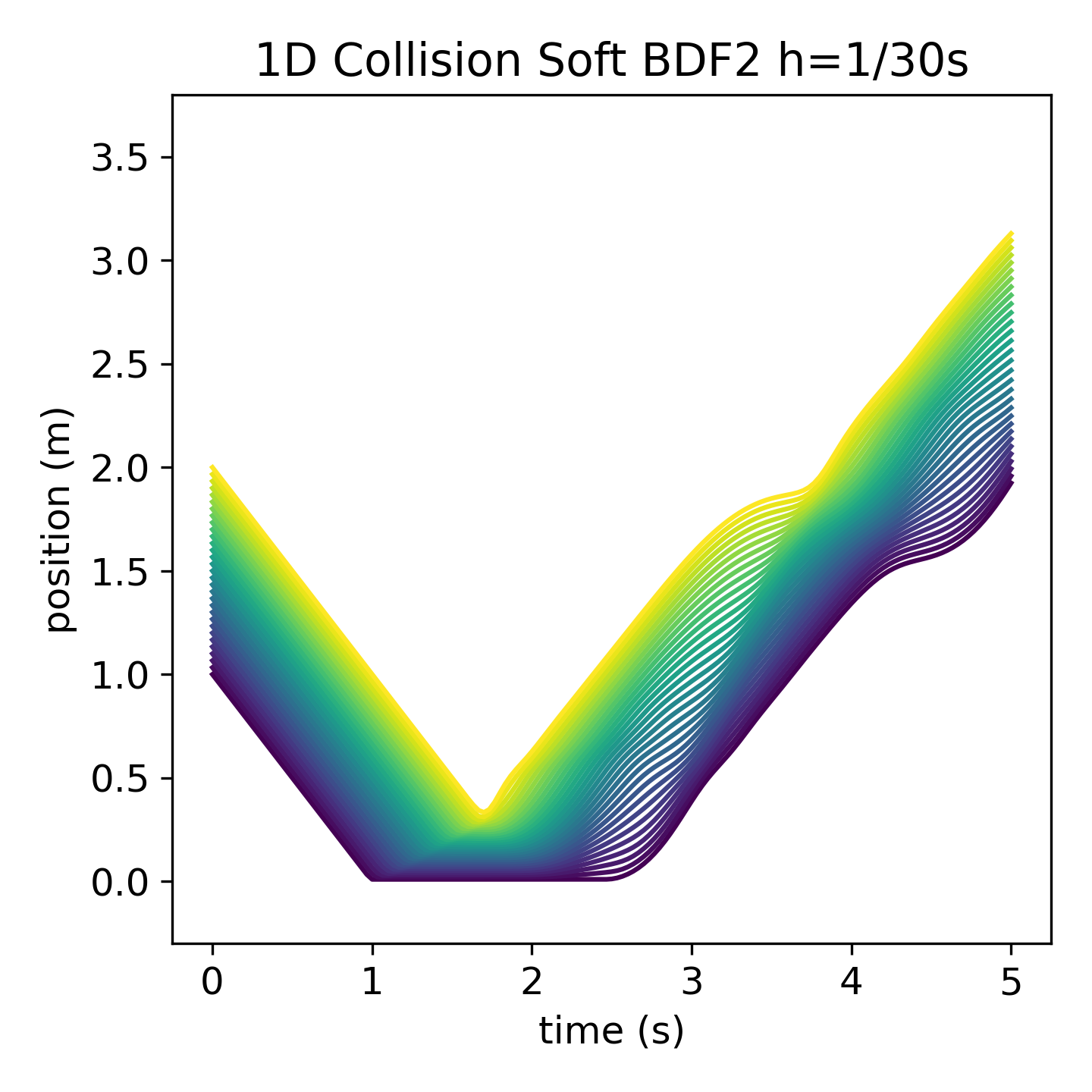}
  \includegraphics[width=0.31\linewidth]{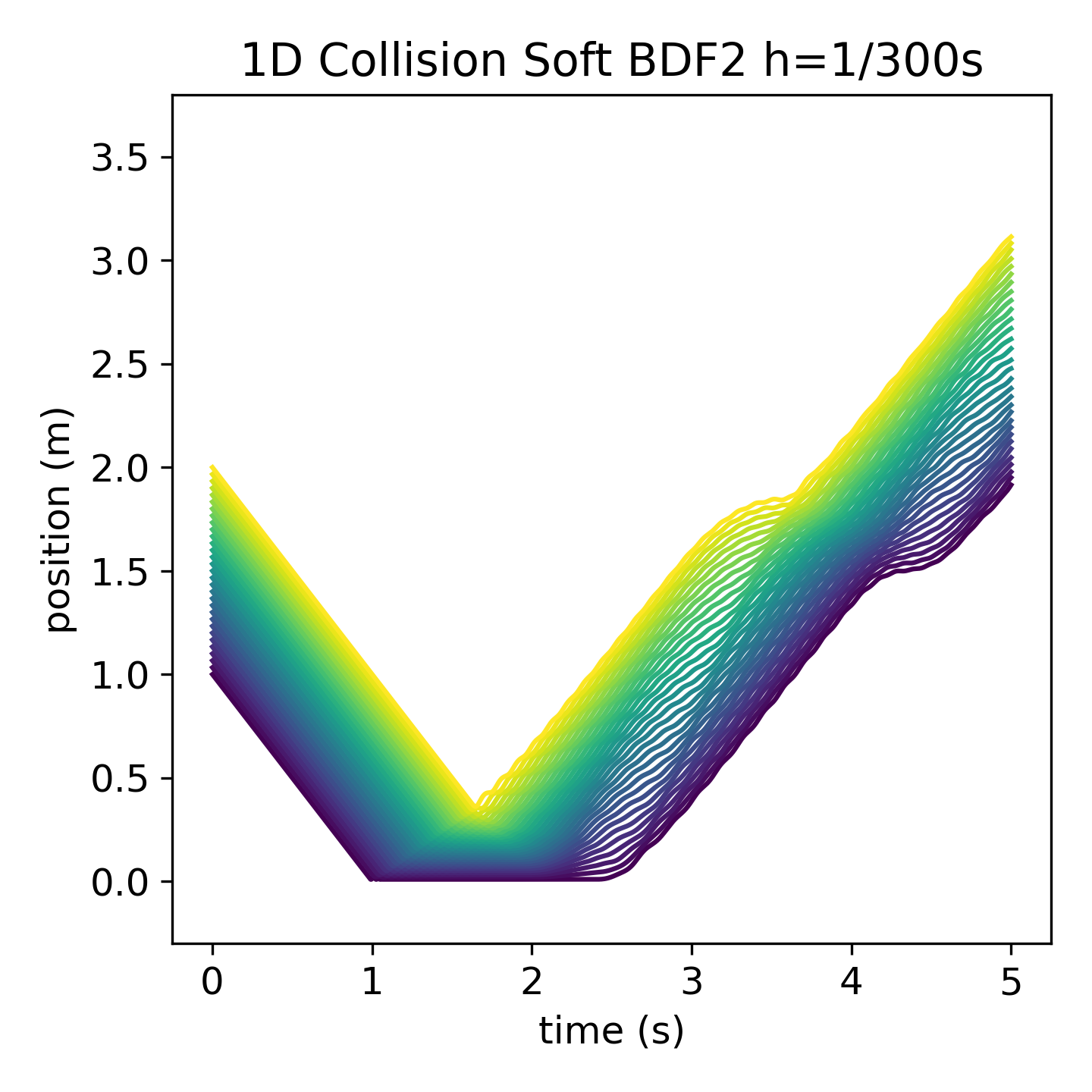}
  \includegraphics[width=0.31\linewidth]{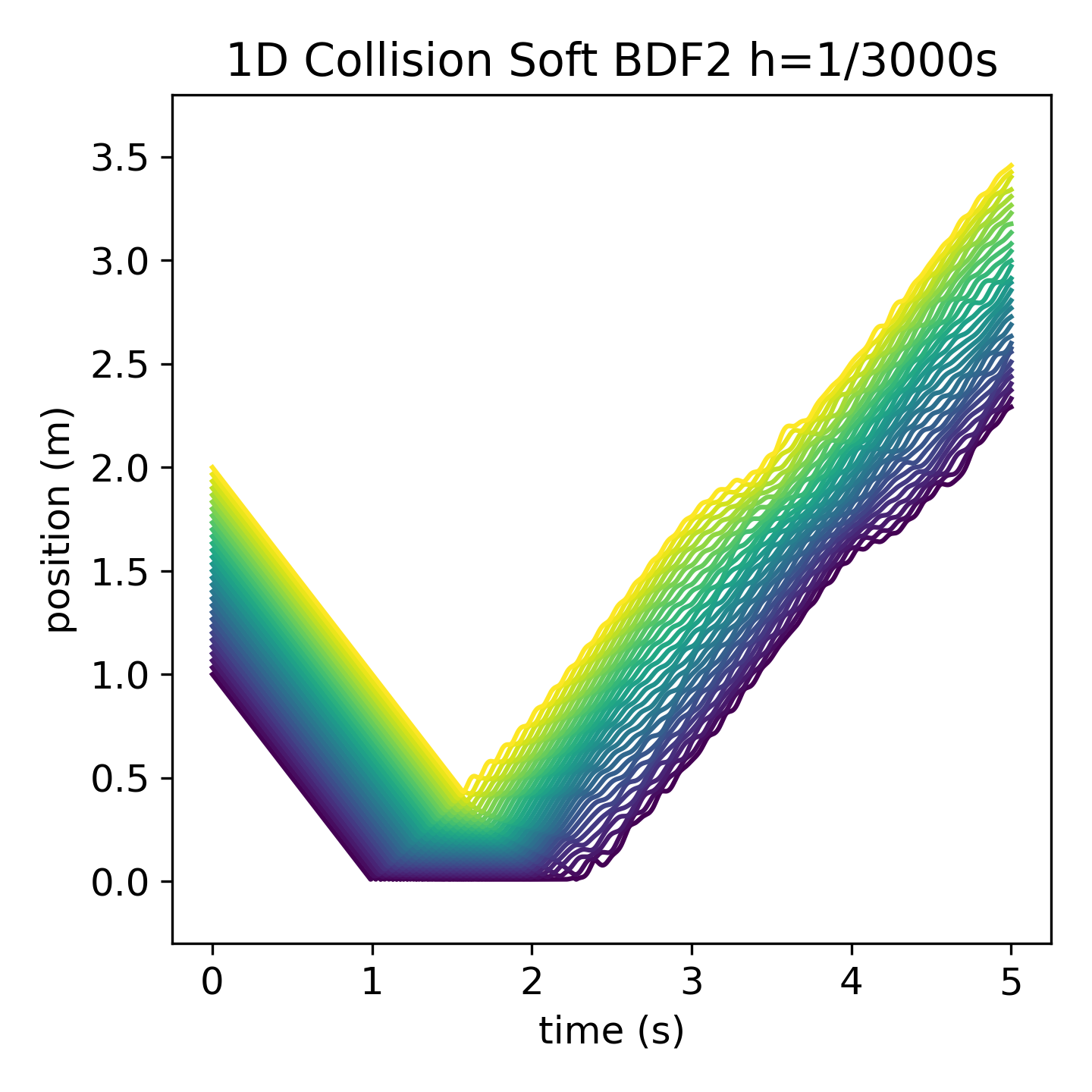}

  \includegraphics[width=0.31\linewidth]{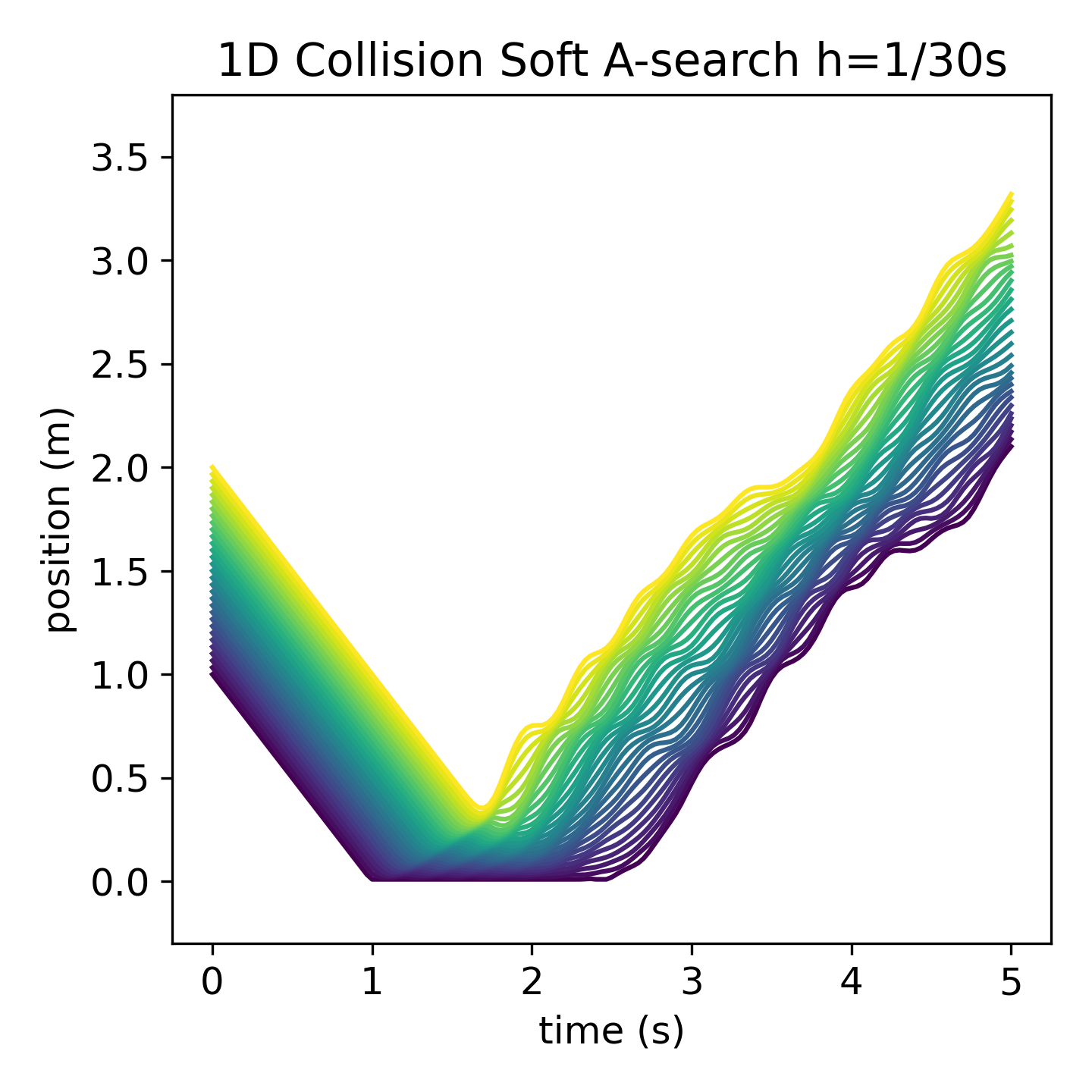}
  \includegraphics[width=0.31\linewidth]{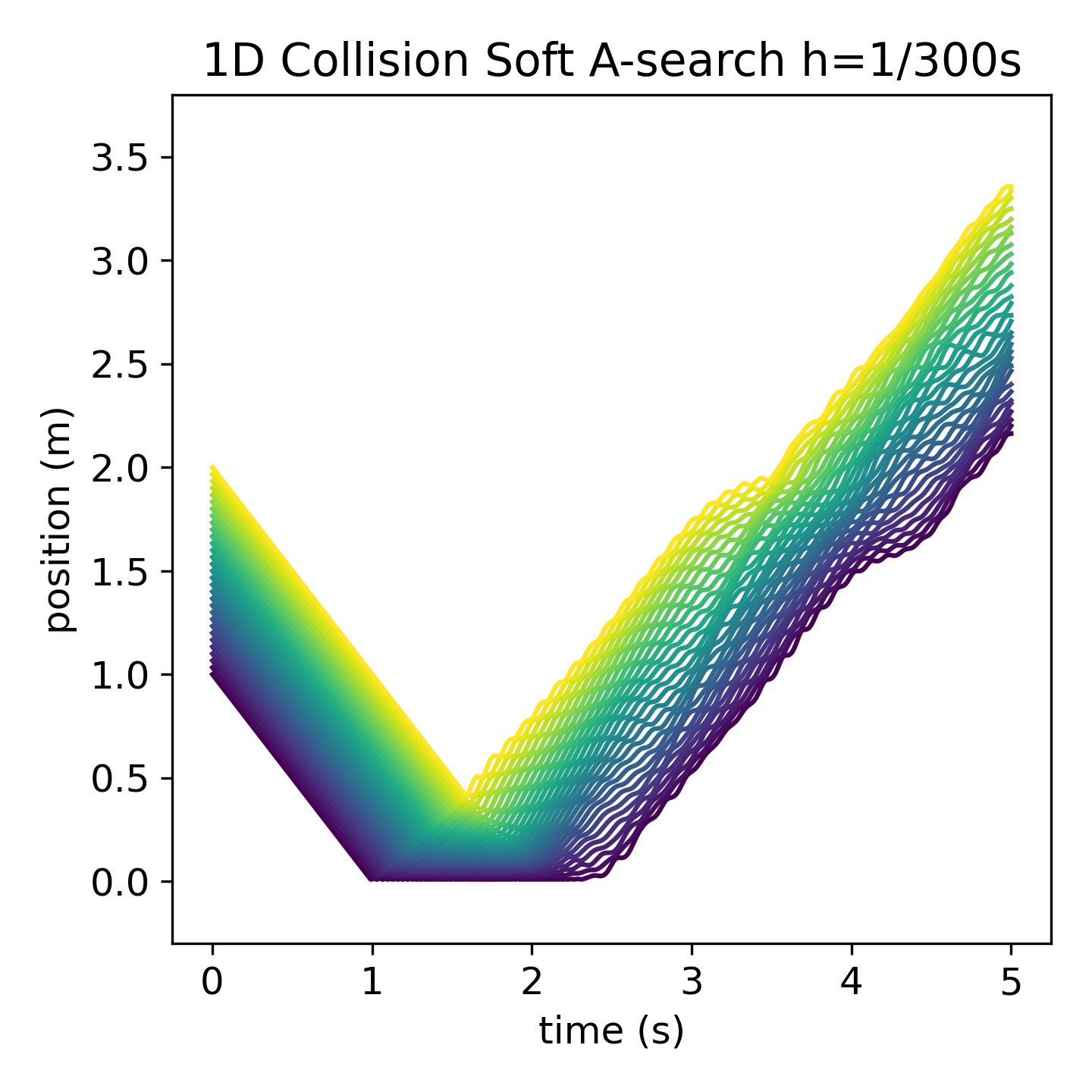}
  \includegraphics[width=0.31\linewidth]{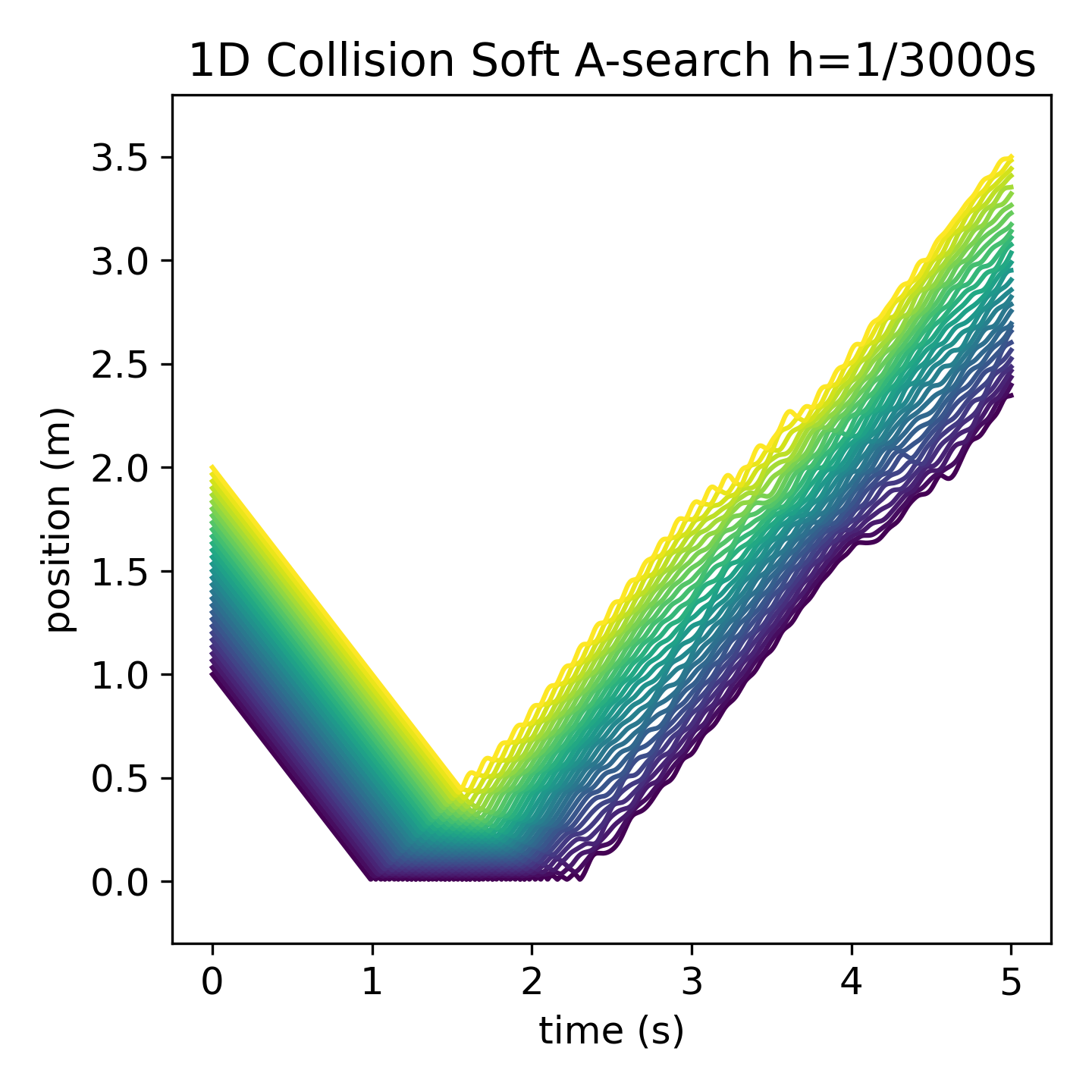}

  \includegraphics[width=0.31\linewidth]{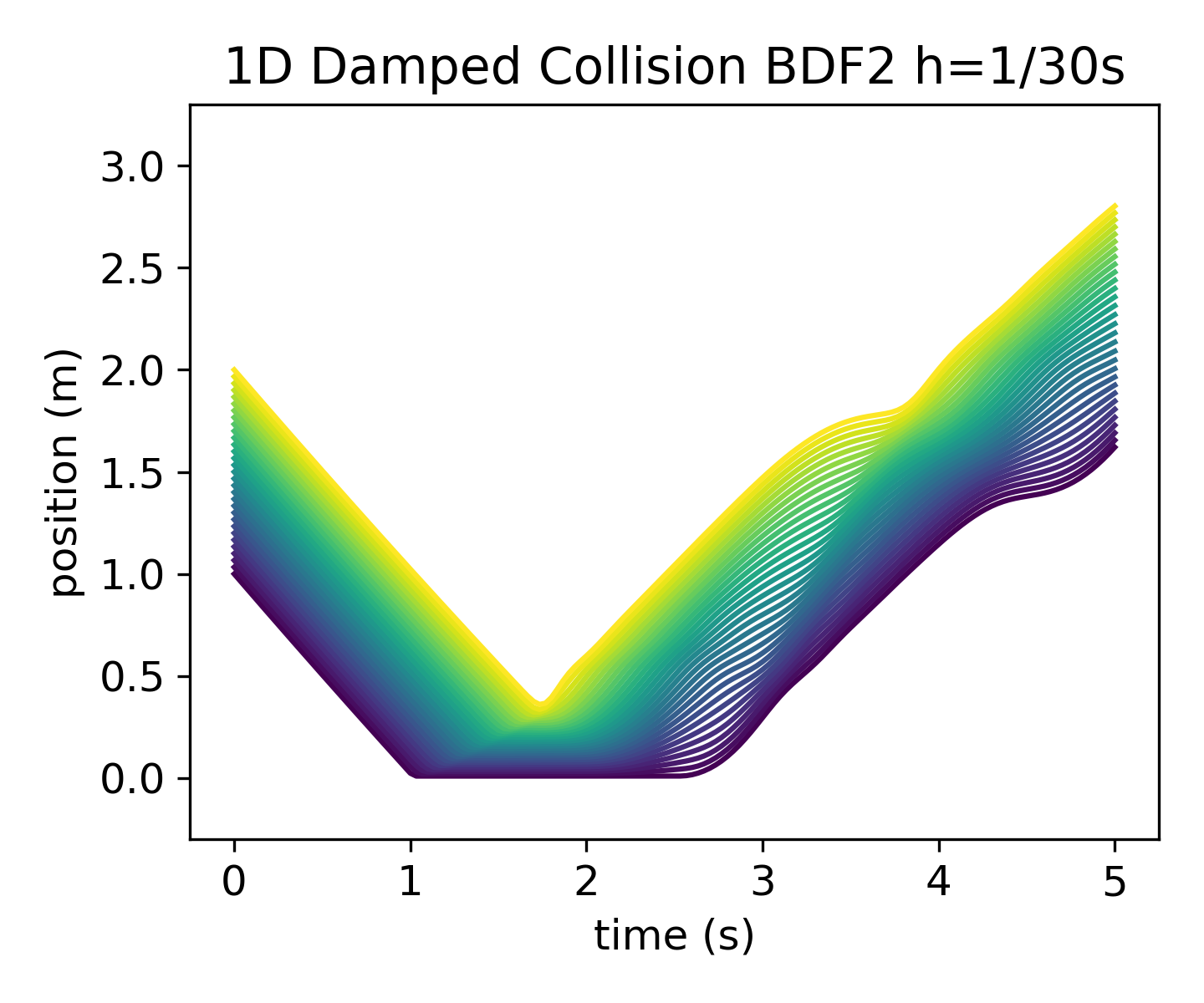}
  \includegraphics[width=0.31\linewidth]{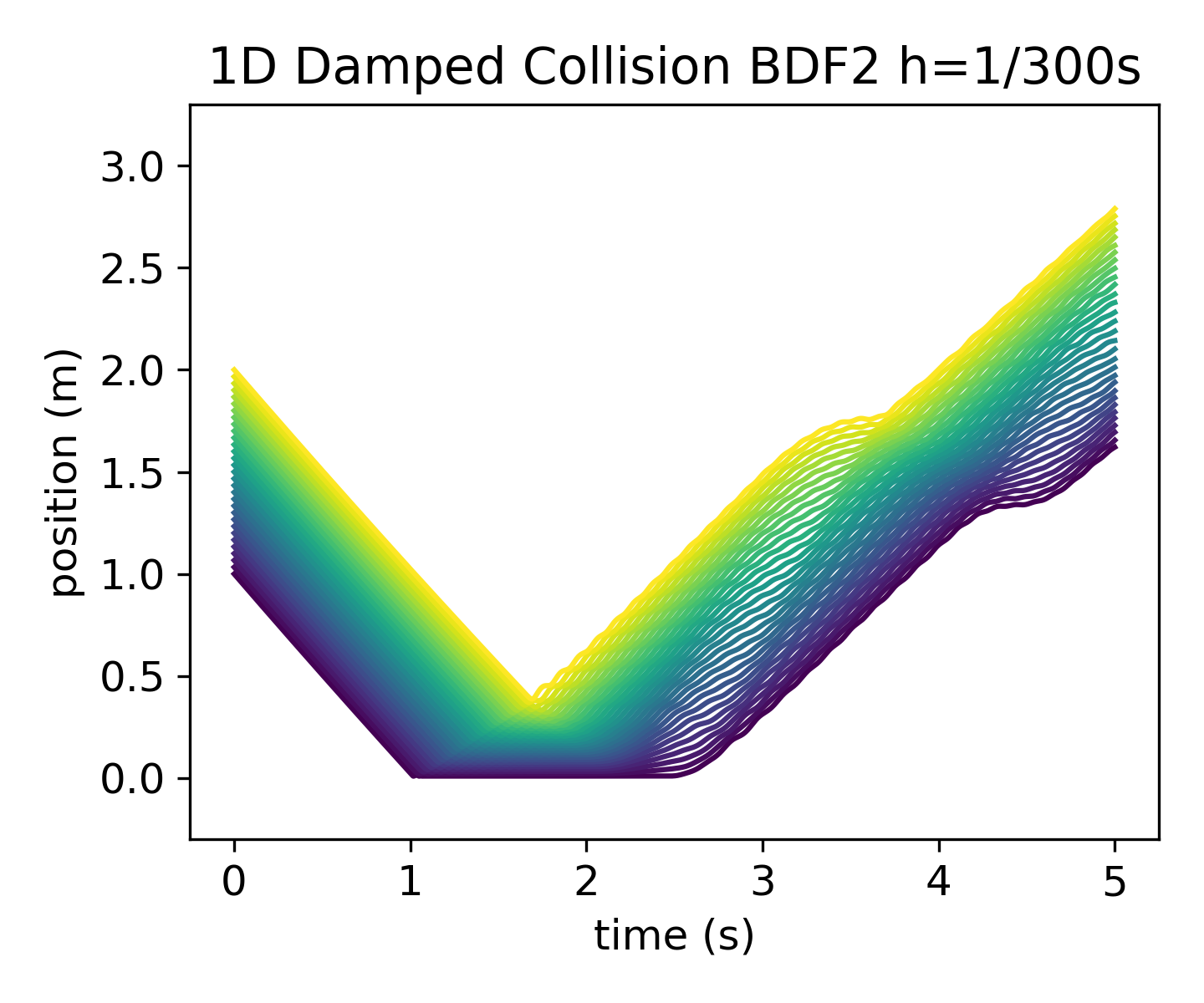}
  \includegraphics[width=0.31\linewidth]{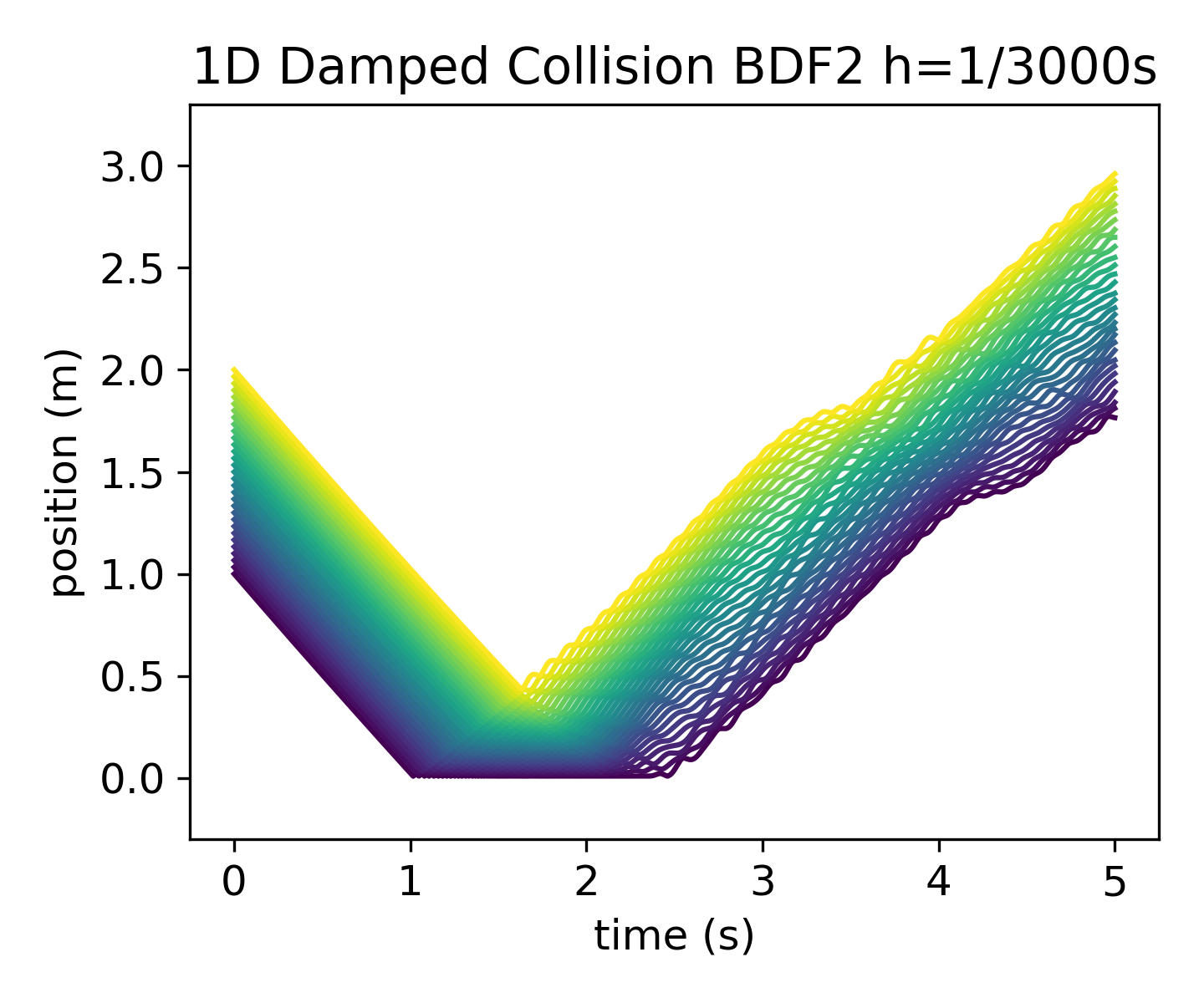}
  \includegraphics[width=0.31\linewidth]{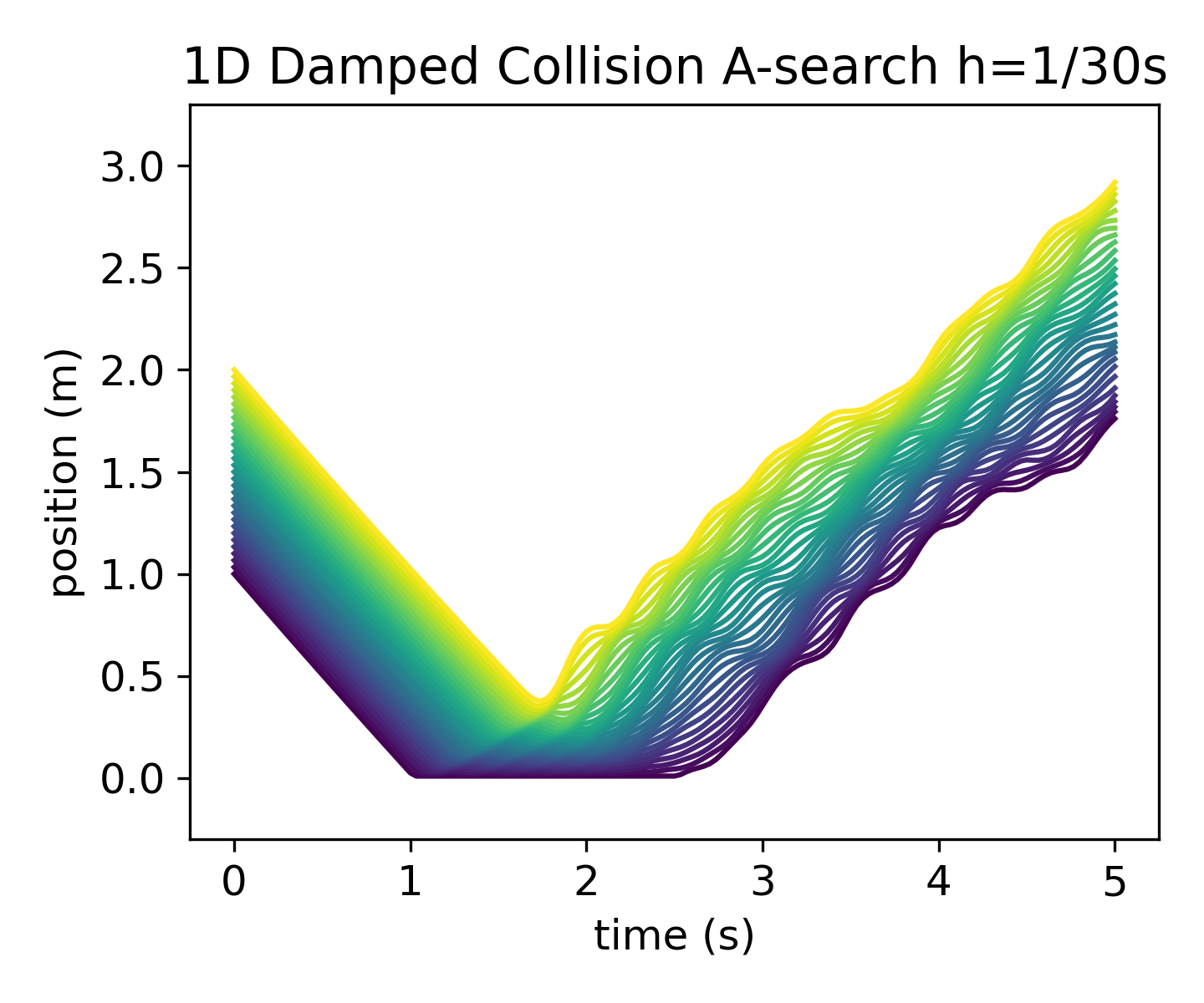}
  \includegraphics[width=0.31\linewidth]{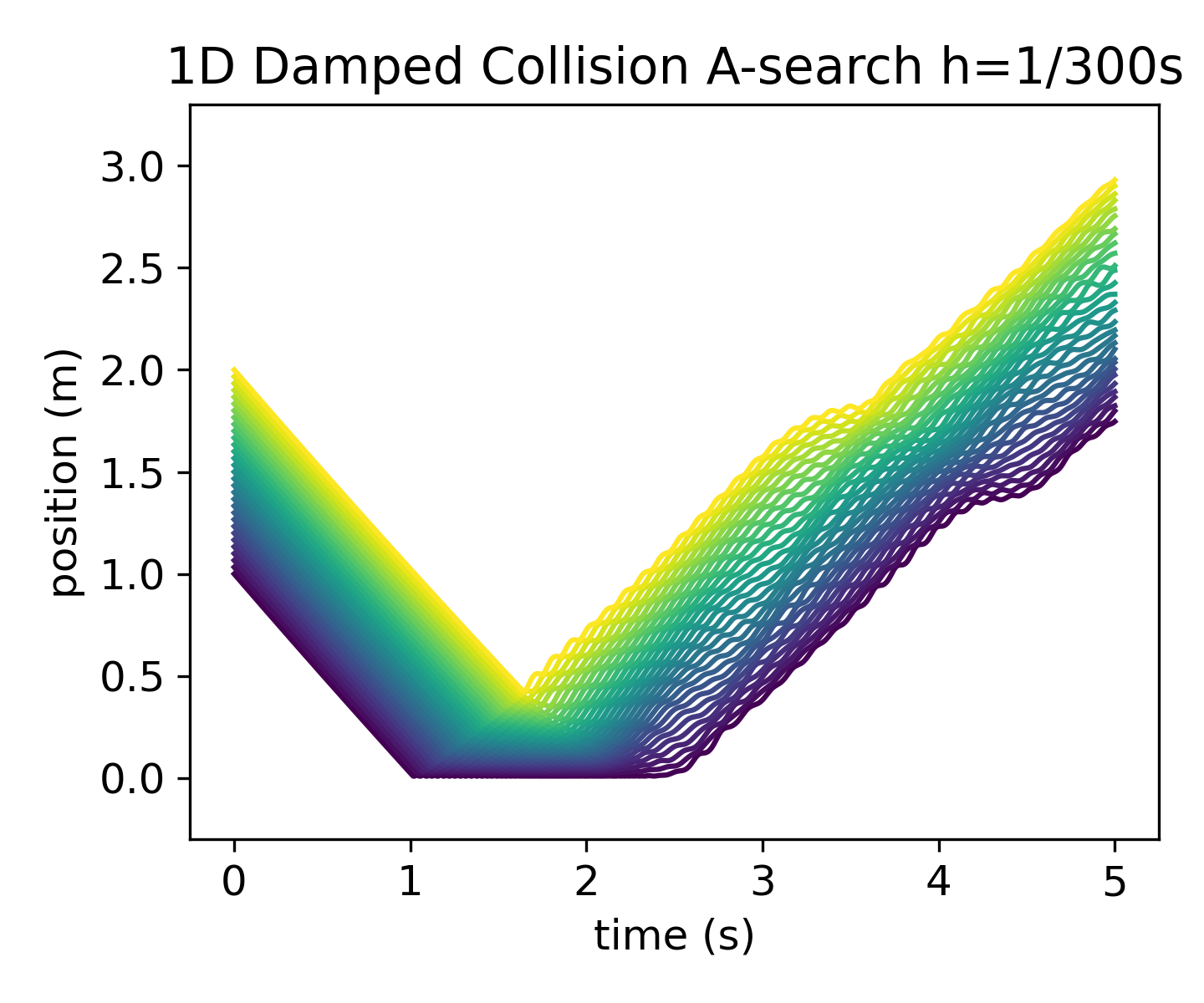}
  \includegraphics[width=0.31\linewidth]{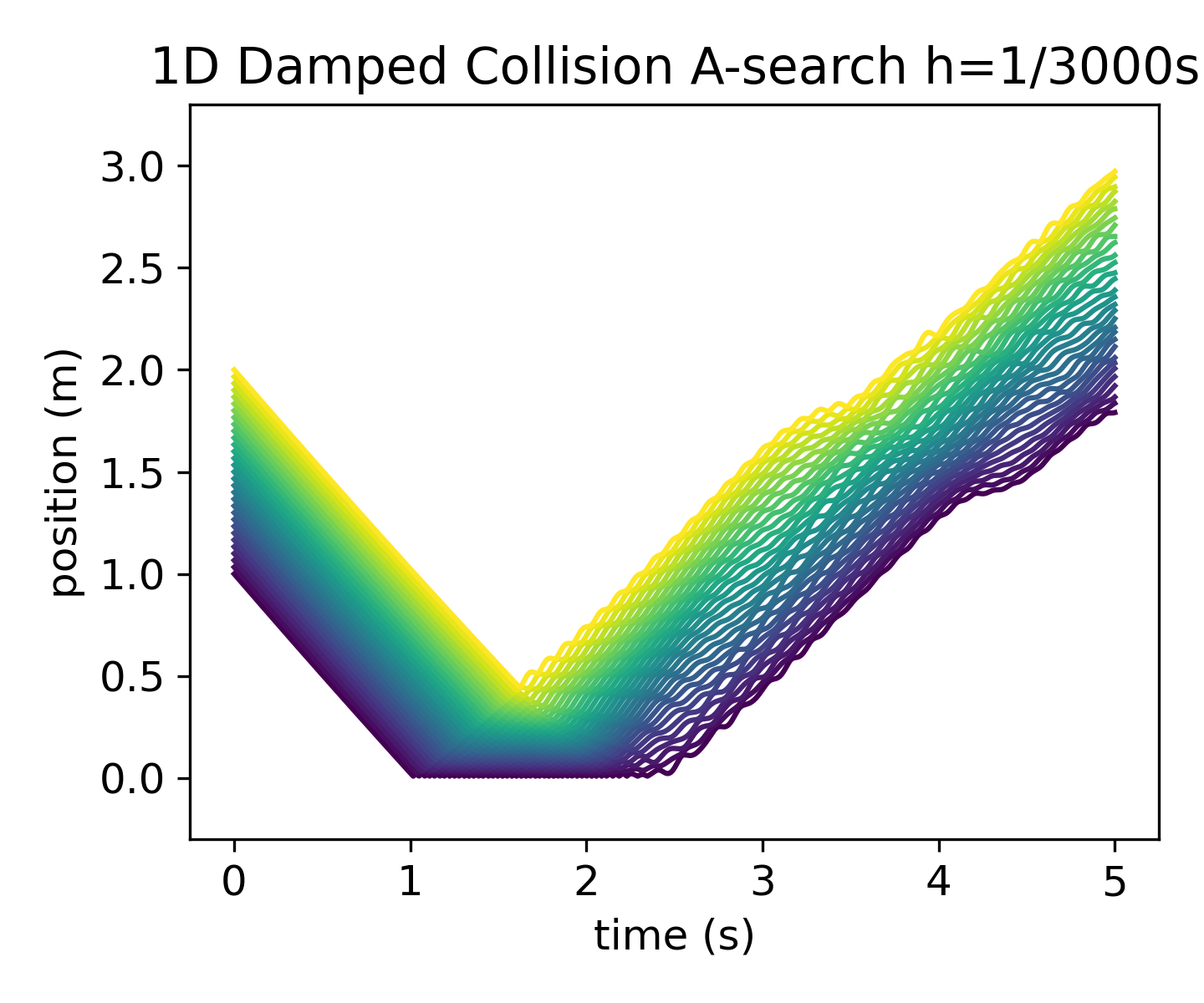}
    \caption{\textbf{1D collision, soft.} As the timestep decreases, BDF2 and A-search converge to the same solution.
  At $h=1/30s$, BDF2 has large dissipation, while A-search kept energies in the low frequencies.
  At $h=1/300s$, A-search have more energy in medium frequencies and a more accurate center of mass velocity. A-1 (not shown) is visually similar to A-search.}
  \label{fig:1dcollisionsoft}
\end{figure*}

We study convergence of \textit{A-1} and \textit{A-search} in one dimension,
with a Neo-Hookean material and IPC barrier force.
We initialize elastic cubes with a length of $1m$,
nodal spacing of $0.033m$, mass $10kg$,
and three varying stiffnesses corresponding to a 
material wave speed of $\sqrt{E/\rho} = 1m/s$,
$10m/s$, and $100m/s$.
We give the cube an initial velocity of $1m/s$, energy $5J$, and let it bounce off a wall (Fig. \ref{fig:1dcollisionsoft}).

We then repeat the above example with semi-implicit Rayleigh damping (Fig. \ref{fig:1dcollisionsoft}).
The kinematic friction coefficient is $\mu k = 0.05 s^{-1}$,
which would lead to a final velocity of $0.78m/s$ for a freely moving body without deformation.
However, the compression during the collision leads to more friction,
and the final velocity of the convergent solution is approximately $0.64m/s$.

Additionally, we compute the linear eigenmodes of the discretized cube at rest,
and project the simulated results on these eigenmodes to obtain an approximate energy spectrum.
The energy in the energy specturm is computed from the linearized system,
so this energy reflects amplitude of the eigenmodes instead of the true Neo-Hookean energy,
hence the spectrum should be interpreted qualitatively and not quantitatively. 

\begin{table*}[]
\begin{tabular}{l|ccc|ccc|ccc|ccc}
\toprule
\multicolumn{1}{c|}{Final velocity} & \multicolumn{3}{c|}{Soft}  & \multicolumn{3}{c|}{Medium}   & \multicolumn{3}{c|}{Stiff} &       \multicolumn{3}{c}{Damped Soft}        \\
                     \multicolumn{1}{c|}{(m/s)} & BDF2 & A1 & A-search & BDF2 & A1 & A-search & BDF2 & A1 & A-search & BDF2 & A1 & A-search \\
                     \midrule
h=1/30s              & 0.776                    & 0.810                  & 0.819                       & 0.742                      & 0.643                  & 0.632                       & 0.795                     & 0.978                  & 0.999                       & 0.641                           & 0.661                  & 0.667                       \\
h=1/300s             & 0.760                    & 0.831                  & 0.816                       & 0.937                      & 0.934                  & 0.931                       & 0.842                     & 0.782                  & 0.788                       & 0.624                           & 0.634                  & 0.636                       \\
h=1/3000s            & 0.849                    & 0.849                  & 0.862                       & 0.965                      & 0.971                  & 0.972                       & 0.993                     & 0.992                  & 0.992                       & 0.639                           & 0.629                  & 0.641                       \\
h=1/30000s           & 0.864                    & 0.866                  & 0.856                       & 0.977                      & 0.977                  & 0.972                       & 0.999                     & 0.999                  & 0.999                       & 0.643                           & 0.641                  & 0.643           \\ \bottomrule 
\end{tabular}
  \caption{\textbf{Final center of mass velocity of 1D collision.} A-1 and A-search noticeably perform better than BDF2 in matching the final velocity in the soft case at $h=1/30s$ and $h=1/300s$. }
  \label{fig:1d}
\end{table*}

As the time step increases, 
and in all cases including friction,
the three numerical integrators BDF2, A-1, and A-search converged to approximately the same solution,
while A-1 and A-search additionally gave reasonable results at small time steps.
The trapezoidal method would instead see a numerical explosion at $h=1/300s$.

\begin{figure*}[!htb]
  \centering
  \includegraphics[width=0.49\linewidth]{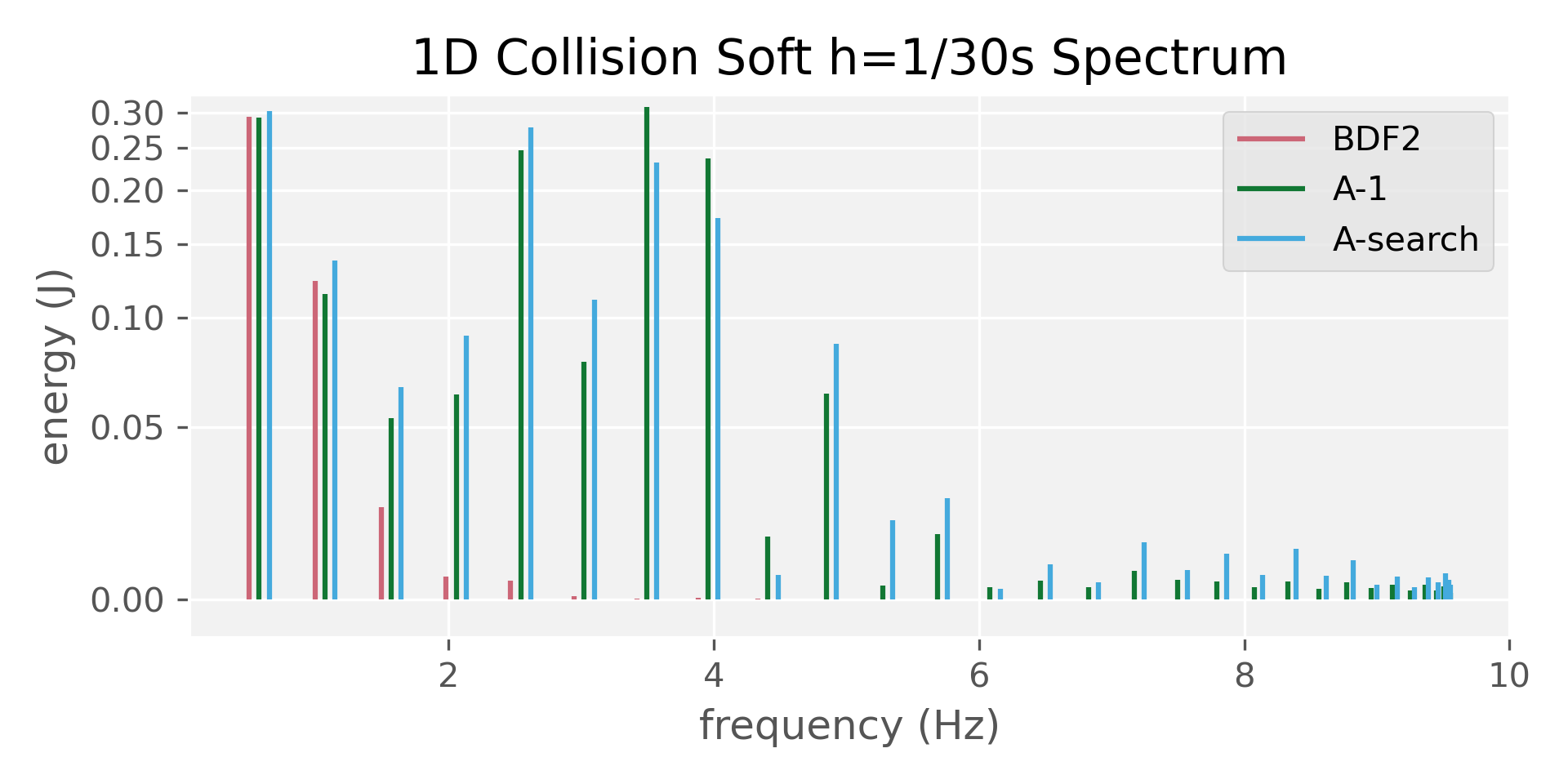}
  \includegraphics[width=0.49\linewidth]{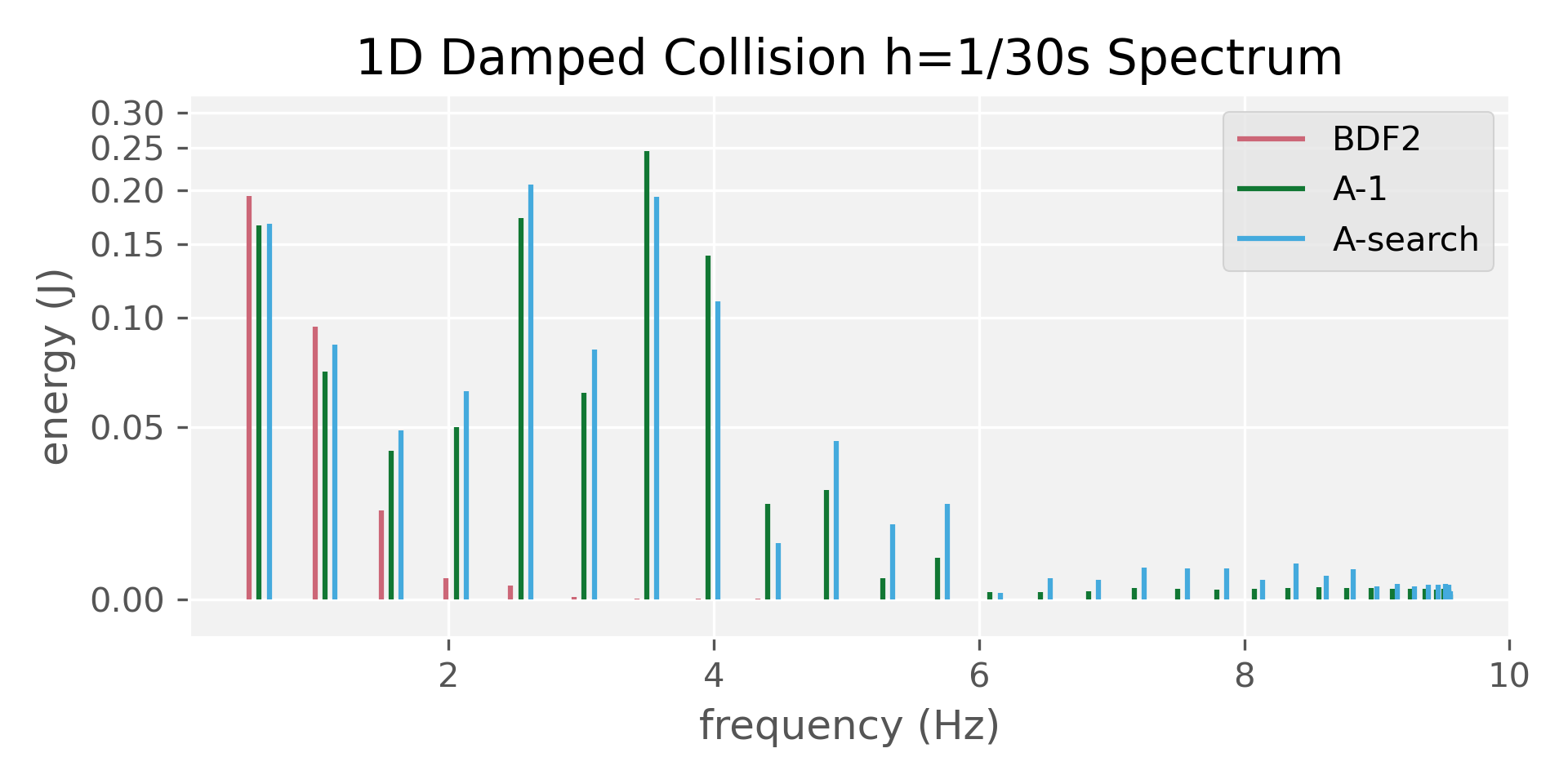}
  \includegraphics[width=0.49\linewidth]{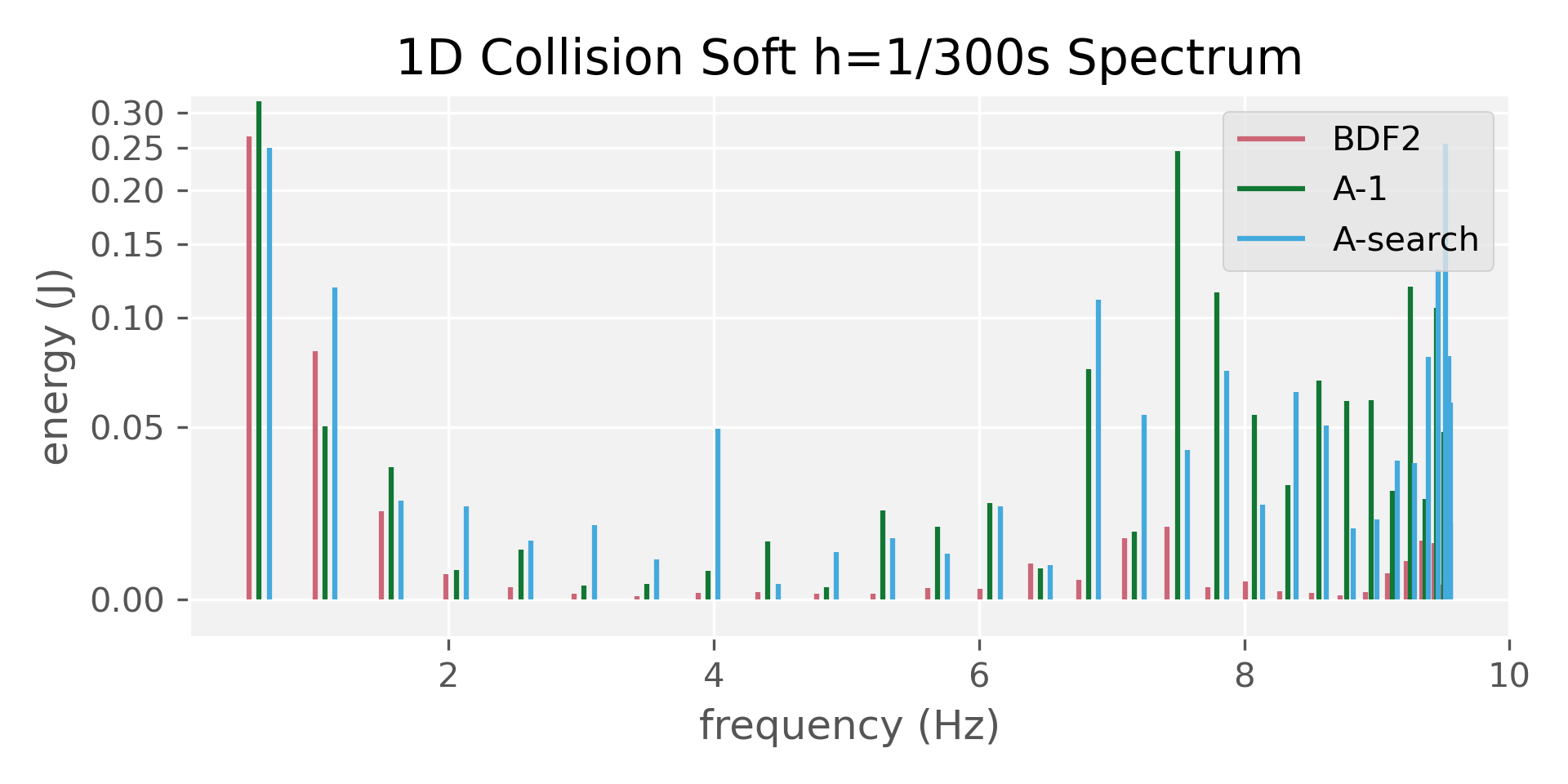}
  \includegraphics[width=0.49\linewidth]{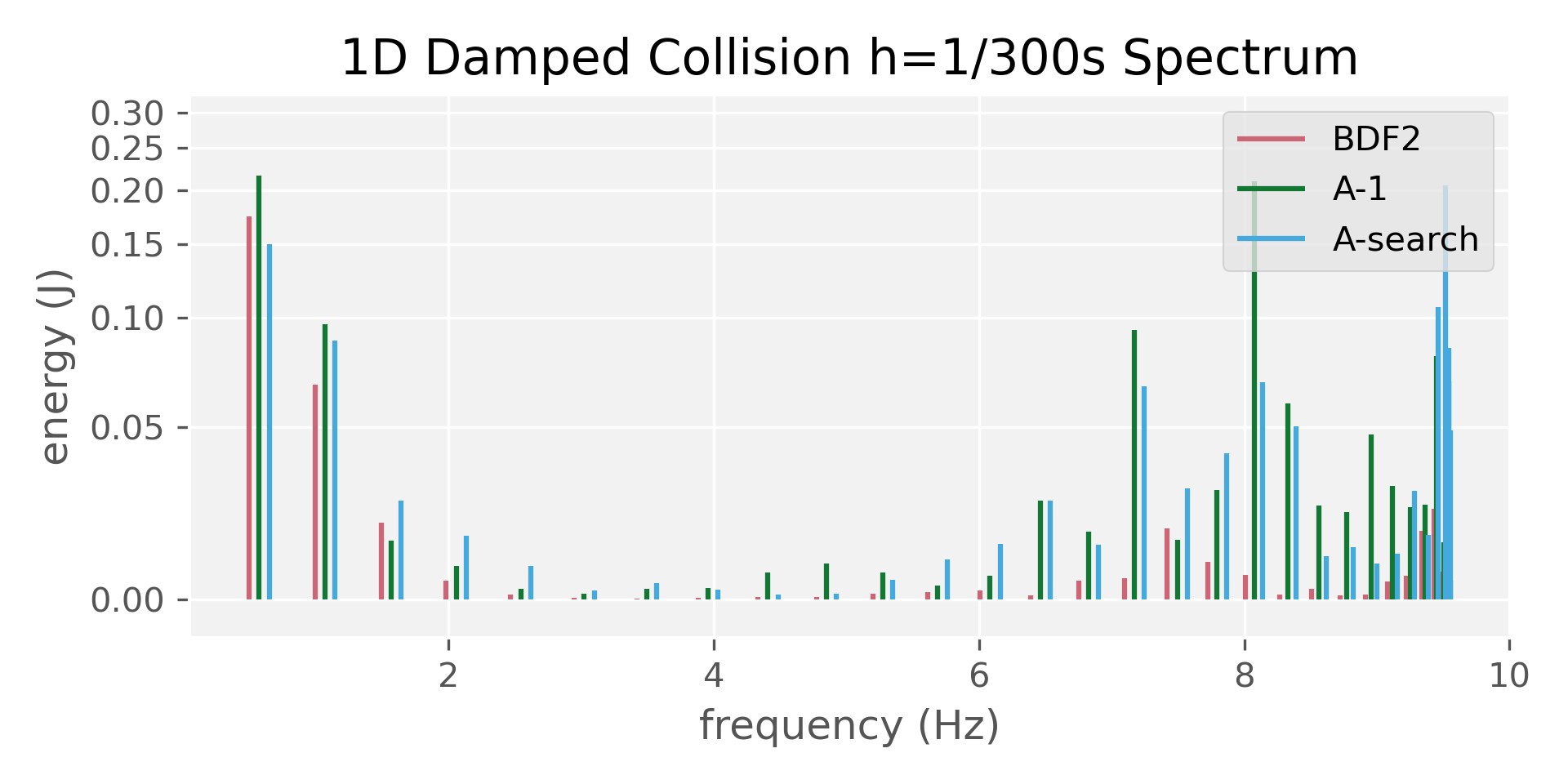}
  \includegraphics[width=0.49\linewidth]{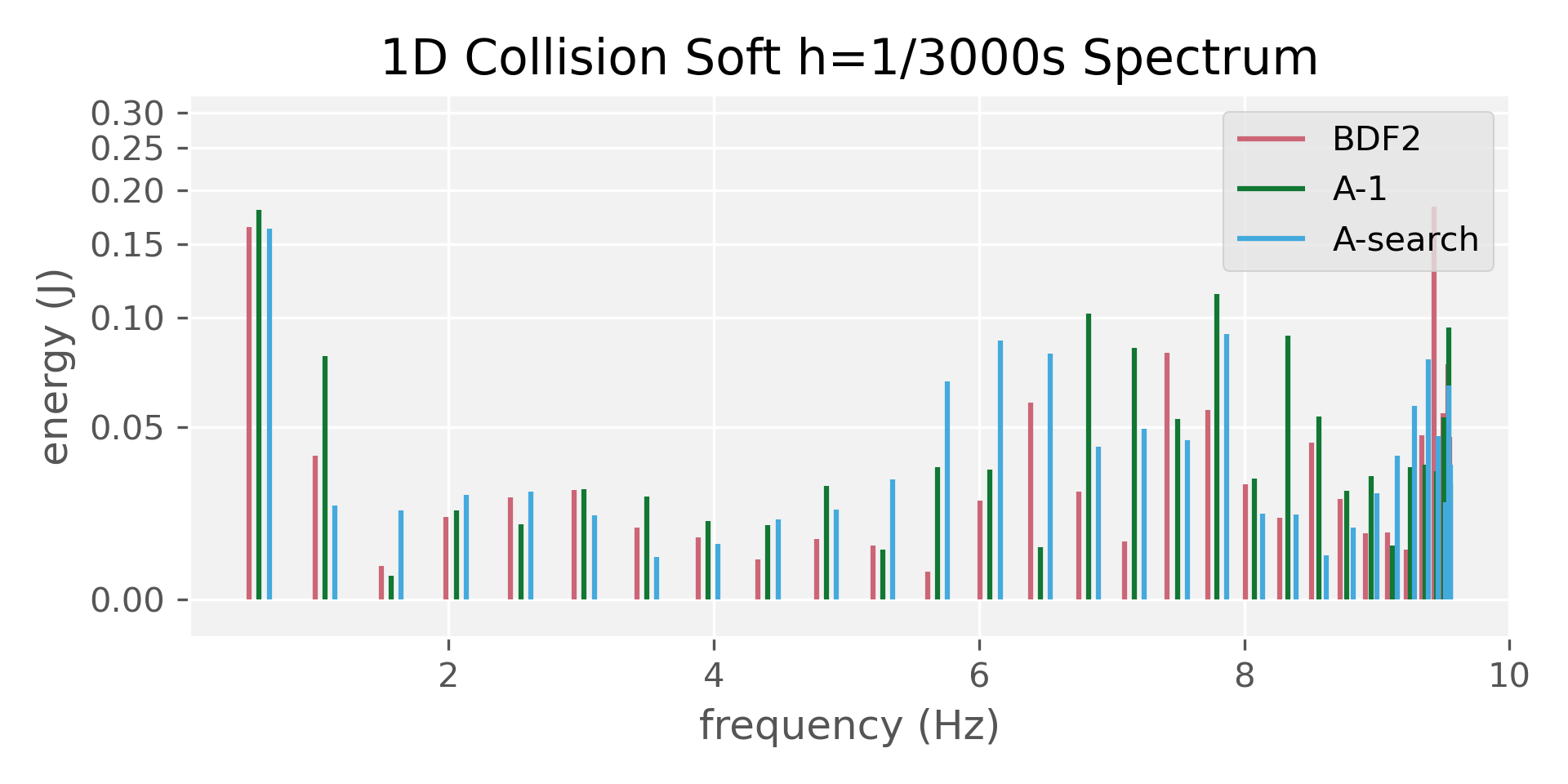}
  \includegraphics[width=0.49\linewidth]{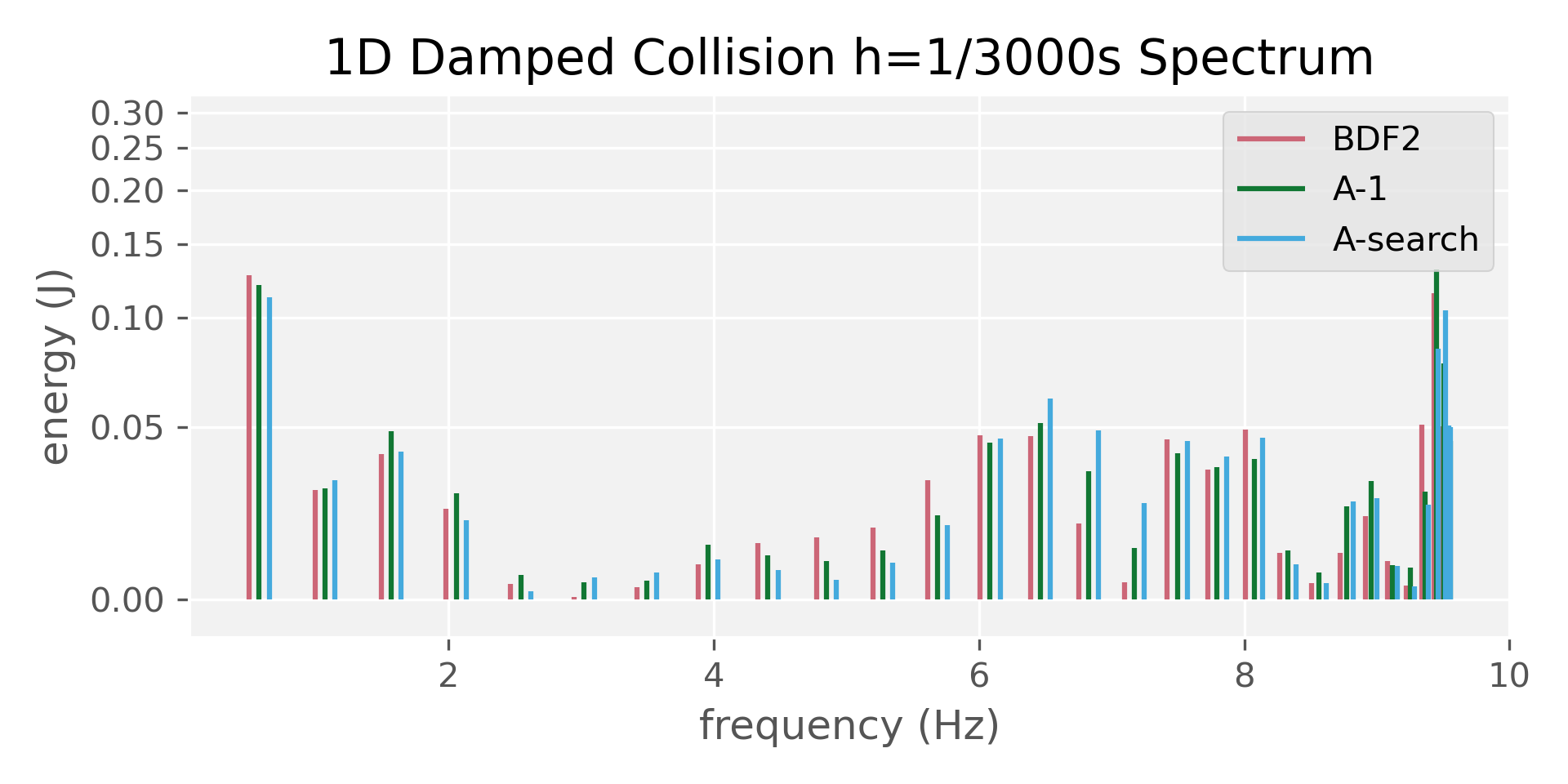}
  \includegraphics[width=0.49\linewidth]{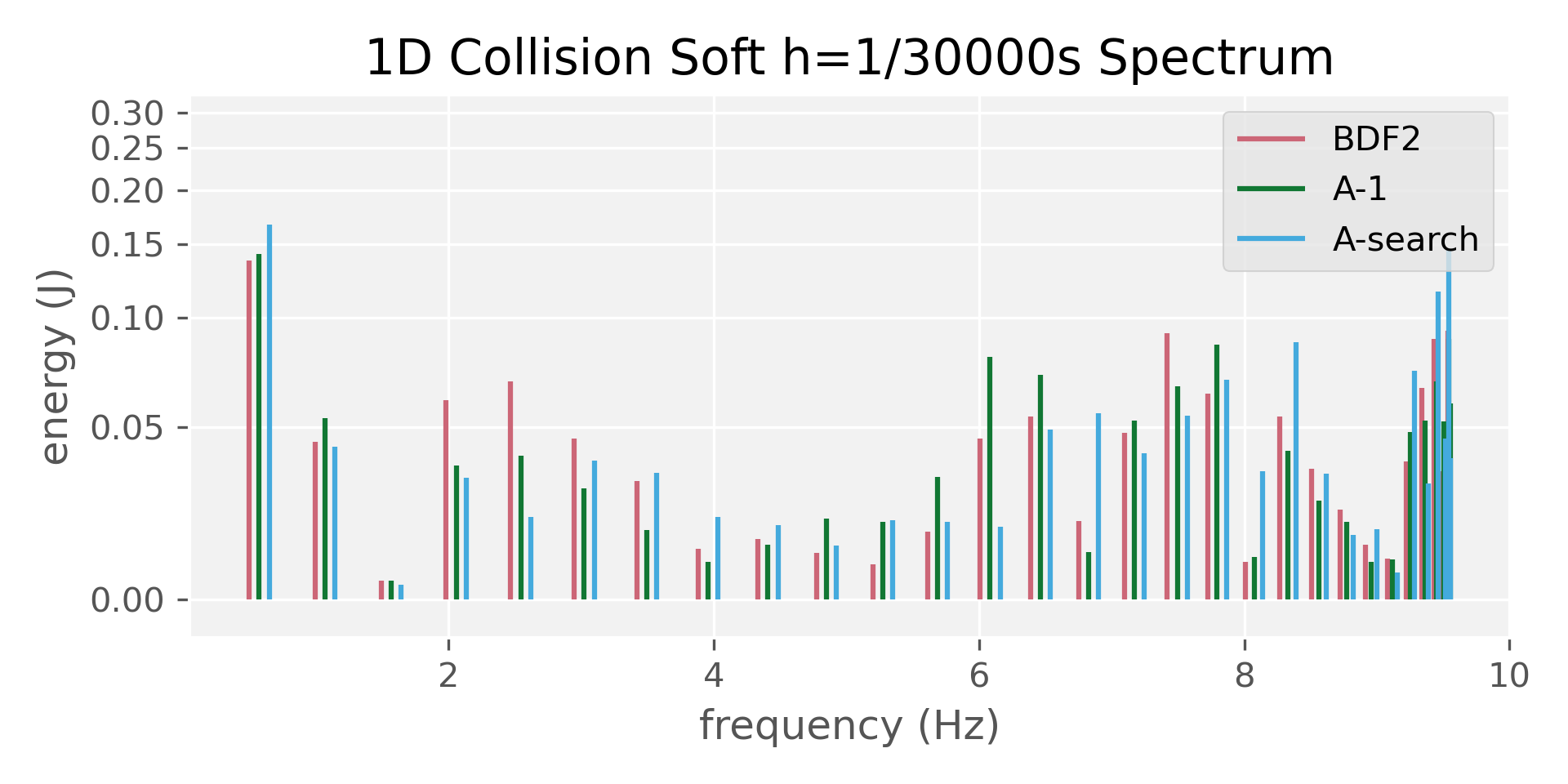}
  \includegraphics[width=0.49\linewidth]{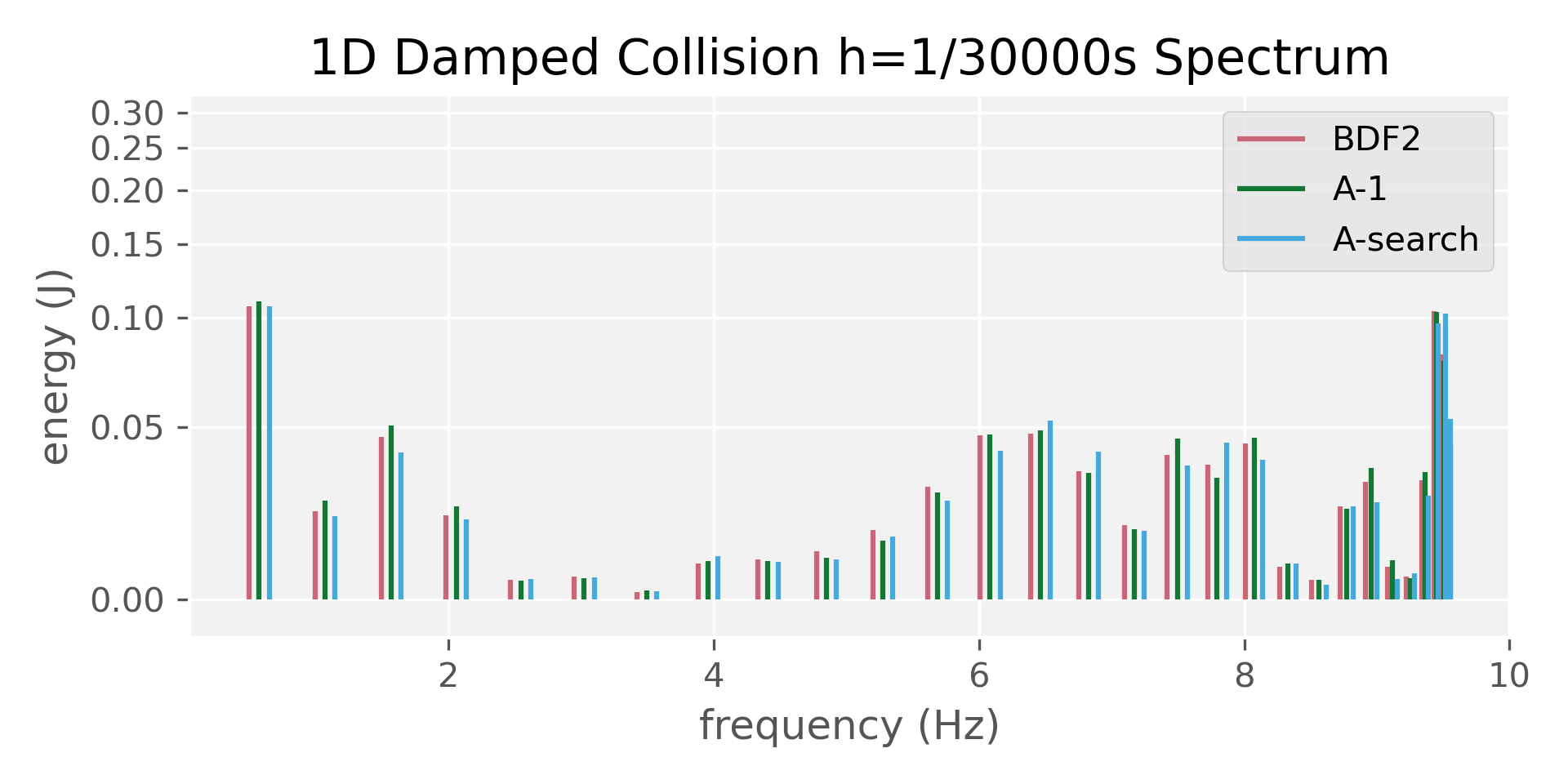}
  \caption{\textbf{1D collision, soft, without and with damping.}
  As the timestep decreases, BDF2, A-1, and A-search converge to the same energy spectrum.
  At $h=1/30s$ and $h=1/300s$, BDF2 has large numerical dissipation, 
  while A-1 and A-search kept energies in the medium frequencies.}
  \label{fig:1dmodessoft}
\end{figure*}

In the soft stiffness case, BDF2 saw an excessive dissipation of energy at $h=1/30s$ and $h=1/300$s.
It only had slow frequencies oscillation in the lowest frequency modes,
and a much smaller center of mass velocity compared to the convergent solution.
On the other hand, A-1 and A-search preferred a more medium frequency vibration (Fig. \ref{fig:1dmodessoft}),
arguably visually closer to the convergent solution, and had a more accurate final center of mass velocity (Table. \ref{fig:1d}).
When friction is added, BDF2 happens to have a more accurate final center of mass velocity.
But again, there is significant lost of energy.
A-1 and A-search kept those energies in the medium frequencies,
producing a more visually desirable solution.
Though this is a collision scene, the same qualitative observations apply to other forms of desirable elastic deformation of low stiffness or frequencies.

On the other hand, for the medium and stiff cases,
the significant numerical stiffness encountered is fairly unique to collision. 
The first observation is that, as the stiffness increases,
A-1 and A-search requires a finer time step size to converge to the correct spectrum (Fig. \ref{fig:1dmodesstiff}).
This is to be expected, especially since A-1 and A-search are first rather than second order accurate. 
However, the amount of vibrational energy also is less significant. 
In the most stiff case, there is a curious resonance effect.
The collision is resolved very well by A-1 and A-search at the very large time step $h=1/30s$,
with negligible energies in any oscillation modes,
unlike BDF2, which sees a loss of velocity (Table. \ref{fig:1d}).
This matches the large time step collision behavior discussed earlier,
as the stiff block effectively behaves as a point mass.
In medium time steps, resonance occurs, as oscillations of the time step frequency is permissible. 
And at small time steps, convergence occurs.
In all cases, the simulation is stable, for both the symplectic A-1 and the energy controlling A-search.
One additional observation is a discrepancy between A-1 and A-search's spectrum, which will be discussed next.

\begin{figure*}[!htb]
  \centering
  \includegraphics[width=0.49\linewidth]{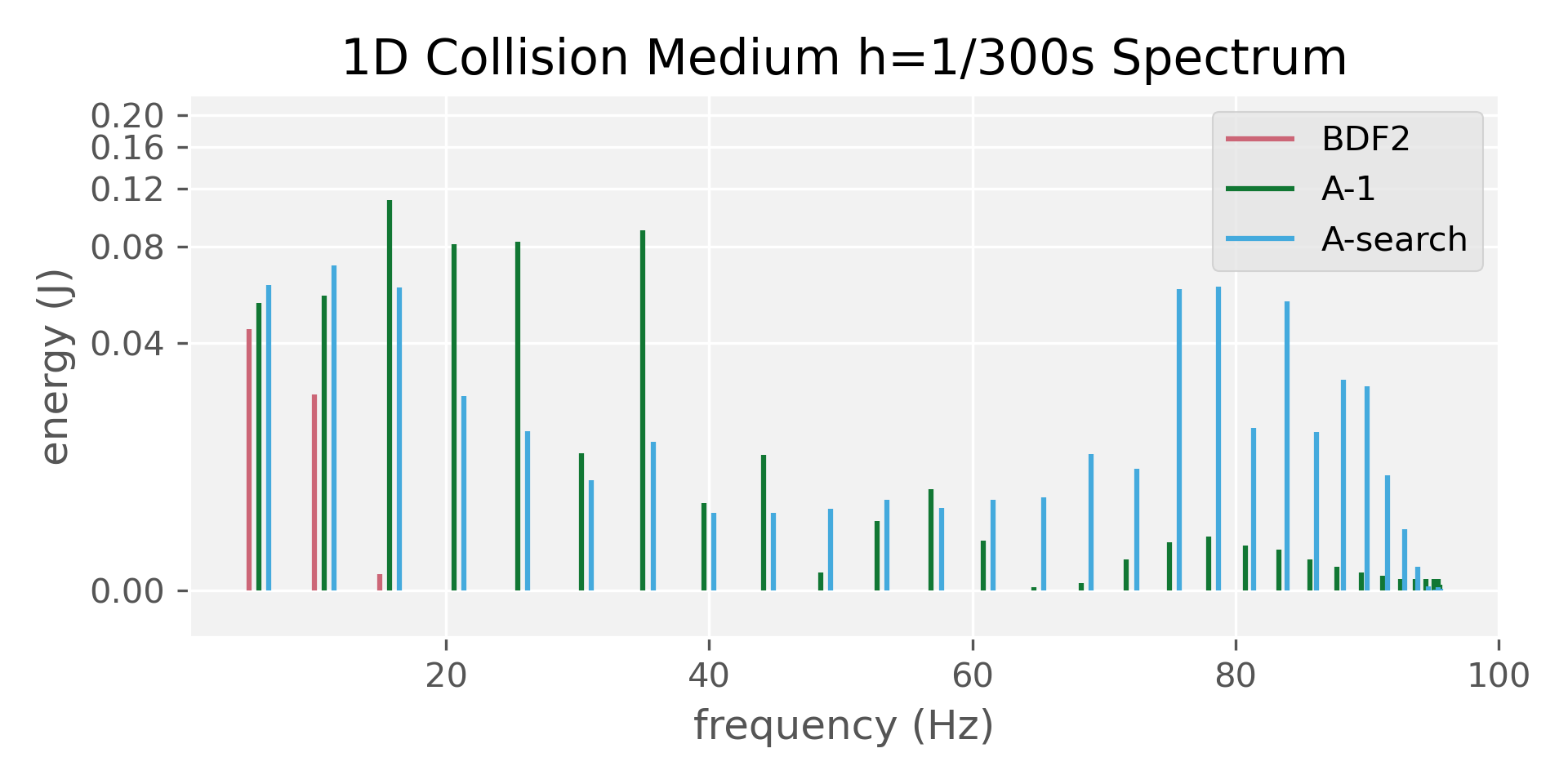}
  \includegraphics[width=0.49\linewidth]{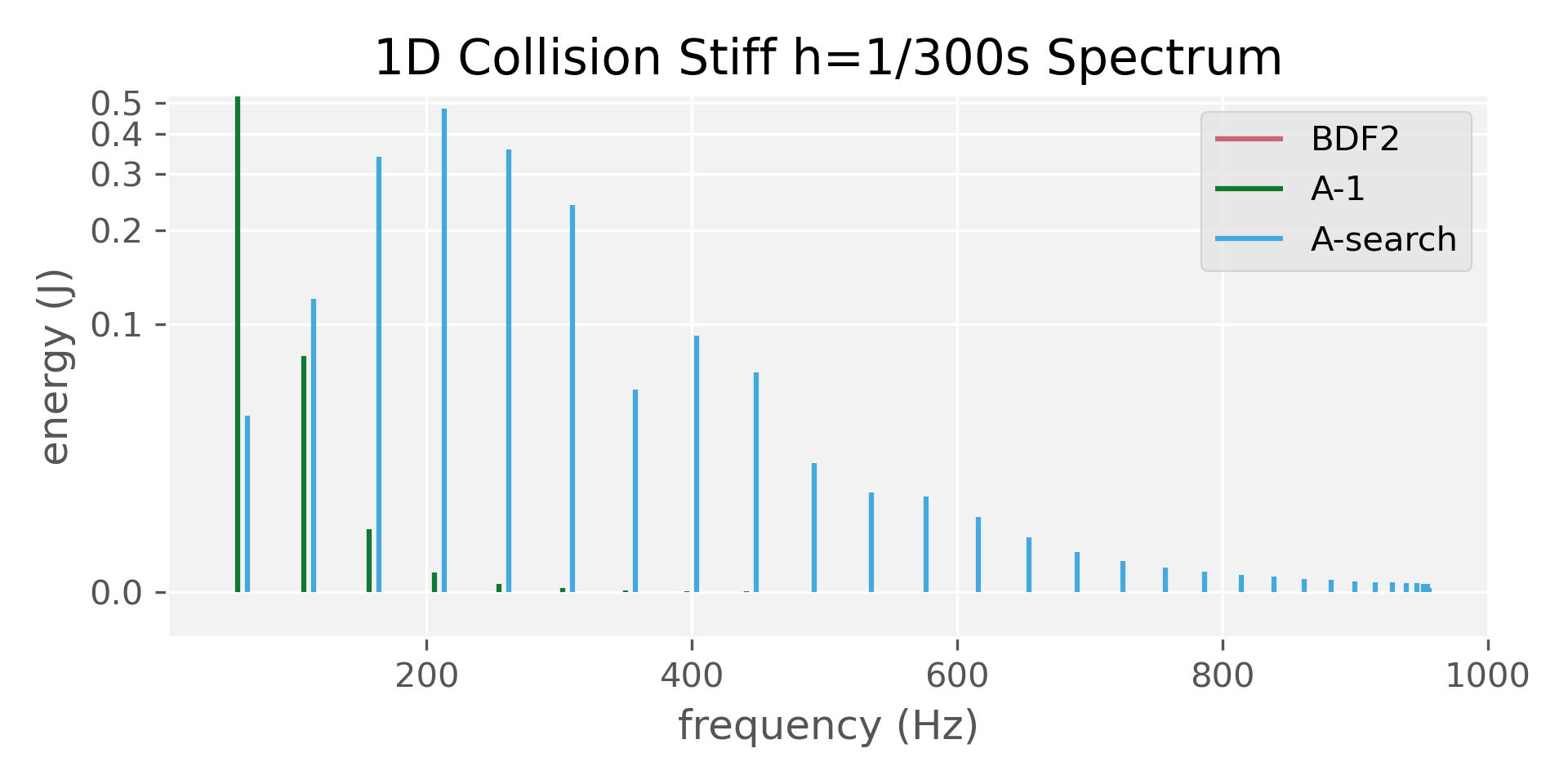}
  \includegraphics[width=0.49\linewidth]{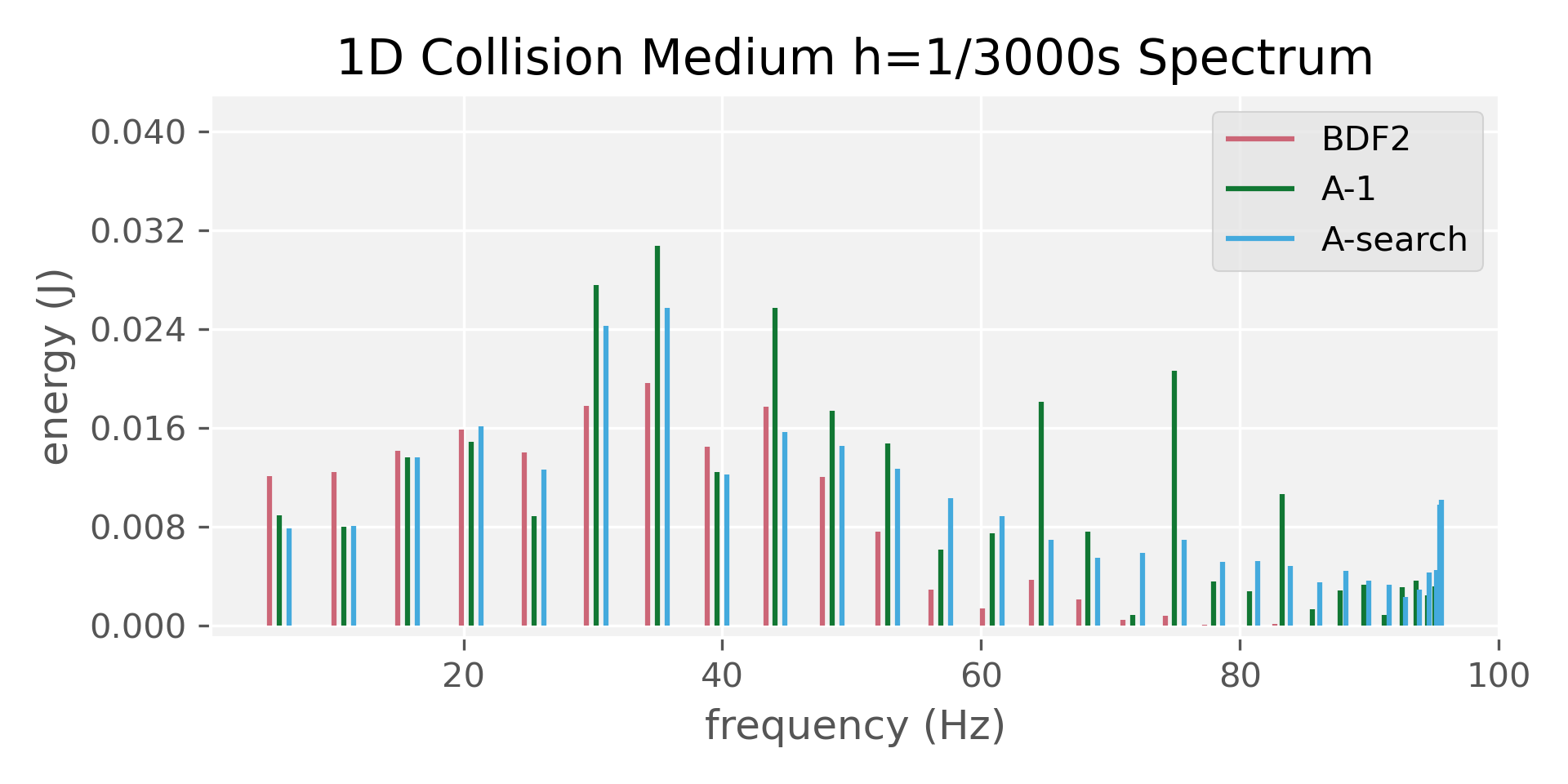}
  \includegraphics[width=0.49\linewidth]{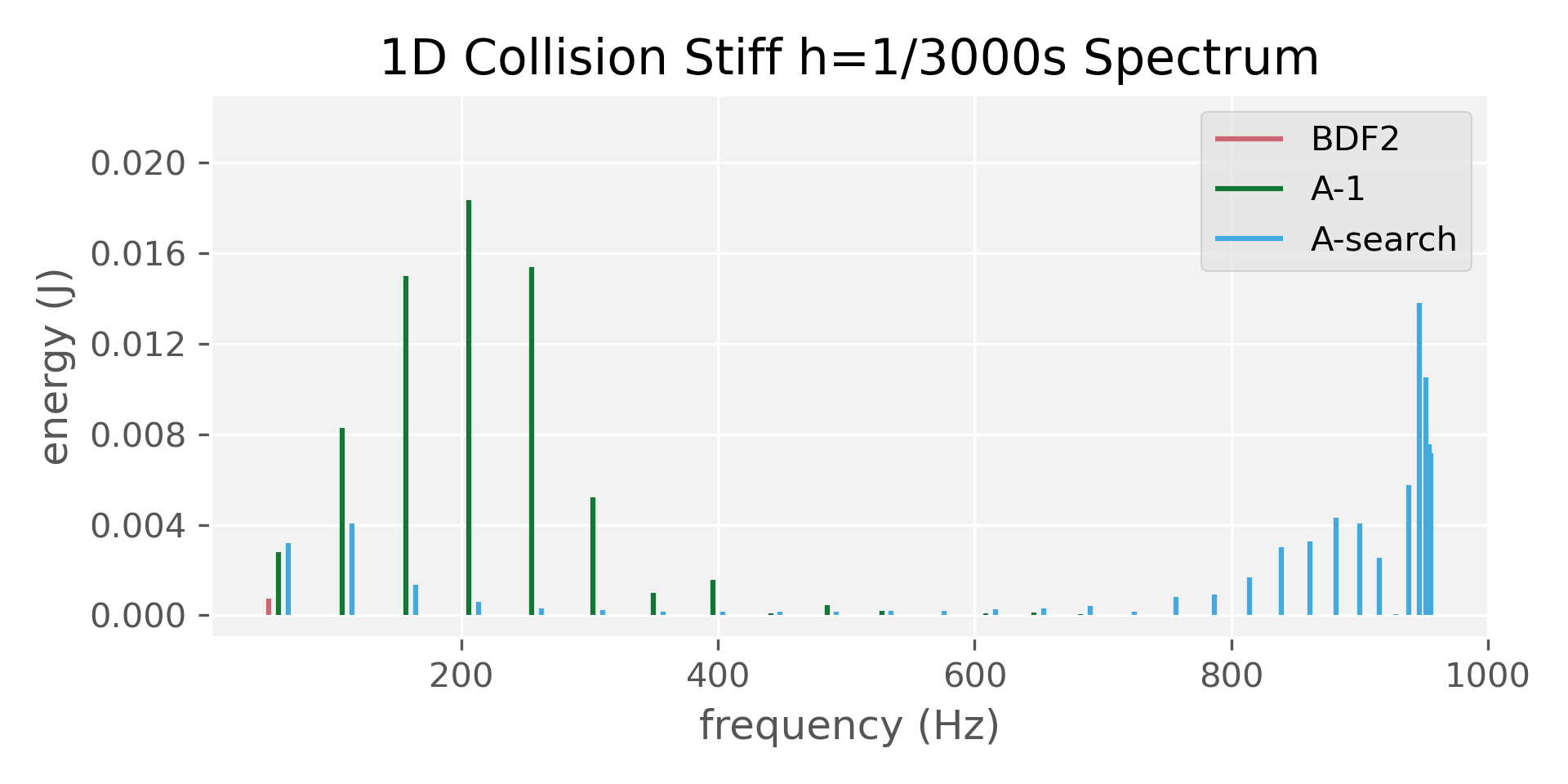}
  \includegraphics[width=0.49\linewidth]{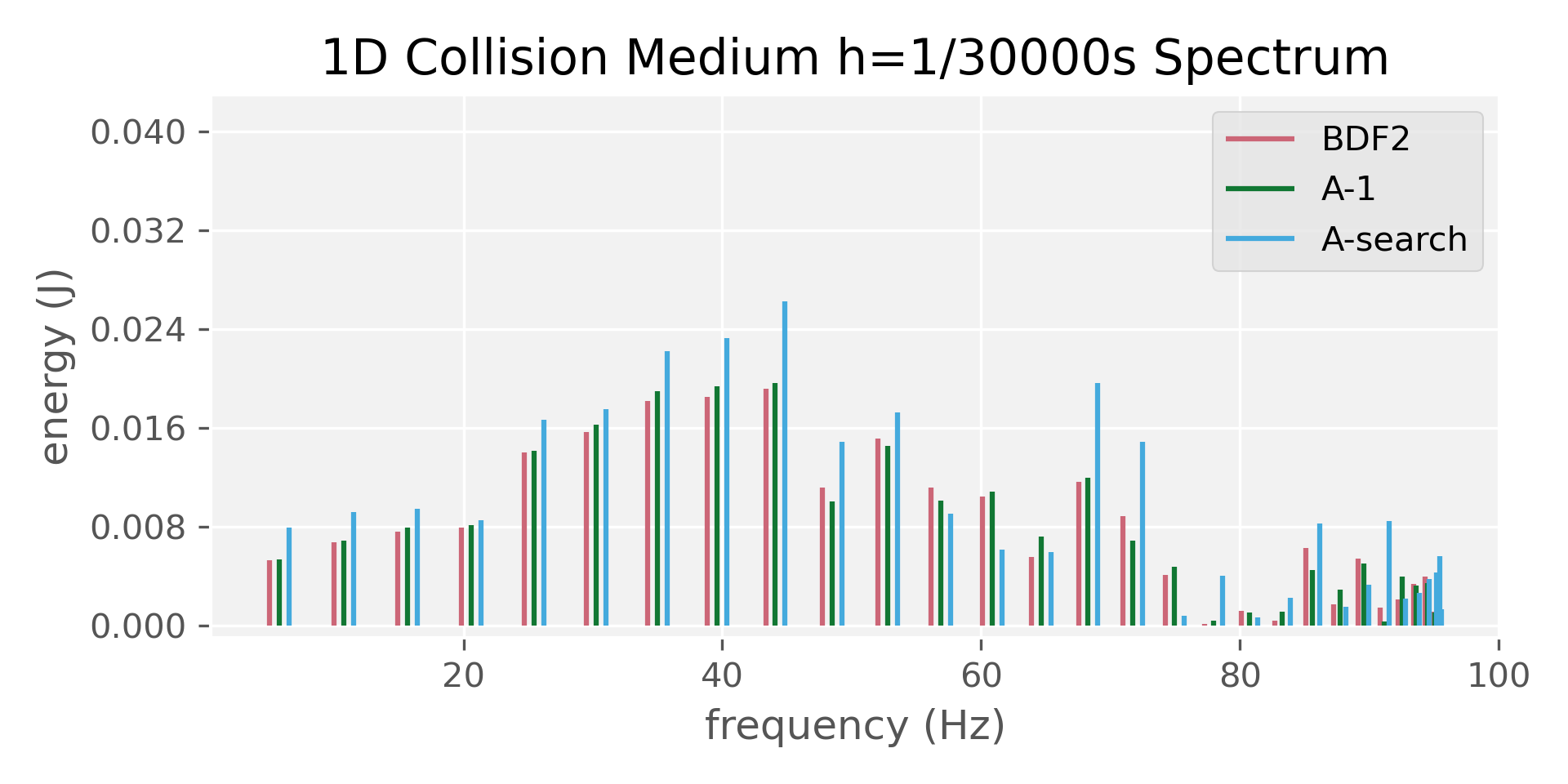}
  \includegraphics[width=0.49\linewidth]{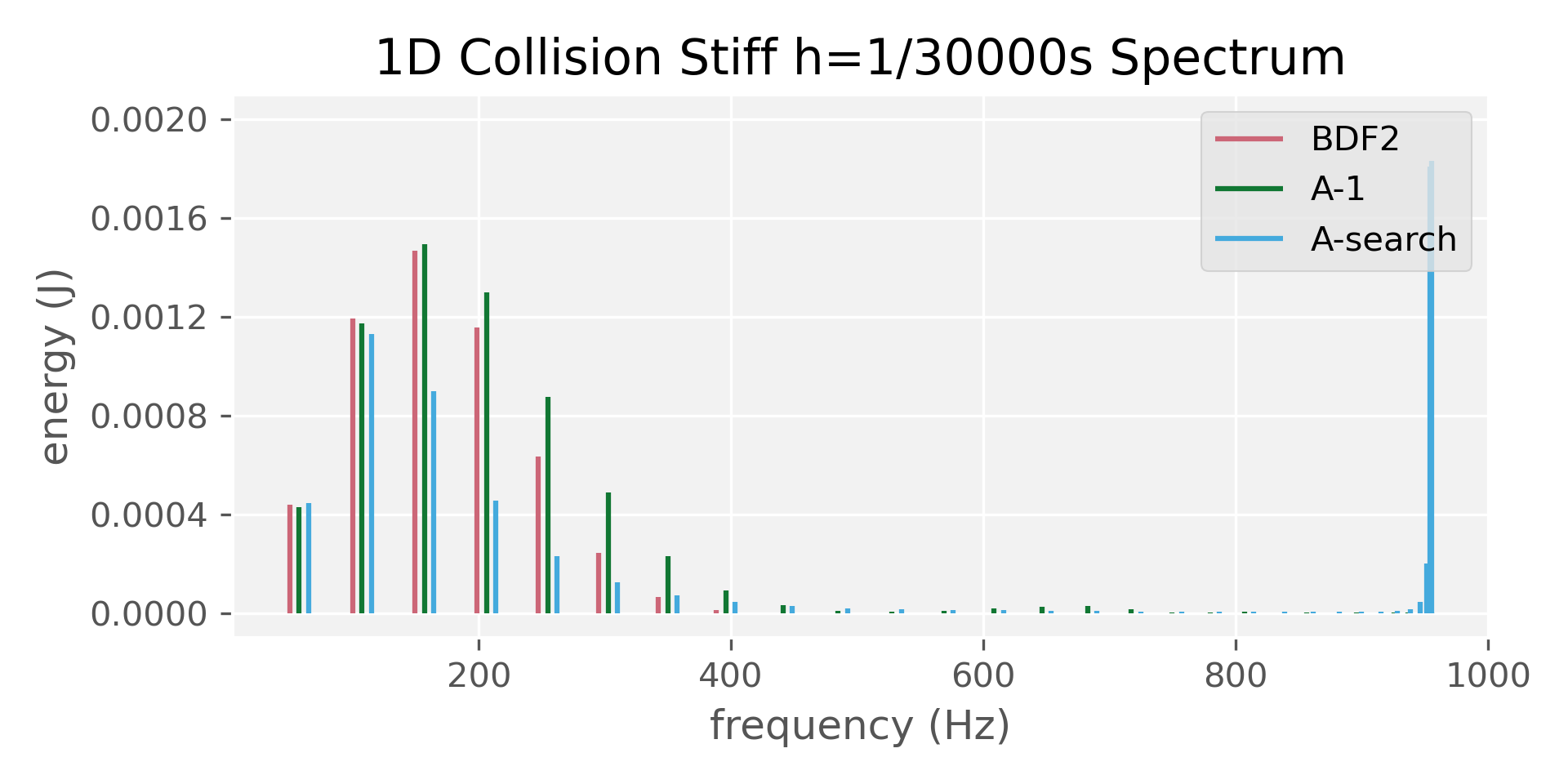}
  \caption{\textbf{1D collision, medium and stiff.} Note the changing axis scales.
  As the scene gets stiffer, A-1 and A-search converge slower due to its first-order accuracy. However, the amount of vibrational energy also becomes less significant. }
  \label{fig:1dmodesstiff}
\end{figure*}

\subsection{Alpha Search} \label{subsec:asearch}

In this subsection, we both perform ablation studies regarding A-search.
Though A-1 is an elegant symplectic integrator that depends on no additional parameters,
the general A-search framework allows for many different ways of determining $\alpha$.
Earlier, we have viewed A-search as an interpolation procedure
with only a single extra, arbitrary hyperparameter $\alpha_{max}$.

We may alternatively take a control theory perspective.
Generally we expect the energy of the symplectic A-1 method to fluctuate due to high-frequency components,
and the exact value of $H_{n+1}$ is not necessarily reflective of the long term behavior of the system.
It suffices for $H_{n+1}$ to be close to $H_0$ for all $n$.
If we denoted the energy change due to \textit{A-1} as
\[ \epsilon_{n+1} = H(\tilde{\vx}_{n+1},\tilde{\vp}_{n+1},\vp_{n+1}^*,1) - H_{n}, \]
and 
\[ u_{n+1}(\alpha) = H(\tilde{\vx}_{n+1},\tilde{\vp}_{n+1},\vp_{n+1}^*,\alpha) - H(\tilde{\vx}_{n+1},\tilde{\vp}_{n+1},\vp_{n+1}^*,1) \]
is the energy correction,
then we have the control problem
\[ E_{n+1} \approx H_{n+1}(\alpha) = H_n + \epsilon_{n+1} + u_{n+1}(\alpha), \]
where $\epsilon_{n+1}$ can be thought of as high-frequency noise, 
possibly biased in the long run.
We have in some sense performed operator splitting on the evolution of the energy,
where $\epsilon_{n+1}$ is implicitly evolved, 
while $u_{n+1}(\alpha)$ is explicitly computable in terms of $\alpha$.

The problem is difficult because the evolution of $H$ is stiff, and the current $\epsilon_{n+1}$ has little relation to prior values of $\epsilon$.
Moreover, the range of $\alpha$ is limited to $[\alpha_{min},\alpha_{max}]$,
leading to a constrained problem.
Ultimately, we utilized a simple error cancellation or deadbeat control strategy to set $\epsilon_{n+1} + u_{n+1}(\alpha) = 0$,
while clipping $\alpha \in [\alpha_{min}, \alpha_{max}]$.
Such an error cancellation strategy is prone to the introduction of extraneous energies at the control frequency.
Indeed, blending suffers from this problem,
as seen in the suspended armadillo (Fig. \ref{fig:suspended}),
where the energy correction settles on an artificial motion at the correction frequency or frame rate frequency.

Rather than introducing increasingly sophisticated control heuristics, which are unlikely to be universal or robust across a wide range of scenarios, our approach focuses on improving the underlying time integration scheme. Because \textit{A-1} is significantly smoother and more robust than the implicit midpoints, we find that \textit{A-search} produces substantially smoother behavior than blending \cite{dinev2018stabilizing}, and that a simple error-cancellation mechanism is often sufficient in practice.

Nevertheless, under sufficiently extreme material parameters, residual high-frequency artifacts may still emerge.
It can be seen in the medium and stiff 1D collisions (Fig. \ref{fig:1dmodesstiff}) that there is a mismatch between the spectrum of A-1 and A-search. 
In particular, A-serach has noticeably more energy around the control frequency in the most stiff case, both at $h=1/300s$ and $h=1/3000s$.
Luckily, this effect is only significant in the most stiff setting,
where the total vibrational energy is small.
Henceforth, the amount of artificial energy introduced by A-search is usually not visible and not a significant problem in practice.

Returning to the subject of $\alpha$ search, 
the $\alpha_{max}$ parameter controls the strength of corrections,
but also the amount of energy introduced by A-search. 
Heuristically, $\alpha_{max} = 1.1$ is plenty of correction.
If $\alpha$ must be larger, this indicates that the physical stiffness push the dynamics beyond the effective resolution of the chosen time step, a regime in which no time integrator can fully eliminate unresolved high-frequency response. 

We test a few simple alternatives to our proposed A-search.
The first two revolves around changing the range of $[\alpha_{min},\alpha_{max}]$,
and thus acts as an ablation study to these parameters.
The last utilizes the fact that A-1 has linear eigenvalues $e^{\pm i \pi/3}$,
hence high frequency energy oscillations with a period of $3$,
and attempts to overlook this oscillation, instead focusing on long term corrections.

\begin{enumerate}
    \item Increase $\alpha_{min} = 0.9$, thus perform $\alpha$ search within $[0.9, 1.1]$.
    This makes A-search closer to A-1 and may decrease noise.
    \item Increase $\alpha_{max} = 2.0$, thus perform $\alpha$ search within $[0.0, 2.0]$. This should both increase energy conservation, but also add more artificial noise, making it closer to blending. 
    \item Perform $\alpha$ search within $[0.0, 1.1]$ only when, for three consecutive time steps including the current step, 
    the energy is all larger or smaller than the target.
    Otherwise, set $\alpha = 1$.
\end{enumerate}

We test the proposed strategies on the medium stiffness 1D collision.
As expected, the sparse A-search method which corrects only at most one out of three steps has a spectrum much closer to A-1.
It avoids the high frequency noises for $h=1/300s$ which A-search had that A-1 didn't have.
On the contrary, the $\alpha_{max} = 2.0$ method has a noticeably larger amount of energy at the highest frequencies for $h=1/3000s$.
The $\alpha_{min} = 0.9$ method appears similar to A-search.

We also test long term behavior on the medium stiffness bouncing ball.
Since the sparse version of A-search does not search every step, 
it may miss the energy conservation right before the ball leaves the ground
and lead to a higher rebound.
This illustrates the difficulty of the control problem.
The $\alpha_{min} = 0.9$ method can also suffer from this to a lesser extent.
In any case, the final visual output is visually fairly similar in for the different strategies,
with much more motion than the highly dissipative BDF2.

\begin{figure}[]
  \centering
  \includegraphics[width=1\linewidth]{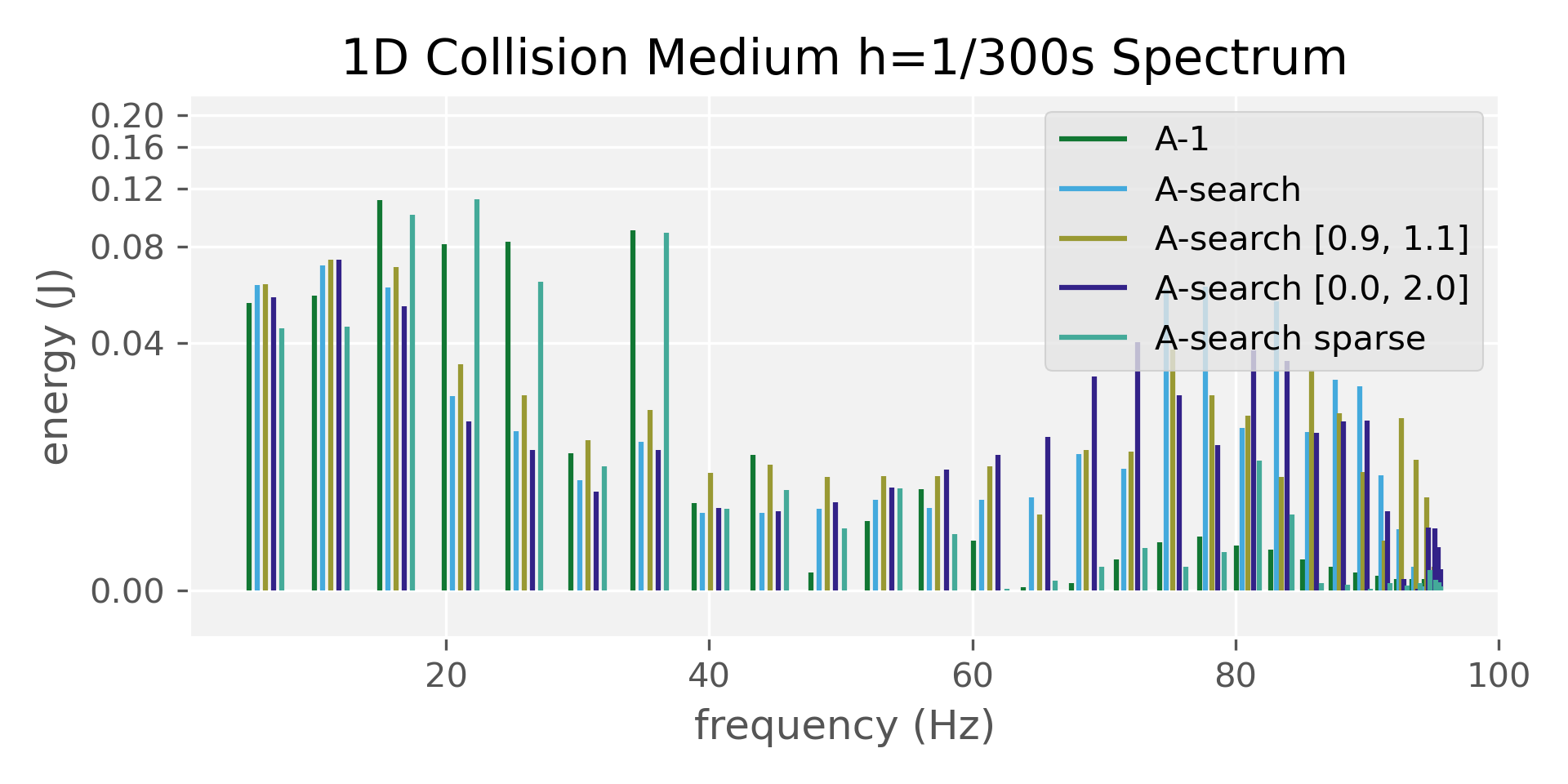}
  \includegraphics[width=1\linewidth]{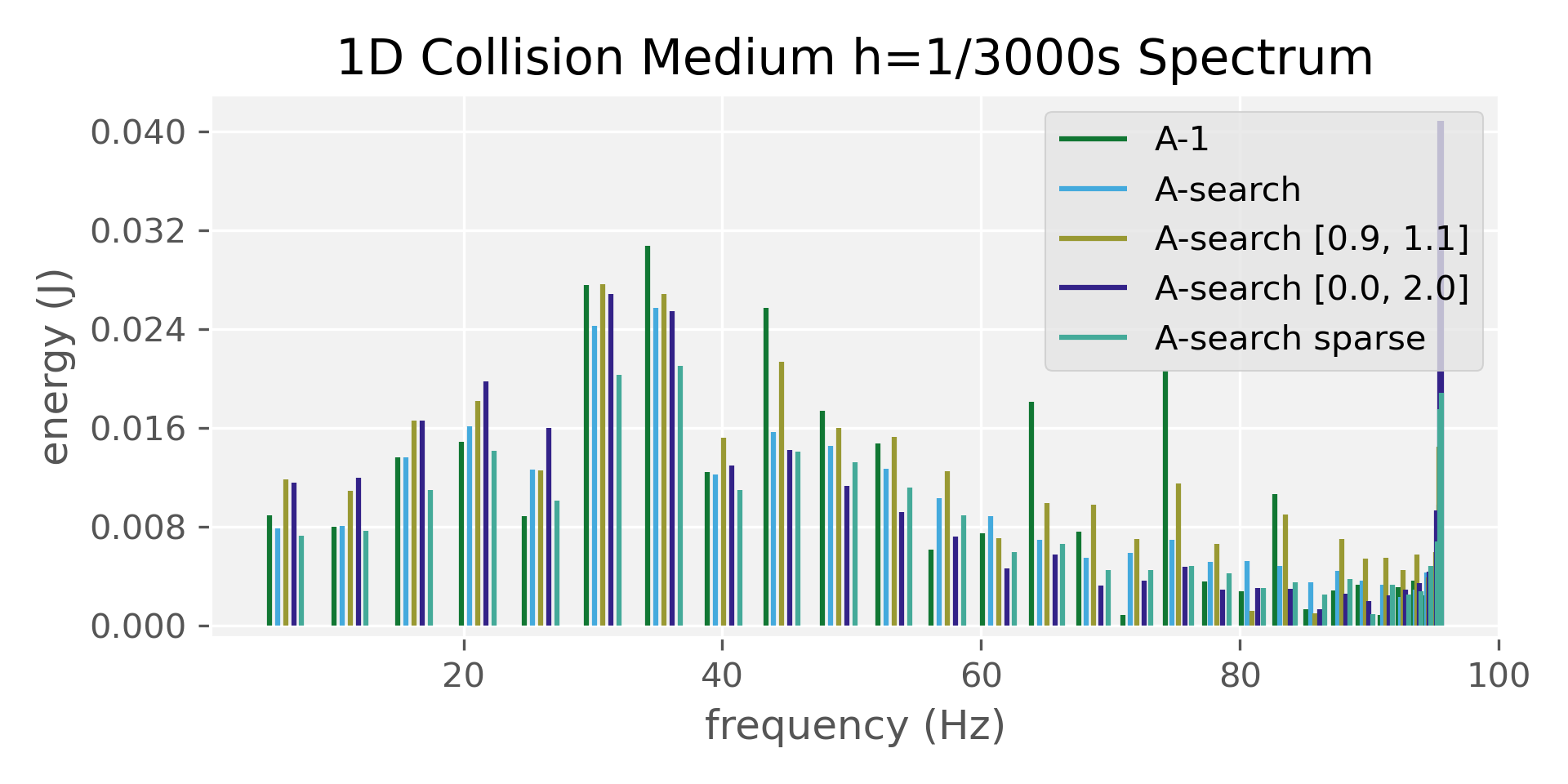}
  \caption{\textbf{Ablation study on A-search (1D collision, medium).}
  The sparse A-search method is noticeably closer to A-1 than A-search at both step sizes, and avoids the high frequency noises for $h=1/300s$.
  The $\alpha_{max} = 2.0$ method has a noticeably larger amount of energy at the highest frequencies for $h=1/3000s$.
  }
  \label{fig:1dmodesstiff}
\end{figure}

\begin{figure}[ht]
  \centering
  \includegraphics[width=1.0\linewidth]{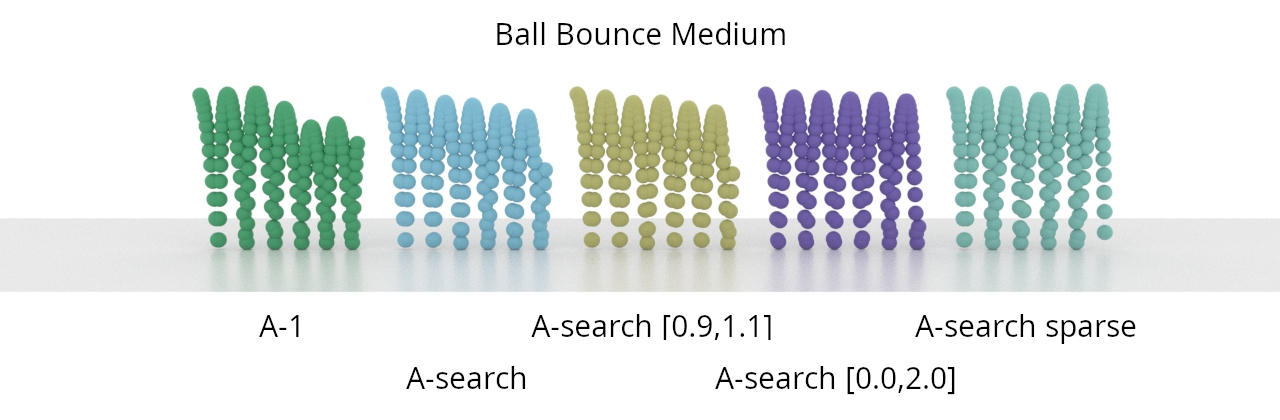}
  \caption{\textbf{Ablation study on A-search (bouncing ball, medium)}. Different search intervals and a sparse variant produce similar rebound behavior, demonstrating robustness to parameter choices.}
  \label{fig:bounceasearch}
\end{figure}


Ultimately, we decided that the ordinary A-search with $\alpha_{max} = 1.1$ is both simple and versatile enough.
With that said, if one is willing to give up strict energy control,
the sparse A-search method could be a viable alternative.
Finally, in all cases, applying modest energy decay can mitigate much of the issues discussed in this section, without altering the core formulation of the method,
and keeping the results still significantly more energetic than full implicit Euler.

\section{Conclusion}

\paragraph{Summary} We introduced a class of linearly symplectic integrators, the decoupled symplectic methods, basing off of existing Runge-Kutta methods.
By dynamically adjusting quadrature points for momentum interpolation, we obtain a class of energy-controllable integrators, the decoupled $\alpha$-methods.
We show that A-1 and A-search, arising from implicit Euler, enables stable and energetic simulations of elastodynamic systems.
A-search provides a bias of low frequency motion over energy dissipation, 
addressing the limitations and of traditional methods such as implicit Euler and BDF2.
Furthermore, A-search's energy controllable mechanism means that the user can directly interpolate between energy conservation and the natural-looking numerical dissipation due to implicit Euler.

Our method demonstrates stability and dynamical fidelity across a range of scenarios, 
even under challenging conditions involving large deformations and collisions,
while maintaining computational efficiency with large time steps. 
Furthermore, the integrator seamlessly supports inversion-free and penetration-free constraints and ensures robust handling of contact, making it particularly well-suited for animation and graphics applications.
The results showcase A-search as a versatile and effective tool for high-quality simulation of elastic objects, offering a step forward in balancing energy control, stability, and performance.

\paragraph{Limitations and Future Work}
While our A-search method demonstrates robust performance and stability, several limitations and opportunities for future work remain. 
First of all, A-1 and A-search are limited to first-order accuracy. 
Investigating higher-order decoupled symplectic integrators via other Runge-Kutta methods presents an exciting direction for extending our approach. 
One may also think about combining multistep methods with the decoupled framework.
Naively, one can derive a decoupled multistep based off of the BDF2 method, which although is linearly symplectic, is not zero stable or consistent.
Other approaches may be possible.
See Appendix \ref{appendix:higher-order} for more details.

Regarding A-1, a better theoretical understanding of the integrator can be investigated.
We have examined a large time step limit $h \rightarrow \infty$ for a point object,
though a large stiffness limit $E \rightarrow \infty$ for an elastic object is also interesting.
Moreover, our theoretical studies have primarily been limited in 1D.
An analysis of non-convex constraints and force singularities in higher dimensions could be valuable for A-1. 
Regarding A-search, while we have demonstrated that controlling $\alpha$ directly is simple and effective,
one can instead use $\alpha$ as a cheap to compute dimensionless error estimate,
and use it to drive adaptive methods, which may be based on implicit Euler or A-search itself.


Finally, as A-search is only a time integrator, there are many ways of alternating other pieces of the simulation pipeline, which may exhibit symbiotic effects with A-search.
Since A-search tends to dampen high-frequency features in simulations, one could directly remove high-frequency modes through force evaluations on a reduced-order model or coarser representation \cite{lan2021medial}, which could further enhance stability.
At the present, only ABD is considered.
Another promising avenue involves exploring solver accelerations in conjunction with A-search. For example, applying quasi-Newton methods and GPU optimization \cite{huang2024gipc,huang2025stiffgipc},
making A-search even faster.
These enhancements could further broaden the applicability and efficiency of A-search, advancing its utility in complex and demanding simulation scenarios.

\bibliographystyle{ACM-Reference-Format}
\bibliography{main}

\appendix 

\section{Material and Scene Parameters}\label{appendix:material}

For the following material parameters, unless otherwise specified, the mass density is \qty{1000}{\kg\per\meter\cubed}, the Poisson ratio is \qty{0.3}{}, 
gravitational acceleration is $g =$ \qty{9.8} {\meter\per\second\squared},
the IPC barrier stiffness is $\kappa =$ \qty{1e5}{\newton\per\meter} and contact distance threshold is $\hat{d} =$ \qty{1}{\milli\meter}.
\begin{enumerate}
    \item \textit{Rotating cube.} The cubes have side length \qty{10}{\centi\meter} and initialized with an angular velocity of \qty{15}{\radian\per\second}. The soft and stiff cubes have Young's modulus \qty{5e4}{\pascal} and \qty{1e6}{\pascal} respectively.

    \item \textit{Twisted bar.} The bar has dimensions 20 x 10 x 5\qty{}{\centi\meter} and Young's modulus \qty{1e5}{\pascal}.

   \item \textit{Suspended armadillo.} The armadillo has height roughly \qty{50}{\centi\meter} and Young's modulus \qty{1e5}{\pascal}. The armadillo is fixed by the back, and the majority of the armadillo is initialized with a downwards velocity linearly increasing to roughly \qty{0.5}{\meter\per\second} at the bottom.
    
    \item \textit{Bouncing ball.} The balls have diameter \qty{10}{\centi\meter}, dropped at a height of \qty{1}{\meter}, and the soft, medium, and stiff balls have Young's modulus \qty{1e5}{\pascal}, \qty{1e6}{\pascal} and \qty{1e7}{\pascal} respectively.

    \item \textit{Vibrating membrane.} The membrane has diameter \qty{20}{\centi\meter}, mass density \qty{400}{\gram\per\meter\squared}, membrane Young's modulus \qty{1e5}{\pascal},
    and an initial velocity amplitude of \qty{1}{\meter\per\second}.

    \item \textit{Dice roll.} The dice have side length \qty{10}{\centi\meter},
    dropped from a height of \qty{30}{\centi\meter},
    with an initial horizontal velocity \qty{0.3}{\meter\per\second}.
    The dice have Young's modulus \qty{3e5}{\pascal}.

    \item \textit{Bunny stair drop.} 
    The bunny has length roughly \qty{25}{\centi\meter} and Young's modulus \qty{2e5}{\pascal}. The barrier distance threshold is $\hat{d} =$ \qty{0.5}{\milli\meter}.

    \item \textit{Armadillo trampoline.} 
    The armadillo has height roughly \qty{50}{\centi\meter} and Young's modulus \qty{1e5}{\pascal}. The membrane has diameter \qty{1}{\meter}, mass density \qty{400}{\gram\per\meter\squared}, membrane and bending Young's modulus \qty{5e6}{\pascal}. The barrier distance threshold is $\hat{d} =$ \qty{0.5}{\milli\meter}.
    
    \item \textit{Chain net.} Each link has length roughly \qty{7}{\centi\meter}. The ball has diameter \qty{15}{\centi\meter}. All objects have Young's modulus \qty{1e8}{\pascal}.

    \item \textit{Bunny splash.} The bunny has length roughly \qty{25}{\centi\meter}, mass density \qty{3000}{\kg\per\meter\cubed}, and Young's modulus \qty{1e5}{\pascal}.
    The small blocks have side length \qty{3}{\centi\meter}, Young's modulus \qty{5e4}{\pascal}, and Poisson ratio \qty{0.45}..
    The barrier distance threshold is $\hat{d} =$ \qty{0.5}{\milli\meter}.
\end{enumerate}

\section{Runge-Kutta Methods}\label{appendix:RK}

In general, an $s$-stage Runge-Kutta method applied to an autonomous system takes the form
\begin{align*}
    \vk_i &= f(\vz_n + h \sum_{j=1}^s a_{ij} \vk_j) \text{ for } 1 \leq i \leq s \\
    \vz_{n+1} &= \vz_n + h \sum_{i=1}^s b_i \vk_i.
\end{align*}
The coefficients $(A,b)$ is the Butcher tableau. On the other hand, a partitioned Runge-Kutta method takes the form
\begin{align*}
    \vk_i &= f(\vx_n + h \sum_{j=1}^s a_{ij} \vk_j, \vp_n + h \sum_{j=1}^s \hat{a}_{ij} \ell_j)   \\
    \vell_i &= f(\vx_n + h \sum_{j=1}^s a_{ij} \vk_j, \vp_n + h \sum_{j=1}^s \hat{a}_{ij} \vell_j)   \\
    \vx_{n+1} &= \vx_n + h \sum_{i=1}^s b_i \vk_i \\
    \vp_{n+1} &= \vp_n + h \sum_{i=1}^s \hat{b}_i \vell_i .
\end{align*}
The corresponding Butcher tableau is $(A,b)$ and $(\hat{A},\hat{b})$.


Special classes of Runge-Kutta methods include diagonally implicit (DIRK) methods where the matrix $A$ is lower triangular,
thus enabling $\vk_i$ to be solved sequentially.
A DIRK method is stiffly accurate if the last row of $A$ equals $b$, hence $\vk_s = f(\vz_{n+1})$.
In the case of our Hamiltonian problem $H(x,p) = \frac{1}{2} \Vt p \Vt_{\vm^{-1}}^2 + P(\vx)$, 
a stiffly accurate method will produce a valid position with $P(\vx_{n+1}) < \infty$.
The trapezoidal method is stiffly accurate,
though implicit midpoint is not.
On the other hand, while implicit midpoint is symplectic,
the trapezoidal method is not,
but it is called conjugate symplectic since it is equivalent to implicit midpoint under a change of coordinates in the phase space.
Both trapezoidal and implicit midpoint are compositions of implicit and explicit Euler in different permutations.

\section{Stability}\label{appendix:stability}

\paragraph{Stability of the decoupled symplectic method}
Recall that our decoupled $\alpha$-method $\Xi(x,v)$ is obtained from an original method $\Phi(z)$ and its adjoint $\Psi(z)$.
For testing simple harmonic motion,
up to rescaling it suffices to consider 
the simple harmonic motion $\dot{x} = v, \dot{v} = -x$,
or $\dot{z} = iz$ if we interpret $z = x - iv$.

The underlying Runge Kutta method $\Phi$ leads to an update rule $(x_{n+1}, v_{n+1}) = Q_R(h) (x_n,v_n)$.
Using $z_n = x_n - iv_n$,
$Q_\Phi(h)$ has elements $\RE \ R_\Phi(ih)$ in the diagonal and $\pm \IM \ R_\Phi(ih)$ off diagonal,
where $R_\Phi(h\lambda)$ is the stability function of the 1D test equation $z = -\lambda z$ due to $\Phi$.
Nothing is special here, this is just due to how complex numbers multiply.

Now, we observe that the update rule of the decoupled symplectic method $\Xi_1$ itself has that of $\Phi$ in the first row,
and $\Psi$ in the second row, thus
\[ \det Q_\Xi = \RE \left( R_{\Phi} (ih) \overline{R_{\Psi} (ih)} \right) \]
and 
\[ \Tr Q_\Xi = \RE \left( R_{\Phi} (ih) + R_{\Psi} (ih) \right). \]
Hence, we have reduced linear stability of $\Xi$ to that of $\Phi$ and $\Psi$.

The stability function of a Runge-Kutta method $A, b$ satisfies
\[ R(z) = \frac{\det (I + zA^\dagger)}{\det(1 - zA)} , \]
where $A^\dagger = eb^T - A$ is the matrix for the adjoint Runge-Kutta method
$(A^\dagger, b)$, and $e = (1,\ldots,1)$.
It is often preferred to swap the intermediates,
thus reflecting elements of $A^\dagger$ and $b$ across the middle,
but this is not necessary for our theoretical analysis.
If $\Phi$ had tableau $(A,b)$, and $\Psi$ being its adjoint,
had $(A^\dagger, b)$, then
\[ \det Q_\Xi (z) = \RE \left( \frac{\det (I + zA^\dagger)}{\det(1 - zA)} \overline{\frac{\det (I + zA)}{\det(1 - zA^\dagger)}} \right) = 1 \]
when $z = ih$ is purely imaginary.
This proves that the decoupled symplectic method is \textit{linearly symplectic}.

While we have been working with the symmetric adjoint, it appears also possible to instead take the symplectic adjoint,
whose linear analysis is equivalent.
It is unclear if there are any benefits of using the symplectic adjoint in the decoupled method.
For implicit Euler, the distinction is irrelevant, explicit Euler is both adjoints.

When looking at the trace, 
there is an asymmetry from taking the real part.
If $\Phi$ is implicit Euler and $\Psi$ explicit Euler,
then $R_\Phi = 1/(1-z)$ and $R_\Psi = 1+z$,
so $\RE \ R_\Phi (ih) = 1/(1+h^2)$ but $\RE \ R_\Psi = 1$. 
Therefore, 
\[ \vt \Tr Q_\Xi \vt  = \vt 1 + \frac{1}{1+h^2} \vt \leq 2, \]
and A-1 is \textit{stably symplectic}.

\paragraph{Stability of other methods} 
We now explicitly compute the update rule $Q$ of various methods.
As we are writing out both the position and velocity, it is useful to return to the problem $m \ddot{x} = -k x$ with constants explicit.
With symplectic Euler we have 
 \[ Q^{sym}(h, k) = \begin{pmatrix} 1 - h^2 k/m & h \\ -hk & 1 \end{pmatrix} . \]
Note that $\det Q = 1$, thus symplectic Euler is \textit{linearly symplectic}.
If $\hbar = h^2 k/m \leq 4$, then it is also stable.
On the other hand, for midpoint/trapezoidal we have
\[ Q^{mid}(h, k) = \frac{1}{1+h^2k/4m} \begin{pmatrix} 1 - h^2 k/4m & h \\ -hk & 1 - h^2 k/4m \end{pmatrix} .  \]
Here $\det Q = 1$ and $\Tr Q = 2(4 - \hbar)(4 + \hbar)$. 
So, trapezoidal is \textit{stably symplectic}, unconditional upon $h$. 

For A-search we can compute that
\[ Q^{\alpha}(h, k) = \frac{1}{1+h^2k/m} \begin{pmatrix} 1 & h \\ -hk(1+\alpha h^2k/m) & 1 + h^2 \alpha k/m \end{pmatrix} . \]
Additionally,
\[ \det Q = 1 + \frac{(\alpha-1)\hbar}{1+\hbar}, 
\Tr Q = \frac{2 + \alpha \hbar}{1 + \hbar} \]
and 
\[ \Tr^2 Q- 4\det Q = - \frac{4\hbar + \hbar^2 \alpha (4-\alpha)}{(1+\hbar)^2}. \]
When $\alpha \in [0,4]$, this quantity is less than or equal to $0$, so eigenvalues are complex, unconditional upon $h$. 
When $\alpha = 1$, the method is \textit{stably symplectic}, unconditional upon $h$.

\section{Higher Order Methods}\label{appendix:higher-order}

For higher order decoupled symplectic methods, 
\textit{linear symplecticity} is always satisfied, and we are concerned with stability. 
It is possible to simplify the trace via the matrix determinant lemma. 
Compute
\[ \Tr Q_{\Xi} =  \frac{\det(I+h^2A^\dagger) + \det(I+h^2A^\dagger)}{\det(I-h^2A A^\dagger)} \frac{1}{1+h^2s^2} \]
where 
\[ s = h^{-2} b^T(h^{-2} I-AA^\dagger)^{-1}e. \]
We can consider the limit $h \rightarrow \infty$.
If both $A, A^{\dagger}$ had full rank,
then the first term of $\Tr Q_{\xi}$ would approach $-2$,
while the second term approaches $1$,
hence generically the eigenvalues of $Q_\Xi$ would be $-1$ as $h \rightarrow \infty$.
However, for stiffly accurate methods,
usually $A, A^{\dagger}$ does not both have full rank,
as is the case for implicit Euler or SDIRK2. 
It is possible to construct two stage Runge Kutta methods $\Phi$ such that $\Tr Q_{\Xi}$ approaches arbitrary constants as $h \rightarrow \infty$.
The decoupled symplectic method constructed by SDIRK2 has limiting trace value $-\sqrt{2}-1/2 \approx -1.914$,
though it has a minimum of approximately $-2.16$ at around $h^2 \approx 44$, making it not unconditionally stable.

The difficulty of analysis comes from requiring both higher order accuracy and stability.
Among all two stage, second order, stiffly accurate Runge Kutta methods (this gives two free parameters), none are always \textit{stably symplectic}, 
except when $\Phi$ itself is symmetric or symplectic.
We do not consider these cases, 
since we want asymmetric methods where $\Phi$ is more stable or dissipative,
while $\Psi$ correspondingly is less stable.
However, using a larger amount of stages could overcome this problem.

One may also wonder if the decoupled idea can be combined with multistep methods. 
One can formally define a forwards differentiation formula which is adjoint to the backward differentiation formula BDF2.
Remarkably, the combined method that, given $\vx_{n-1}, \vv_{n-1}, \vx_n, \vv_n$, 
computes $\vx_{n+1}$ via backward differentiation and $\vv_{n+1}$ via forward differentiation is \textit{linearly symplectic} on the phase space of two time steps.
That is, if $\vu_n = (\vx_n, \vv_n, \vx_{n-1}, \vv_{n-1})$, then $\vu_n \mapsto \vu_{n+1}$ is symplectic.
However, as a multistep method, the forward difference formula is not zero stable, and fails to be convergent. 

This issue is generic and holds for all general linear methods.
Indeed, generically a general linear method applied to $\dot{z} = f(z)$ has form
\[ \vu_{n+1} = R \vu_n + h \vphi(f, h, \vu) \]
where $\vu_n = (\vz_n,\ldots,\vz_{n-k+1})$. 
For BDF2 the matrix $R$ is
\[ R_{BDF2} = \begin{pmatrix} 4/3 & -1/3 \\ 1 & 0 \end{pmatrix}. \]
The adjoint must have the matrix $R^{-1}$. 
However, zero stability is equivalent to $\Vt R \Vt \leq 1$,
hence for both the original and adjoint method to be stable,
one must have $\Vt R \Vt = 1$.
Thus conflicts with our desire that the original method is strictly stable.
A multistep method is strictly stable if $R$ only has a single eigenvalue with magnitude $1$, and all other eigenvalues have magnitude smaller than $1$, 
which is necessary for perturbations to be dissipated.

The condition of being symplectic in the higher dimension phase space might also seem too strong to ask,
and one may instead consider the underlying one step method of a multistep method. 
All strictly stable methods have an underlying one step method,
such that the sequence $\vz_{n}$ computed by the multistep method asymptotically approaches the sequence $\vz_{n}$ due to a one-step method \cite{hairer3}.
However, this underlying one step method cannot be symplectic \cite{TANG199383}.
It is unclear whether it is possible to repeat our construction so that some kind of underlying one step method is somehow symplectic.

\section{Single-Particle Collision Tests}\label{appendix:collision}

In this section we will use a toy model of contact to deduct properties of our new integrators against previous integrators. 
From these models we make the following observations:
\begin{enumerate}
    \item Among the non-adaptive methods considered, A-1 is the only method that handles elastic restitution at large time step sizes. 
    \item When performing energy search it is necessary to allow the energy to decrease for two time steps when the collision is being resolved.
\end{enumerate}

We consider a single particle with mass $m$ facing the one-sided barrier potential 
\[ P(x) = 1_{x > 0} \frac{1}{2} kx^2, \] 
and let $k/m = \omega^2$. 
The particle has initial position $x = \beta h$ where $\beta \in [0,1]$ is the phase of the collision and velocity $v = -1$.
The true time the particle spends in the barrier is $\pi/\omega$.
The behavior of the simulation is completely characterized (up to time and position rescaling) by the dimensionless quantity $\hbar = \omega^2 h^2$. 
It is desirable in this example for the final velocity of the particle to be $v \approx 1$, 
or in other words for energy to be conserved.
Implicit Euler only conserves energy well when $\hbar \ll 1$ and otherwise loses a significant portion of energy. 
Symplectic Euler conserves energy well up to $\hbar \leq 4$, 
but not when $\hbar > 4$, as discussed earlier about the eigenvalues. 

\paragraph{Midpoint/Trapezoidal}
For convenience we work with trapezoidal, but the two are equivalent up to a phase shift of $\beta = 1/2$. 
The integrator is symplectic and performs fine when $\hbar \ll 1$,
but if $\hbar \gg 1$, the method suffers from the force discontinuity. 
As $h \rightarrow \infty$, if the potential was symmetric, the integrator consistently takes two steps to solve one period of oscillation,
and the behavior is symplectic.
However, if the potential is only one-sided, 
then there is not enough dampening when the particle flies out, 
and there is a tendency to gain velocity. 
After the first step 
\[ (x_1,v_1) = \left( \frac{4 \beta - 4}{h\omega^2}, 1-2 \beta \right) . \]
After the second step, if $\beta < 2/3$, then the particle leaves the potential with
\[  (x_2,v_2) = \left( 2 h - 3 \beta h, 3-4\beta \right) . \]
If $\beta > 2/3$, then 
\[  (x_2,v_2) = \left(  \frac{8 - 12\beta}{h\omega^2}, 2 \beta - 1 \right). \]
One more step is necessary, where
\[  (x_3,v_3) = \left( 5 \beta h - 3 h, 8 \beta - 5 \right). \]
Note that in both cases the velocity is very wrong.
Moreover, the abrupt change in position while the collision is still being resolved tends to lead to instabilities when simulating a full object instead of a point.
In particular, three-quarter of all possible phases leads to energy gain. 

\paragraph{A-1} A-1 is symplectic and it performs well when $\hbar \ll 1$.
Additionally when $\hbar \gg 1$, the collision consistently takes four steps to resolve.
After the first step, the particle moves slightly into the barrier with 
\[ (x_1, v_1) = \left(  \frac{\beta - 1}{h\omega^2}, -1 \right).  \]
Due the explicit velocity evaluation, no change in velocity so far.
After the second step, 
\[ (x_2, v_2) = \left(  \frac{- 1}{h\omega^2}, -\beta \right).  \]
After the third step, 
\[ (x_3, v_3) = \left(  \frac{-\beta}{h\omega^2}, 1-\beta \right).  \]
Finally, after the fourth step, the particle moves fully out the barrier with 
\[ (x_4, v_4) = \left(  h - \beta h, 1 \right).  \]

While the collision does take longer to resolve, the final velocity is correct.
There is one very important feature of A-1, and it is that energy is not conserved throughout the process.
The potential energy is small,
and the kinetic energy is temporarily lost as seen in the decrease in speed.

\paragraph{A-search}
For our adaptive A-search integrators which search based on energy conservation, 
this may be a problem. 
In order to force energy conservation on all time steps independent of $\beta$, 
it is necessary to allow for arbitrarily large $\alpha$.
This is problematic when simulating an entire scene, 
where a huge $\alpha$ can completely throw off the balance of other objects.
If we instead cap $\alpha$ by $\alpha_{max}$, 
then in general there will be one or two steps where the energy is lost,
but after the fourth time step the energy will return to normal.
It turns out that $\alpha_{max}$, does not matter. 
Still working in $h \rightarrow \infty$,
one can show that for A-search 
\[ v_2 = - \min(b \alpha_{max}, 1), v_3 = 1+v_2,\] 
and always $v_4 = 1$. 
What is more interesting is that this behavior persists even when \textit{A-search} does not estimate the correct energy. Suppose \textit{A-search} assumes that the incoming speed exceeds unity, taking some value $V > 1$. Upon collision with the barrier, and in the limit $h \rightarrow \infty$, the intermediate velocities satisfy
\[
v_2 = -\min(\beta \alpha_{\max}, V), \qquad
v_3 = \max(1 + v_2, -V),
\]
after which the final velocity is still $v_4 = 1$. Remarkably, the outgoing velocity is independent of the energy level assumed by \textit{A-search}.

Therefore, A-search has the natural effect of only dampening bounces and not increasing bounces,
which is very important in a complex scene. 
In practice, for not a point mass, some energy may still be gained during collisions from elasticity,
so it is best to keep $\alpha_{max}$ not too much larger than $1$.

\paragraph{IPC Barrier.}
Now we examine the IPC barrier which is used in practice. Recall that 
\[ P(x) = -\kappa (\frac{x}{\hat{d}} - 1)^2 \log \left( \frac{x}{\hat{d}} \right) 1_{x < \hat{d}}, \]
where $\hat{d}$ is the barrier distance threshold and $\kappa$ the barrier stiffness.
Once again, up to time and position rescaling we may consider, without loss of generality,
$\hat{d} = 1$ and $v = -1$. 
The constant $\kappa$ does effect the quality of the trajectory.
The behavior of the numerical simulation is moreover dependent on the time step $h$. 

Curiously, when midpoint/trapezoidal and A-1 integrators are applied to the IPC barrier,
in the limit $h \rightarrow \infty$,
very similar behavior is observed in comparison to the quadratic barrier,
independent of $\kappa$, whether $\kappa \approx 1$ leads to a roughly cubic barrier potential,
or if $\kappa$ is decreased while $h$ is increased so that the particle penetrates deeply in the logarithmic part of the barrier.
For trapezoidal the collision takes two or three steps to resolve and the final velocity is dependent on the phase of the object but goes up to $3$.
For A-1 the collision takes four steps to resolve and the final velocity is always 1.
Moreover, there are two steps where energy loss occurs.
For A-search one or two steps of energy loss occurs.
It seems like there is some type of universality that may apply for the integrators for all convex barriers, 
not just the two tested above.

\end{document}